\documentclass[11pt,tightenlines,eqsecnum,floats,aps,amsfonts,amsmath,amssymb,nofootinbib,prd,shownopacs,floatfix]{revtex4}
\usepackage[english]{babel}
\usepackage[utf8]{inputenc}
\usepackage{amsmath,amsfonts,amssymb,extarrows,nicefrac}
\usepackage[Symbolsmallscale]{upgreek}
\usepackage{xcolor}
\usepackage{graphicx}
\usepackage[hyperfootnotes=false]{hyperref} 
\usepackage{natbib,doi} 
\usepackage{bbm}
\usepackage{pdfpages}
\usepackage{makecell}
\usepackage{mathtools}
\usepackage{mathrsfs}
\usepackage{stmaryrd}
\usepackage[scr=dutchcal]{mathalfa}
\usepackage{textgreek}
\usepackage{titlesec}

\usepackage{tikz}
\usetikzlibrary{fit}
\usetikzlibrary{patterns}
\usetikzlibrary{decorations.pathreplacing}
\tikzset{
  highlight/.style={rectangle,rounded corners, fill=white, pattern color=gray, pattern=crosshatch dots, fill opacity=0.4,draw,thin,inner sep=0pt}
}
\newcommand{\tikzmark}[2]{\tikz[overlay,remember picture,baseline=(#1.base)] \node (#1) {#2};}

\newcommand{\Highlight}[1][submatrix]{
    \tikz[overlay,remember picture]{
    \node[highlight,fit=(left.north west) (right.south east)] (#1) {};}
}

\newcommand{\nocontentsline}[3]{}
\newcommand{\tocless}[2]{\bgroup\let\addcontentsline=\nocontentsline#1{#2}\egroup}

\renewcommand{\thesection}{\Roman{section}}
\renewcommand{\thesubsection}{\Alph{subsection}}
\renewcommand{\thesubsubsection}{\arabic{subsubsection}}

\makeatletter
\renewcommand{\p@subsection}{\thesection.} 
\renewcommand{\p@subsubsection}{\thesection.\thesubsection.} 
\makeatother

\def\clap#1{\hbox to 0pt{\hss#1\hss}}
\def\mathclap{\mathpalette\mathclapinternal}
\def\mathclapinternal#1#2{\clap{$\mathsurround=0pt#1{#2}$}}

\newcommand{\be}{\begin{equation}}
\newcommand{\ee}{\end{equation}}

\let\originalleft\left
\let\originalright\right
\renewcommand{\left}{\mathopen{}\mathclose\bgroup\originalleft}
\renewcommand{\right}{\aftergroup\egroup\originalright}
\newcommand{\lr}[1]{\left(#1\right)} 
\newcommand{\lrabs}[1]{\left\lvert #1 \right\rvert} 
\newcommand{\norm}[1]{\left\lvert\!\left\lvert #1 \right\rvert\!\right\rvert} 
\newcommand{\e}[1]{\ensuremath{\mathrm{e}^{#1}}}
\newcommand{\es}[1]{\ \ensuremath{\mathrm{e}^{#1}}} 
\newcommand{\ex}[1]{\exp\left(#1\right)}

\renewcommand{\d}{\mathrm{d}}
\newcommand{\pdif}[2]{\ensuremath{\frac{\partial#1}{\partial#2}}}
\newcommand{\dif}[2]{\ensuremath{\frac{\d#1}{\d#2}}}
\newcommand{\infsum}[1]{\sum_{#1=-\infty}^{\infty}} 
\newcommand{\zeroinfsum}[1]{\sum_{#1=0}^{\infty}} 
\DeclareMathOperator{\tr}{tr}
\newcommand{\trb}[1]{\tr\lr{#1}}
\DeclareMathOperator{\sgn}{sgn}
\newcommand{\inv}{{}^{-1}} 
\newcommand{\transp}{{}^{\mathrm{T}}} 

\newcommand{\infint}{\int\limits_{-\infty}^{\infty}\d }

\renewcommand{\Re}[1]{\mathrm{Re}\lr{#1}}

\renewcommand{\i}{\ensuremath{\mathrm{i}}} 
\newcommand{\kchf}[1]{\ensuremath{{}_1\text{F}_1\!\left(#1\right)}} 
\newcommand{\snf}{T_{\mathrm{\gamma}}} 
\newcommand{\cnf}{T^{\mathrm{c}}_{\gamma}} 
\newcommand{\Ptg}{\psi^t_g} 
\newcommand{\cs}{\Psi_m} 
\newcommand{\csnorm}{\norm{\cs}^2} 
\newcommand{\pcsnorm}{\norm{\cs}^2_+} 
\newcommand{\mcsnorm}{\norm{\cs}^2_-} 
\newcommand{\csnormalise}{\norm{\cs}^{-2}} 
\newcommand{\qr}{{\hat q}^{i_0}_{I_0}(r)}

\newcommand{\V}{{\hat V}} 
\newcommand{\Vr}{{\hat V}^r} 
\newcommand{\lp}{\ell_{\mathrm{P}}}
\newcommand{\lpz}{\lp{}^{3r}|Z|^{\frac{r}{2}}}

\newcommand{\rhalf}{\frac{r}{2}}
\newcommand{\opr}{\frac{1+r}{2}}
\newcommand{\sumset}[1]{\sum_{\left\{#1^{i}_{I}\right\}\in\mathbbm{Z}}}
\newcommand{\isumset}[1]{\sum_{\left\{#1_{i}\right\}\in\mathbbm{Z}}}
\newcommand{\pzero}{p^{i_0}_{I_0}}
\newcommand{\pmax}{p_{\text{max}}}
\newcommand{\Gsum}{\Gamma\!\lr{\tfrac{1+r}{2}}}
\newcommand{\Ghalf}{\Gamma\!\lr{\tfrac{1}{2}}}
\newcommand{\rrfour}{\frac{r(1-r)}{4}}
\newcommand{\rreight}{\frac{r(1-r)}{8}}
\newcommand{\nk}{\mathscr{n}_{\mathscr{k}}}
\newcommand{\nik}{\mathscr{n}^i_{I,\mathscr{k}}}
\newcommand{\mik}{\mathscr{m}^i_{I,\mathscr{k}}}

\renewcommand{\pi}{\text{\textpi}} 

\begin{document}

\titleformat{\section}[block]{\large\bfseries\filcenter}{\thesection.}{1em}{\MakeUppercase} 
\titleformat{\subsection}[block]{\bfseries\filcenter}{\thesection.\thesubsection}{1em}{} 
\titleformat{\subsubsection}[block]{\filcenter}{\thesection.\thesubsection.\thesubsubsection}{1em}{\textit} 

\title{Analysing (cosmological) singularity avoidance in loop quantum gravity using {U$(1)^3$ coherent states and Kummer's functions}}

\author{Kristina Giesel}
\thanks{kristina.giesel@gravity.fau.de}
\author{David Winnekens}
\thanks{david.winnekens@fau.de}
\affiliation{Institute for Quantum Gravity, Theoretical Physics III, \\ Department of Physics, FAU Erlangen-N\"urnberg, \mbox{Staudtstr.~7}, 91052 Erlangen, Germany}
\begin{abstract} 
Using a new procedure based on Kummer's Confluent Hypergeometric Functions, we investigate the question of singularity avoidance in loop quantum gravity (LQG) in the context of U$(1)^3$ complexifier coherent states and compare obtained results with already existing ones. 
Our analysis focuses on the dynamical operators, denoted by $\qr$,  whose products are the analogue of the inverse scale factor in LQG and also play a pivotal role for other dynamical operators such as matter Hamiltonians or the Hamiltonian constraint. For graphs of cubic topology and linear powers in $\qr$, we obtain the correct classical limit and demonstrate how higher order corrections can be computed with this method. This extends already existing techniques in the way how the involved fractional powers are handled. We also extend already existing formalisms to graphs with higher-valent vertices. For generic graphs and products of $\qr$, using estimates becomes inevitable and we investigate upper bounds for these semiclassical expectation values. Compared to existing results, our method allows to keep fractional powers involved in $\qr$ throughout the computations, which have been estimated by integer powers elsewhere. Similar to former results, we find a non-zero upper bound for the inverse scale factor at the initial singularity. Additionally, our findings provide some insights into properties and related implications of the results that arise when using estimates and can be used to look for improved estimates.
\end{abstract}

\maketitle
\newpage
\tableofcontents
\newpage
\section{Introduction}
Motivated by results in loop quantum cosmology (LQC) on the cosmological singularity avoidance, it is of big interest to find out whether singularities --- and especially the Big Bang singularity --- are resolved in the framework of full loop quantum gravity (LQG) too. LQC is a quantum mechanical toy model of LQG with a finite amount of degrees of freedom as it quantises general relativity (GR) not as a whole but only its symmetry reduced, cosmological sector. It was introduced by Bojowald in a series of papers~\cite{LQC1,LQC2,LQC3,LQC4} based on former work with Kastrup \cite{Bojowald:1999eh} and evolved quickly into an active field of research; see for instance \cite{Agullo:2016tjh,Ashtekar:2011ni,Bojowald:2006da} for reviews and the references therein. While the results of~\cite{Bojowald1,Bojowald2,Bojowald3,Bounce1,Bounce2} are indeed very promising concerning the avoidance of the Big Bang singularity and replacing it by a big bounce accordingly, there is a lot of discussion on how LQC is embedded into full LQG~\cite{Engle:2005ca,Brunnemann:2007du,Engle:2008in,Brunnemann:2010qk,Engle:2016hei,Engle:2016zac,Bodendorfer:2015hwl}. Hence, it is of importance to also approach the possibility of singularity avoidance from the side of full LQG --- guided by the seminal results of LQC where possible. In order to proceed into this direction, an analysis of the operators describing the quantum dynamics in LQG is necessary. One approach that addresses this question follows the strategy to obtain cosmological models from full LQG as for instance in \cite{Alesci:2016gub,Dapor:2017rwv}, where the latter relies on semiclassical techniques in order to obtain cosmological models from LQG. In general, the semiclassical sector of the theory provides a framework where this question is of interest. Entering this realm, in turn, requires that we have appropriate semiclassical states for the theory, which we can use for computing expectation values of the relevant operators such as for instance the inverse scale factor in this sector. For loop quantum gravity, SU(2) complexifer coherent states were constructed in \cite{ComplexifierCoherent}, based on a complexification method introduced in \cite{Hall1,Hall2}, which was later generalised to diffeomorphism invariant gauge theories \cite{ALMMT}. In a series of papers  \cite{GCS1,GCS2,GCS3,GCS4}, it was shown that these complexifier coherent states fulfil the desired properties such as peakedness in the configuration, momentum and phase space representation or the Ehrenfest theorems, i.e. that they do reflect the behaviour of classical physics in zeroth order in $\hbar$, and that the commutator of two operators (divided by $\i\hbar$) resembles the Poisson bracket of the corresponding classical functions. Accordingly, those states are also referred to as semiclassical states and one can use them to perform a semiclassical analysis. One class of dynamical operators relevant in this context, denoted as ${\hat q}^j_e(r)$, constitutes the main ingredient of many quantum operators that are important for describing the theory's dynamics. As the inverse scale factor can be constructed from it as well, it is furthermore the main object of investigation when looking into singularity avoidance in cosmology.  One of the reasons why this class of operators, however, is not easy to handle is that they contain the volume operator to the power of $r\in\mathbbm{Q}$, making many analytical calculations impossible since the full spectrum of the volume operator is not known. This is the reason why we are forced to use estimates, approximations or simplified models if we want to proceed. As far as semiclassical investigations are concerned, in \cite{AQG3} so-called semiclassical perturbation theory was introduced that allows to replace the volume operator by a power series expansion in terms of operators that involve only integer powers of the flux operators. For those, semicassical expectation values can be analytically computed if one uses the SU(2) complexifier coherent states as has been shown in \cite{GCS1,GCS2,GCS3,GCS4}. Another possibility is to replace the SU(2) coherent states by U(1)$^3$ coherent states, which have the advantage that they diagonalise the volume operator. 

In previous work \cite{Towards1,Towards2}, Sahlmann and Thiemann presented i.a. a procedure for calculating expectation values with respect to U(1)$^3$ complexifier coherent states of the operators that constitute the Hamiltonians of various fields being coupled to gravity, including the operator ${\hat q}^j_e(r)$. They were able to perform their calculations without the usage of estimates to solve the occurring integrals in the expectation value as they restricted their analysis to cubic graphs. However, their approach involves a Taylor expansion of quantities with fractional powers, which was crucial for obtaining analytical expressions as final results. Later, Brunnemann and Thiemann \cite{Brunnemann1,Brunnemann2} analysed the question of singularity avoidance in LQG. In their work, they showed that the analogue of the inverse scale factor operator in LQG is unbounded from above on zero volume states. However, they found indeed an upper bound for the expectation value of the inverse scale factor operator with respect to U(1)$^3$ coherent states at the Big Bang. In order to obtain this result, they applied a chain of estimates that allowed them to circumvent the evaluation of the initial integrals and instead replace them by ones that can be integrated analytically with the methods they used. The details will be discussed below (cf. section~\ref{subsec:LikeBrunnemannThiemann}). They conclude that singularity avoidance in LQG, if existing, has to be addressed differently than in LQC, but the existence of an upper bound of the inverse scale factor's expectation value with respect to coherent states at the Big Bang can be seen as a strong indication for the respective singularity's resolution at least in the semiclassical sector of the theory.

The present article aims at revisiting the semiclassical analysis of the inverse scale factor operator by applying methods introduced in our companion paper~\cite{paper1} using also U$(1)^3$ coherent states.  In our previous work, we showed that Kummer's confluent hypergeometric functions (KCHFs) can be used to analytically compute expectation values of fractional powers of the momentum operator with respect to U(1) coherent states or the well-known quantum mechanical coherent states. This technique also applies to U$(1)^3$ coherent states frequently used in the earlier analysis of LQG. The reason why KCHFs fit well into this framework is that the KCHFs of first and second kind are in some sense dual to each other under the Fourier transform, which was shown in \cite{Pichler} and heavily used and discussed in detail in \cite{paper1}. The integrals at hand for the semiclassical expectation values result essentially in Kummer's confluent hypergeometric function of the first kind. If being interested in the semiclassical limit, one can subsequently make use of the KCHF's asymptotic expansion for large arguments to obtain in zeroth order the result one would expect from the classical calculation. This expansion is performed in terms of the classicality parameter encoded in the coherent states, which is $\hbar$ in the case of the quantum mechanical coherent states. The aim of the present work now is to apply the KCHF-procedure in the LQG-framework and calculate semiclassical expectation values there. An advantage of this method compared to the former analysis in \cite{Towards1,Towards2,Brunnemann1,Brunnemann2} is that we are able to avoid estimates at certain stages of the computation since we can integrate fractional powers against Gaussians, whereas in the former work either Taylor expansions or estimates were necessary to substitute the involved fractional powers. In our analysis here, depending on the scenario we consider, we can perform our computations either even without the additional usage of estimates or with ones that are slightly better adapted to the fractional power of the operator. The scenarios we consider differ by the choice of the underlying graph, particularly its valence, or by the power of the operator ${\hat q}^j_e(r)$. 
We will show that we can on the one hand extend the method used by Sahlmann and Thiemann also to more general graphs than cubic ones and on the other hand adapt the procedure of Brunnemann and Thiemann with the use of the KCHFs and compare their results to ours. The latter allows us further to discuss limitations and generalisations of existing estimates in this context. 
\newline
\newline
The paper is structured as follows: First, we will briefly introduce the relevant operators of LQG and the mathematical tools in subsection~\ref{subsec:LQG}. Subsequently, in subsection~\ref{subsec:KCHFs}, we summarise how semiclassical expectation values can be computed using KCHFs and then apply this procedure in order to compute the expectation value of fractional powers of the momentum operator with respect to coherent states in~\ref{subsec:mtmoproots} as an illustrating example.

In section~\ref{sec:woEstimates}, we apply this method to the volume operator and perform semiclassical computations without the additional usage of estimates. Starting with graphs of cubic topology in~\ref{subsec:cubicgraphs}, we first calculate the basic building block of their semiclassical  expectation value of $\qr$ with respect to U(1)$^3$ coherent states in~\ref{subsubsec:rigorous3edges}. We only illustrate the general procedure, while the details of this calculation can be found in Appendix~\ref{Appendix9integrals}. The final result can be found in \eqref{3edgesFinalGeneral} on page \pageref{3edgesFinalGeneral}. Its specialisation to the case $p=0$ that corresponds classically to the cosmological singularity is shown in  \eqref{eq:resultRigorousP=0General} on page \pageref{eq:resultRigorousP=0General}. Afterwards, we discuss the semiclassical limit in~\ref{subsec:SemClassCubic}, where again the detailed derivation is presented in Appendix~\ref{sec:AppSemLimitCubic}. In our next step in subsection~\ref{subsec:highervalent}, we then proceed to higher valent vertices  and, after a general introduction, apply the procedure of Sahlmann and Thiemann to these not necessarily cubic graphs in~\ref{subsubsec:SahlmannThiemann}. The final result for the semiclassical expectation values of $ \prod_{k=1}^N {\hat q}^{j_k}_{J_k}(r)$  via a generalised form of the procedure of Sahlmann and Thiemann can be found on  page  \pageref{eq:FinResSTManyEdes} in \eqref{eq:FinResSTManyEdes}. 

For more complicated and general scenarios, the relevant integrals cannot be solved by the methods discussed in the former section and therefore, in section~\ref{sec:wEstimates}, we discuss semiclassical computations for the volume operator that also rely on estimates. The first calculation of~\ref{subsec:LikeBrunnemannThiemann} follows the route of Brunnemann and Thiemann and --- likewise to their result --- also yields an upper bound for the inverse scale factor, but in our case is adopted in such a way that we use different estimates as we can then evaluate the integrals at hand by means of KCHFs.  The semiclassical expectation values of $\qr$ using a generalised form of the estimates of Brunnemann and Thiemann is shown in  \eqref{eq:ResultALaBrunnemann} on page  \pageref{eq:ResultALaBrunnemann}. The case $p=0$ is separately discussed and can be found in \eqref{eq:ResultALaBrunnemannP=0} on page \pageref{eq:ResultALaBrunnemannP=0}. A generalisation of the above procedure for $ \prod_{k=1}^N {\hat q}^{j_k}_{J_k}(r)$ is given in  \eqref{eq:ResultALaBrunnemannThiemannGeneral} on page \pageref{eq:ResultALaBrunnemannThiemannGeneral} and for the specific case $p=0$ in \eqref{eq:ResultALaBrunnemannThiemannGeneralP=0} on page \pageref{eq:ResultALaBrunnemannThiemannGeneralP=0}. The comparison of the way different estimates enter into the final result allows us to discuss the limitations of such estimates as well as finding new estimates that potentially improve the results, or at least let us understand the problems that arise due to the utilisation of them. These are illustrated in subsections \ref{subsec:towards} and \ref{subsec:newEstimates}, followed by a comparison with the initial Brunnemann and Thiemann approach in \ref{subsec:comparison}. The final result for the semiclassical expectation values of $\qr$ using a new kind of estimate is presented in  \eqref{finaltowards} on page \pageref{finaltowards} and in \eqref{finalTowardsp=0} on page \pageref{finalTowardsp=0} for the case $p=0$. Considering a further new kind of estimate, the semiclassical expectation value of $\qr$ has been recalculated in  \eqref{eq:resultImprovedEstimateGeneral} on page \pageref{eq:resultImprovedEstimateGeneral}. The final conclusion is then found in section~\ref{sec:conclusion}, followed by an appendix that i.a. covers the full calculation behind subsections \ref{subsubsec:rigorous3edges} and \ref{subsec:SemClassCubic}, an overview over the essential estimates we used in section \ref{sec:wEstimates} as well as a comparison between our method and the one used by Sahlmann and Thiemann for a simple example.

\section{Brief review of the relevant LQG operators and semiclassical techniques} \label{sec:backgroundANDtools}

\subsection{Relevant LQG operators} \label{subsec:LQG}

Loop quantum gravity (LQG) is a theory of SU(2) Yang--Mills type, but with the difference that it is not built upon a background metric and instead treats spacetime as a dynamical object itself. It is based on a formulation of general relativity in terms of Ashtekar--Barbero variables that consist of pairs of an SU(2) connection  $A^j_a$, where $j$ denotes the SU(2) and $a$ the spatial index, and a so-called `electric field' $E^a_j$ as the conjugate variable. After quantisation, the theory's kinematical Hilbert space $\mathcal{H}_{\text{kin}} = L^2\lr{\overline{\mathcal{A}},\d \mu_{\text{AL}}}$ is based on the set of  generalised connections $\overline{\mathcal{A}}$ together with the Ashtekar--Lewandowski measure $\mu_{\rm AL}$. An orthonormal basis in  $\mathcal{H}_{\text{kin}}$ is provided by so-called spin network functions (SNFs) $\snf$. These are defined for a given graph $\gamma$, which is embedded into the spatial manifold of the classical theory and consists of finitely many edges $M$ that only join at the vertices. Given the connection $A$, we can define an associated holonomy $h_{e_i}(A)$ for each edge $e_i$, which is the path-orderd exponential over the connection integrated along the edge. SNFs can then be understood as cylindrical functions of the holonomies $h_{e_i}(A)$ for a given $\gamma$ composed of $M$ edges $e_i$:
\begin{equation}
    \snf(A) = \snf\lr{h_{e_1}(A),h_{e_2}(A),\ldots,h_{e_M}(A)}, \quad \snf \colon \mathrm{SU}(2)^M \rightarrow \mathbbm{C}.
\end{equation}
The conjugate momenta of the holonomies are the fluxes $E_{S}(E)$, which are the aforementioned fields $E$ smeared over appropriate surfaces $S$. In the kinematical Hilbert space, these become holonomy and flux operators --- the analogue of the position and momentum operator in quantum mechanics. For more detailed introductions to LQG, see for instance the reviews~\cite{LQG1,LQG2,LQG3} or the books~\cite{LQG4,LQG5} for an even more comprehensive overview.

As mentioned in the introduction, an important class of operators is
\begin{equation}
    {\hat q}^j_e(r) \coloneqq \trb{ \tau_j{\hat h}_e\left[ \lr{{\hat h}_e}^{-1}, \Vr \right]} \label{eq:qSU2},
\end{equation}
since products 
\begin{equation*}
\prod_{k=1}^N {\hat q}^{j_k}_{e_k}
\end{equation*}
of them constitute basic building blocks of various Hamiltonian operators ~\cite{Towards1,Towards2} within LQG. If in turn interested in studying singularity avoidance, the configuration $r=\frac{1}{2}$, $N=6$ that results in the operator corresponding to the inverse scale factor after additionally multiplying by $\nicefrac{1}{\lp{}^{12}}$ is of particular interest.
To explain the individual building blocks of the operator in more detail, we denote for each given graph $\gamma$ its set of edges by $E(\gamma)$ and its set of vertices by $V(\gamma)$. In the equation above, $\tau_j$ is a basis of $\mathfrak{su}(2)$ and $\hat h_e$ the holonomy operator acting on edge $e$ of the SNF. The operator ${\hat V}^r$  is the volume operator~\cite{VolRovelli,VolAshtekar} to the power of $r\in\mathbbm{Q}$, where the volume operator itself is given by
\begin{equation}
    {\hat V} =\beta^3\lp{}^3\sum_{v_I\in V(\gamma)} \sqrt{\left|{\widehat Q_{v_I}}\right|}.
\end{equation}
Therein, $\beta$ denotes the Barbero--Immirzi parameter and $\lp$ the Planck length. This form of the volume operator illustrates the fact that it collects all the contributions of vertices $v_I\in V(\gamma)$ that lie in the region the volume is to be calculated of. The operator that acts at each vertex $v_I$ and actually evaluates those contributions reads
\begin{equation} \label{eq:defQ}
    {\widehat Q}_{v_I} = \frac{1}{48} \sum_{e\cap e'\cap e''=v_I} \epsilon_{ijk}\epsilon(e,e',e''){\hat X}^i_{v_I,e}{\hat X}^j_{v_I,e'}{\hat X}^k_{v_I,e''} \ ,
\end{equation}
where the right-invariant vector fields ${\hat X}^{i}_{v_I,e}$ act on SNFs as
\begin{align}
    \lr{{\hat X}^i_e\snf}(h) \coloneqq \dif{}{t} \snf\lr{\e{t\tau_i}h}\Big\vert_{t=0}
\end{align}
and $\epsilon(e,e',e'')$ is the so-called sign factor, which is sensitive to the orientation of the triple of edges that meet at $v_I$ and takes the values $\pm1$ or $0$ if the edges are linearly dependent. Thus, ${\hat Q}_{v_I}$ collects all contributions of all right-invariant vector fields ${\hat X}^{i}_{v_I,e}$  on configurations of all triples of edges $e,e',e''$ that meet at $v_I$, taking care of their orientation via $\epsilon(e,e',e'')$.

As mentioned above and like \cite{Towards1,Towards2,Brunnemann1,Brunnemann2}, we will replace SU(2) by U(1)$^3$ as a simplification for our semiclassical analysis.  Replacing SU(2) by U$(1)^3$ can be done without changing the obtained results quantitatively, as~\cite{GCS2,GCS3,AQG3} showed, while simplifying the calculations drastically due to U$(1)^3$ being Abelian. 
The holonomy flux algebra for U$(1)^3$ on the Hiblert space ${\cal H}_e=L_2(\text{U}(1)^3,\d\mu_H)$ associated with each edge $e$, where $\d\mu_H$ denotes the corresponding Haar measure, reads
\begin{eqnarray*}
\left[\hat{h}^j_e,\hat{p}^k_{e'}\right]= -\i\frac{\lp{}^2}{a^2}\delta_{ee'}\delta^{jk}\hat{h}^k_{e'},\quad \left[\hat{h}_e^j,\hat{h}^k_{e'}\right]=0 \quad\text{and}\quad \left[\hat{p}^j_e,\hat{p}^k_{e'}\right]=0.
\end{eqnarray*}
Here, we introduced a generic length scale $a$ in order to work with dimensionless fluxes for later convenience. The holonomy operators $\hat{h}^j_e$ act by multiplication and the flux operators can be expressed in terms of right invariant vector fields of U$(1)$, that is $\hat{p}^j_e=\frac{1}{a^2}\hat{X}^j_e$ with $\hat{X}^j_e\coloneqq \i h^j_e\frac{\partial}{\partial h^j_e}$.

Ultimately, the important feature of using U$(1)^3$ is that the U$(1)^3$ equivalents of the SNFs diagonalise the volume operator and, accordingly, the U$(1)^3$ version of ${\hat q}^j_e(r)$. In the U$(1)^3$ version of LQG, the SNFs are replaced by so-called charge network functions (CNFs). This name is caused by the fact that by working with U$(1)^3$, the states are now graphs $\gamma$ consisting of $M$ edges $e_I \in E(\gamma)$, meeting other edges in vertices $v_J\in V(\gamma)$, and where all those edges $e_I$ are each equipped with three U(1)-charges $n^i_I, i\in\{1,2,3\}$. As analogue of the operator~(\ref{eq:qSU2}) in U$(1)^3$, we use
\begin{equation}
    \qr \coloneqq \frac{1}{a^{3r}} {\hat h}^{i_0}_{I_0}\left[\left({\hat h}{}^{i_0}_{I_0}\right)^{-1},{\hat V}^r\right] , \label{eq:qr}
\end{equation}
where $ {\hat h}^{i_0}_{I_0}$ now is the holonomy operator acting on edge $e_{I_0}$ and U(1)-copy $i_0$ of the CNF by increasing the charge $n^{i_0}_{I_0}$ by 1, $\left({\hat h}{}^{i_0}_{I_0}\right)^{-1}$ the inverse thereof and ${\hat V}^r$ the volume operator to the power of $r\in\mathbbm{Q}$.
To make future calculations dimensionless right from the start, we additionally included the prefactor $\nicefrac{1}{a^{3r}}$, where $a$ is a length scale that links the classicality parameter $t$ and the Planck length $\lp$ via $t = \nicefrac{\lp{}^{2}}{a^2}$. This length scale $a$ is originally introduced to work with dimensionless fluxes in the course of the construction of the coherent states and its corresponding complexifier. 

Following the notation of~\cite{Brunnemann2}, the volume eigenvalues of $\hat V$ acting on charge network states $\cnf$ are given by
\begin{equation} \label{eq:voleigen}
    {\hat V}\, \cnf \eqqcolon \sum_{v} \lambda\lr{\left\{n^j_J\right\}}\cnf = \sum_{v} \lp{}^3\sqrt{\left| Z\sum_{I,J,K}\epsilon_{ijk}\epsilon(IJK)n^i_I n^j_J n^k_K \right|} \, \cnf ,
\end{equation}
i.e. they diagonalise the volume operator. In this formula, $Z \coloneqq \frac{\beta^3}{48}$ was chosen in accordance with \cite{VolGiesel} and we absorbed the Immirzi paramter also in the definition of $Z$. The composition of the above expression follows closely the definition of the volume operator, which is a logical consequence of the CNFs being eigenfunctions thereof: We can read $\epsilon_{ijk}n^i_I n^j_J n^k_K$ as the determinant of a ``charge matrix'', with one label
being the three U(1)-copies $\{i\}$ and the other one representing the three edges $e_I, e_J, e_K$ meeting at $v$. Then, $\sum_{I,J,K}\epsilon(IJK)$ sums over all (oriented) configurations of three edges meeting at $v$, while $\sum_{v}$ lastly collects the contributions of all vertices lying inside the region of interest, $v\in V(\gamma)$. This perspective will become more important when we establish estimates for the expression above in subsection~\ref{subsec:highervalent}. The reason why we will need estimates can also be seen at this level already: Facing expectation values of $\qr$ with respect to coherent states, one ultimately has to integrate expressions like~\eqref{eq:voleigen} against Gaussian functions, which is in general analytically not possible.

Following the notation of~\cite{Brunnemann2} again, the action of $\qr$ on the charge network state $\cnf$ is
\begin{align}
    a^{3r}\cdot \qr \cnf &= \lr{\Vr -{\hat h}^{i_0}_{I_0}\Vr\left({\hat h}{}^{i_0}_{I_0}\right)^{-1} } \cnf \nonumber\\
    & = \lr{\lambda^r\lr{\left\{n^i_I\right\}} - \lambda^r\lr{\left\{n^i_I-\delta^{ii_0}\delta_{II_0}\right\}}} \cnf .
\end{align}
The second contribution means that $\lambda^r$ is evaluated not on the initial charge matrix but a modified one with $n^{i_0}_{I_0} \mapsto n^{i_0}_{I_0} -1$, reflecting the action of $\left({\hat h}{}^{i_0}_{I_0}\right)^{-1}$ before the one of the volume operator.

The complexifier coherent states~\cite{ComplexifierCoherent} can similarly be constructed in the U$(1)^3$-framework. They are labelled by a point $m=\lr{A^{(0)},E^{(0)}}$ in phase space, around which the coherent state is peaked. Furthermore, they involve a so-called classicality parameter denoted by $t$, which provides access to the classical limit of the theory if being sent to zero. For complexifier coherent states, this is usually chosen to be $t=\frac{\lp{}^2}{a^2}\sim \hbar$, with the generic length parameter $a$, as mentioned above. The analogue in quantum mechanics is just $\hbar$. In the notation of~\cite{Brunnemann2}, the coherent states have the form
\begin{equation}
    \cs(A) =  \prod_{\overset{e_I\in E(\gamma)}{i=1,2,3}} \, \sum_{n^i_I\in\mathbbm{Z}} \e{-\frac{t}{2}\lr{n^i_I}^2+n^i_I p^i_I(m)} \left[ \e{\i \theta^i_I(m)}\e{-\i \theta^i_I(A)} \right]^{n^i_I} , \label{eq:coherentstate}
\end{equation}
where we assumed the dimensionless classicality parameter $t$ to be the same for all edges. This is not a necessary choice but rather taken in order to keep the formulae more concise. The square bracket's second exponential function denotes the inverse of the U(1)-holonomy
\begin{equation}
    h^i_I(A) = \e{\i \int_{e_I} A^i_a(e_I(t)) \, {\dot e_I}^a(t) \,\d t } \eqqcolon \e{\i \theta^i_I(A)},
\end{equation}
while the first one is one part of the so-called complexified holonomy,
\begin{equation}
    h^i_I(Z(m)) = \e{p^i_I(m)} h^i_I(A|_m) \eqqcolon \e{p^i_I(m)} \e{\i \theta^i_I(A|_m)},
\end{equation}
which enters into the construction of the complexifier coherent states.
Therein, $m=(A,E)\mapsto Z(m)$ can be obtained from a complexifier that is quadratic in the fields $E^a_j$ and leads to a complexification of the connections $A^j_a$, see \cite{Brunnemann2} for the exact form. In our notation, we have $\{A^j_a(x),E^b_k(y)\}=\kappa\delta^b_a\delta^j_k\delta^{(3)}(x-y)$ with $\kappa\coloneqq 8\pi G_N$, where $G_N$ is Newton's constant. For a more detailed treatment and construction of these coherent states see~\cite{GCS1,GCS2,GCS3,GCS4,ComplexifierCoherent,Towards1,Towards2,Brunnemann2}.

With the coherent states being linear combinations of the conjugate charge network states $\overline{T}_c$, we need to evaluate $\qr$ with respect to those and get for the expectation value of interest
\begin{align}
    \langle \qr \rangle_{\cs} &= \frac{1}{\csnorm a^{3r}} \sum_{\left\{n^i_I\right\}\in\mathbbm{Z}} \e{\sum_{i,I}\lr{-t\lr{n^i_I}^2+2p^i_In^i_I}}\lambda^r\lr{\left\{n^{i_0}_{I_0}\right\}} \nonumber\\
    & \coloneqq \frac{1}{\csnorm a^{3r}}  \sum_{\left\{n^i_I\right\}\in\mathbbm{Z}} \e{\sum_{i,I}\lr{-t\lr{n^i_I}^2+2p^i_In^i_I}} \lr{\lambda^r\lr{\left\{n^i_I\right\}} - \lambda^r\lr{\left\{n^i_I+\delta^{ii_0}_{II_0}\right\}}} \nonumber\\
    &= \frac{1}{\csnorm a^{3r}}  \sum_{\left\{n^i_I\right\}\in\mathbbm{Z}} \e{\sum_{i,I}\lr{-t\lr{n^i_I}^2+2p^i_In^i_I}} \lpz \lr{\left| \sum_{IJK} \epsilon(IJK)\epsilon_{ijk}n^i_In^j_Jn^k_K \right|^{\frac{r}{2}} - \right. \nonumber\\
    & \qquad\qquad \left. -\left| \sum_{IJK} \epsilon(IJK)\epsilon_{ijk}\lr{n^i_I+\delta^{ii_0}\delta_{II_0}}\lr{n^j_J+\delta^{ji_0}\delta_{JI_0}}\lr{n^k_K+\delta^{ki_0}\delta_{KI_0}}\right|^{\frac{r}{2}}}. \label{eq:startingExpectationvalue}
\end{align}
Note that there is now a + in the commutator's second $\lambda^r$ due to the action of $\left({\hat h}{}^{i_0}_{I_0}\right)^{-1}$ on the conjugate CNF $\bar{T}_c$. We used $\delta^{ii_0}_{II_0}$ as a shorthand notation for $\delta^{ii_0}\delta_{II_0}$.

Typically, as a next step, one performs a Poisson resummation according to~(\ref{PoissonResum}) in order to transform the above expression into one that is converging much faster for $t\to0$: As we are interested in the semiclassical limit of small $t$, the Gaussian in $t\lr{n^i_I}^2$ of~\eqref{eq:startingExpectationvalue} will become wider and wider, hence forcing us to include more and more terms of the sum over $n^i_I$. After the application of the Poisson resummation, however, it becomes a rapidly decreasing function of new $N^i_I \in \mathbbm{Z}$ and it turns out that the $N^i_I=0$ contribution dominates. To perform the Poisson resummation, we define $x^i_I\coloneqq Tn^i_I \coloneqq \sqrt{t} \, n^i_I$. First, we apply it to the squared norm of the states  and obtain (like in~\cite{Brunnemann2})
\begin{equation}
    \norm{\cs}^2 = \prod_{\substack{e_I \in E(\gamma) \\ i \in 1,2,3}} \sum_{N^i_I\in\mathbbm{Z}} 2\pi \sqrt{\frac{\pi}{t}}\es{\frac{\lr{p^i_I}^2}{t}} \e{-\frac{\pi^2\lr{N^i_I}^2+2\pi\i N^i_I p^i_I}{t}} \eqqcolon \prod_{\substack{e_I \in E(\gamma) \\ i \in 1,2,3}} 2\pi \sqrt{\frac{\pi}{t}}\es{\frac{\lr{p^i_I}^2}{t}} \lr{1+K_t} \label{eq:norm}.
\end{equation}
$K_t$ therein is of order $\mathcal{O}\lr{t^\infty}$, implying it can be neglected at any point where one considers the limit $t\to 0$. Given this, the expectation value takes the form\footnote{Note that we can use $\nicefrac{\lp}{a} = \sqrt{t} = T$.}
\begin{align}
    \langle \qr \rangle_{\cs} &= \frac{\lrabs{Z}^{\rhalf}T^{3r}}{\csnorm} \sum_{\left\{N^i_I\right\}\in\mathbbm{Z}} \lr{\frac{2\pi}{T}}^{3M} \; \int\limits_{-\infty}^\infty \d^9x^i_I \es{\sum_{Ii}\lr{-\lr{x^i_I}^2+2\frac{p^i_I-\pi\i N^i_I}{T}x^i_I}}\frac{1}{T^{3r}} \nonumber\\
    & \hspace{-1.1cm} \cdot \left(\left| \sum_{IJK} \epsilon(IJK)\epsilon_{ijk} x^i_I x^j_J x^k_K \right|^{\frac{r}{2}} - \right. \nonumber\\
    & \left. - \left| \sum_{IJK} \epsilon(IJK)\epsilon_{ijk}\lr{x^i_I+T\delta^{ii_0}\delta_{II_0}}\lr{x^j_J+T\delta^{ji_0}\delta_{JI_0}}\lr{x^k_K+T\delta^{ki_0}\delta_{KI_0}}\right|^{\frac{r}{2}} \right). \label{eq:Expectationvalue}
\end{align}

Our aim is now to apply new methods for determining semiclassical expectation values of the kind above. As an explicit, analytical calculation for the general case is not feasible, we will rely either on estimates or less general applications which simplify the expression in (\ref{eq:Expectationvalue}). In section~\ref{subsubsec:rigorous3edges}, we show that we can in fact directly calculate the case of cubic graphs by means of Kummer's confluent hypergeometric functions without the need of estimates. We will also use a procedure similar to the one of Brunnemann and Thiemann~\cite{Brunnemann1,Brunnemann2}, who used a series of inequalities to circumvent the elaborate integrals of~\eqref{eq:Expectationvalue} via estimating the root of the absolute value of the sum of determinants by the sum over the square of the single charges. However, we will demonstrate in section~\ref{subsec:LikeBrunnemannThiemann} that with the help of KCHFs we do not need to go thus far, as we will be able to integrate (basic) root-expressions by means of the formalism introduced in the next section~\ref{subsec:KCHFs}. This allows us to conserve the information encoded in the root's exponent, allowing to ultimately find an estimate with correct powers in the variables.

First of all, however, we present in the next subsections an introduction and overview of the methods we will use in our later computations.

\subsection{Kummer's confluent hypergeometric functions (KCHFs)} \label{subsec:KCHFs}

Since this is a key technique of the present work, we will introduce Kummer's confluent hypergeometric functions (KCHFs) in a bit more detail, also including a concise example of calculating expectation values of roots or more general fractional powers of the momentum operator $\hat p$ for U(1) coherent states in the subsequent subsection~\ref{subsec:mtmoproots}. 

Kummer's confluent hypergeometric functions $\kchf{a,b,z}$ arise in particular via integrals of the form
\begin{align}
\int_{-\infty}^{\infty}\e{-\rho^2(x-\mu)^2}\,\left|x\right|^r
\text{d}x & = \lrabs{\rho}^{-1-r}\,\Gamma\left(\frac{r+1}{2}\right)
\kchf{-\frac{r}{2}, \frac{1}{2},-\mu^2\rho^2} \label{KCHF} \quad\mathrm{ or } \\
\int_{-\infty}^{\infty}\e{-\rho^2x^2+2\rho^2\mu x}\,\left|x\right|^r
\text{d}x & = \left| \rho\right| ^{-1-r} \Gamma \left(\frac{r+1}{2}\right)  \, \kchf{\frac{r+1}{2}, \frac{1}{2}, \mu^2\rho^2} ,
\end{align}
where we call $z$ the \textit{argument} of $\kchf{a,b,z}$ and $\Re{r}>-1 \land \Re{\rho^2}>0$ needs to be fulfilled. Note that~\cite{paper1} showed a generalisation of the above identities by using a theorem from~\cite{Pichler} that proved that the Kummer functions of first and second kind (see next paragraph) are dual, in a certain sense, with respect to the Fourier transformation.

Comparing the two identities, we see that we can proceed from the left hand side of the second one to the left hand side of the first one by completing the square, yielding an $x$-independent prefactor $\exp{\lr{\mu^2\rho^2}}$. Comparing the respective right hand sides, we can use one of Kummer's transformations~\cite[13.2.39 therein]{dlmf}:
\begin{align}
    \kchf{a,b,z} &= \e{z}\,\kchf{b-a,b,-z} \\
    \Rightarrow \kchf{-\frac{r}{2}, \frac{1}{2},-\mu^2\rho^2} &= \e{-\mu^2\rho^2}\kchf{\opr,\frac{1}{2},\mu^2\rho^2}. \label{KummerTrafo}
\end{align}

In general, the confluent hypergeometric functions of the first and second kind are solutions of Kummer's differential equation~\cite{Kummer}
\begin{equation}
z\frac{\d^2w}{\d z^2}+(b-z)\frac{\d w}{\d z}-aw=0 , \label{eq:KummerDGL}
\end{equation}
reading
\begin{align}
    \kchf{a,b,z} &\coloneqq \sum_{n=0}^\infty \frac{(a)_n}{(b)_n n!} z^n \ \text{and} \label{eq:defKCHF1}\\
    U(a,b,z) &\coloneqq \frac{\Gamma(1-b)}{\Gamma(1+a-b)}\kchf{a,b,z} + \frac{\Gamma(b-1)}{\Gamma(a)}z^{1-b} \kchf{1+a-b,2-b,z} \label{eq:defKCHF2}
\end{align}
respectively. Therein, $(a)_n$ denotes the Pochhammer symbol or raising factorial\footnote{Note that the raising factorial is sometimes also denoted by $a^{(n)}$. To make things even worse, Pochhammer himself used $(a)_n$ for the binomial coefficient $\binom{a}{n}$ and $[a]_n^+$ for the raising factorial~\cite[p.\,80-81]{Pochhammer}. We use the above notation as it became the established standard for hypergeometric functions.}
\begin{align}
    (a)_0 & = 1 \ , \nonumber \\
    (a)_1 & = a \ \text{ and}\nonumber \\
    (a)_n & = a (a+1) (a+2)\cdots(a+n-1).
\end{align}
$ \kchf{a,b,z} $ is originally denoted by $ \varphi(\alpha,\beta,x) $ in eq.\,1. of~\cite{Kummer}, while in modern literature, mainly the notation $M(a,b,z)$ is used besides $\kchf{a,b,z}$. However, note that there also exists the regularised KCHF $\text{M}(a,b,z)\,\Gamma(b) \coloneqq M(a,b,z)$, symbolised by an upright M, that avoids the singularities of $\kchf{a,b,z}$ whenever $b$ is 0 or a negative integer.

KCHFs include a vast amount of functions, such as
\begin{align}
\kchf{0,b,z} &=1, \\
\kchf{a,a,z} &=\e{x}, \\
\kchf{-n,\tfrac{1}{2},z^2}&=\lr{-1}^n\frac{n!}{(2n)!}H_{2n}(z) \ \text{and} \\
\kchf{-n,\tfrac{3}{2},z^2}&=\lr{-1}^n\frac{n!}{(2n+1)!\,2z}H_{2n+1}(z),
\end{align}
with the Hermite polynomials $H_n(z)$ as well as other identities for Bateman's function, Bessel functions, Laguerre polynomials and more~\cite{abramowitz+stegun,dlmf}. The Kummer (confluent hypergeometric) functions of the first kind, $\kchf{a,b,z}$, can also be understood as a special limit of the ordinary (or Gaussian) hypergeometric functions
\begin{equation}
    {}_2\text{F}_1\left(a,b;c;z\right) \coloneqq \sum_{n=0}^\infty \frac{(a)_n(b)_n}{(c)_n n!} z^n
\end{equation}
via
\be
\kchf{a,c,z}=\lim_{b\to\infty}{}_2\text{F}_1\!\left(a,b;c;\frac{z}{b}\right) .
\ee

As we are ultimately interested in considering expansions in the classicality parameter, one of the KCHFs' properties that we will often make use of is the asymptotic expansion for large arguments $|z|\to\infty$, which reads~\cite{abramowitz+stegun}
\begin{align}
\kchf{a,b,z} \overset{\lrabs{z}\to \infty}{\approx} \Gamma(b) & \left[ 
\frac{\e{\pm\pi\i a}z^{-a}}{\Gamma(b-a)} \sum_{n=0}^{\infty}
\frac{(a)_n(1+a-b)_n}{n!}(-z)^{-n} +\right. \nonumber \\
& \left.\quad\! +\frac{\e{z} z^{a-b}}{\Gamma(a)} \sum_{n=0}^{\infty} \frac{(b-a)_n(1-a)_n}{n!}z^{-n} \right] , \label{expansion}
\end{align}
where the minus sign in $\exp\lr{{\pm\pi\i a}}$ is chosen if $z$ lies in the right half plane.

\subsection{Expectation values of fractional powers the momentum operator} \label{subsec:mtmoproots}

In the spirit of our companion paper~\cite{paper1}, we now consider the expectation values of fractional powers (involving also the case of roots) of the momentum operator with respect to U$(1)^3$ coherent states. But first of all, we recapitulate the procedure of~\cite{paper1} for determining these expectation values for U(1) coherent states~\cite{GCS2}
\begin{equation}
    \Ptg = \sum_{n\in\mathbbm{Z}} \e{-\frac{t}{2}n^2+pn} \lr{\e{\i\theta(m)}\e{-\i\theta}}^{n},
\end{equation}
where $m$ is the point in phase space the coherent state is peaked around, as before, while they use $g=\e{p+\i\theta\lr{m}}$ to denote the complexified holonomy.
We then obtained in~\cite{paper1}\footnote{Note that we used a different convention for the Fourier transform therein, namely $\frac{\d x}{\sqrt{2\pi}}$ compared to $\d x$ here.}, via~\cite{GCS3}, for $\Re{r}>-\tfrac{1}{2}$
\begin{align}
    \langle |\hat p|^r \rangle_{\Ptg} &= \norm{\Ptg}^{-2} \infsum{n}|tn|^r \e{-tn^2+2np}\nonumber\\
    &= \norm{\Ptg}^{-2} \frac{2\pi}{T}T^r\infsum{N} \; \int\limits_{-\infty}^{\infty}\d x |x|^r \e{-x^2+\frac{2p}{T}x-\frac{2\pi\i N}{T}x} \nonumber\\
    &= \norm{\Ptg}^{-2}  \frac{2\pi}{T^{1-r}}\infsum{N}\Gamma\lr{\tfrac{r+1}{2}}\kchf{\frac{r+1}{2},\frac{1}{2},\lr{\frac{p-\pi\i N}{T}}^2}\label{RecapU1laststep}\\
    & \!\! \overset{t\to 0}{\approx} \lrabs{p}^r \lr{1-\frac{r(1-r)}{4}\frac{t}{p^2}-\frac{r(1-r)(2-r)(3-r)}{32}\frac{t^2}{p^4}+\mathcal{O}\lr{t^3}}. \label{RecapU1final}
\end{align}
Note that as far as only the expectation value is considered $r>-1$ would be allowed, but as discussed in \cite{paper1} we want to allow only $r>-\frac{1}{2}$ for which also the norm $\norm{|\hat p|^r \Ptg}$ is well-defined. The sequence of steps of calculation above  shows the route we follow when tackling expectation values via KCHFs: We performed the Poisson resummation via $x\coloneqq \sqrt{t}\,n \eqqcolon Tn$ from the first to the second line, which leads to an integral of a Gaussian against the $r$-th root of the absolute value. This integral then results in a KCHF and we finally need to perform the asymptotic expansion for large arguments of a KCHF according to~(\ref{expansion}), which ultimately results in a power series in $t$.

The last step of~(\ref{RecapU1final}) is in fact more elaborate and we will illustrate it in more detail. Inserting the formula for the asymptotic expansion together with the state's norm
\begin{equation}
\norm{\Ptg}^2 = 2\pi\sqrt{\frac{\pi}{t}}\infsum{n}\e{\frac{p^2}{t}}\e{-\frac{-\pi^2 n^2}{t}}\e{2\pi\i n \frac{p}{t}} \eqqcolon 2\pi\sqrt{\frac{\pi}{t}}\pi\es{\frac{p^2}{t}} \lr{1+\mathscr{K}_t},
\end{equation}
where we again introduced a $\mathscr{K}_t \overset{t\to 0}{=}0+\mathcal{O}\lr{t^\infty}$, we get
\begin{align}
    (\ref{RecapU1laststep}) &= \frac{\e{-\frac{p^2}{t}}}{2\pi\sqrt{\frac{\pi}{t}}\lr{1+\mathscr{K}_t}}   \frac{2\pi}{T^{1-r}}\infsum{N}\Gsum\Ghalf \cdot\nonumber\\
    &\quad \cdot \left[ \frac{\e{\pm\pi\i\frac{r+1}{2}} }{\Gamma\lr{-\rhalf}}\lr{\lr{\frac{p-\pi\i N}{T}}^2}^{-\frac{r+1}{2}}\sum_{n=0}^{\infty}\frac{\lr{\frac{r+1}{2}}_n\lr{1+\rhalf}_n}{n!} \lr{-\lr{\frac{p-\pi\i N}{T}}^2}^{-n}+ \right.\nonumber\\
    &\qquad \left. + \frac{1}{\Gamma\lr{\frac{r+1}{2}}}\e{\lr{\frac{p-\pi\i N}{T}}^2}\lr{\lr{\frac{p-\pi\i N}{T}}^2}^{\rhalf} \sum_{n=0}^{\infty} \frac{\lr{-\rhalf}_n\lr{\frac{1-r}{2}}_n}{n!}\lr{\lr{\frac{p-\pi\i N}{T}}^2}^{-n} \right].
\end{align}
Via the state's norm, an additional Gaussian $\exp{\lr{-\nicefrac{p^2}{t}}}$ entered. It suppresses the expansion's first series when we consider $t\to 0$, while the ``inverse'' Gaussian in front of the second one just annihilates it. However, as that one is in $\frac{p-\pi\i N}{T}$, there is i.a. $\exp{\lr{-\nicefrac{\pi^2 N^2}{t}}}$ left. This allows us to consider only the $N=0$ contribution and discard the $N$-sum when considering the limit $t\to 0$. Thereby, we in particular also lose all imaginary parts and $\lr{\lr{\nicefrac{p-\pi\i N}{T}}^2}^{\nicefrac{r}{2}} \to \lrabs{\nicefrac{p}{T}}^r$ gives rise to the desired zeroth order with the following power series in $t$ as in~(\ref{RecapU1final}). 

Proceeding now to the U$(1)^3$ case, in order to calculate the corresponding momentum operator $\hat{p}^i_I$ on edge $e_I$ and U(1)-copy $i$, we first take a look at the U$(1)^3$ coherent state~(\ref{eq:coherentstate}) and realise it is the $3M$-fold product of U(1) coherent states --- one for each combination of one of the three copies of U(1) and one of the $M$ edges. The calculation of expectation values of the basic operators is then straightforward and yields
\begin{align}
    \langle\lrabs{\hat{p}^{i_0}_{I_0}}^r\rangle_{\cs} &= \csnormalise  \sumset{n} \lrabs{tn^{i_0}_{I_0}}^r \e{\sum_{i,I}\lr{-t\lr{n^i_I}^2+2n^i_Ip^i_I}} \nonumber\\
    &= \csnormalise  \lr{\frac{2\pi}{T}}^{3M}T^r\sumset{N} \; \int\limits_{-\infty}^\infty \d^{3M}x^i_I \, |x^{i_0}_{I_0}|^r \es{\sum_{i,I}\lr{-\lr{x^i_I}^2+\frac{2p^i_I}{T}x-\frac{2\pi\i N^i_I}{T}x^i_I}} \nonumber\\
    &\!\!\overset{t\to 0}{\approx} \lrabs{p^{i_0}_{I_0}}^r \lr{1-\frac{r(1-r)}{4}\frac{t}{\lr{p^{i_0}_{I_0}}^2}-\frac{r(1-r)(2-r)(3-r)}{32}\frac{t^2}{\lr{p^{i_0}_{I_0}}^4}+\mathcal{O}\lr{t^3}}.
\end{align}
We took into account that all the ``new'' $3M-1$ many integrals --- besides the $x^{i_0}_{I_0}$ one --- were just of standard Gaussian type, resulting in $\sqrt{\pi}\exp{\lr{-\lr{\nicefrac{p^i_I-\pi\i N^i_I}{T}}^2}}$. With the same reasoning as before, this allows us to consider only the dominant term $N^i_I=0\,\forall\, i,I$ in the sum over all $N^i_I$ and the remaining contributing factors are cancelled via the state's norm.

We therefore have at hand a succinct way of calculating expectation values of roots of the momentum operator with respect to coherent states by means of the KCHFs, yielding directly the desired classical limit for zeroth order in $t$. Note that~\cite{GCS3} obtained this zeroth order for $r=\frac{n}{2}, n\in\mathbbm{N}$, via the Hamburger moment problem, but did not provide information about the corrections in higher than zeroth order.

\section{Semiclassical analysis of operators involving the volume operator without estimates
} \label{sec:woEstimates}

The aim of this section is to calculate semiclassical expectation values of the kind of~\eqref{eq:Expectationvalue} in such a way that we do not use estimates. As this is not possible in the general case, we start with the simplification to cubic graphs in subsection~\ref{subsec:cubicgraphs}. For this restriction, it turns out that the computation can be reduced to considering a single \textit{basic building block} for determining the semiclassical expectation value shown in \eqref{eq:Expectationvalue}. The calculation of this semiclassical expectation value is then performed in~\ref{subsubsec:rigorous3edges}, where the detailed steps are provided in Appendix~\ref{Appendix9integrals}. More elaborate scenarios with higher valent vertices are subsequently discussed in~\ref{subsec:highervalent}. For all these cases, the calculations are conducted in a way that in the final result the classical limit --- that is sending the classicality parameter $t\to 0$ --- can be taken as long as we choose the momenta $p$, which label the coherent states, as non-vanishing. In contrast, in the context of~\cite{Brunnemann1,Brunnemann2}, this limit leads to a diverging result. If we are interested in the choice $p \to 0$, which is the one that needs to be considered in the case of a cosmological singularity and the one the work of~\cite{Brunnemann1,Brunnemann2} focused on, then also with our method we need to apply some estimates and --- as a consequence --- run into a similar problem if we in addition want to apply the limit $t\to 0$.

\subsection{Cubic graphs} \label{subsec:cubicgraphs}

\subsubsection{General setup} \label{sec:CubicGraphSetup}

In~\cite{Towards2}, expectation values of $\qr$ are calculated for cubic graphs. This restriction allows for important simplifications also when it comes to the evaluation of the eigenvalues of $\qr$.
Concerning prefactors, they used a slightly different form of $\hat V$ than we did:
\begin{align}
    {\hat V}_{\gamma,v} = \lp{}^3 \sqrt{ \lrabs{ \epsilon_{jkl} \frac{{\hat X}^j_{v,e^+_1}-{\hat X}^j_{v,e^-_1}}{2} \cdot \frac{{\hat X}^k_{v,e^+_2}-{\hat X}^k_{v,e^-_2}}{2} \cdot \frac{{\hat X}^l_{v,e^+_3}-{\hat X}^l_{v,e^-_3}}{2} } }. \label{eq:SahlmannThiemannVolumeop}
\end{align}
Therein, ${\hat X}^j_{v,e} = \i h^j_{e} \pdif{}{h^j_{e}} $ corresponds to \cite{Towards2}'s ${\widehat Y}^e_j$ and is the right-invariant vector field of U(1). The formula above illustrates the topology of the cubic graph as we do not have a sum over edges anymore. Instead, we have left six edges, which can be understood as three pairs of one ingoing and one antiparallel outgoing edge each: $e^{\pm}_{I=1,2,3}$. Then, the eigenvalue of $\frac{1}{a^{3r}}\hat{V}^r$ has the form
\begin{align}
    \lambda^r\lr{\{x^j_J\}} = t^{\frac{3r}{4}} \sqrt{ \lrabs{ \epsilon_{jkl} \frac{x^j_{+,1}-x^j_{-,1}}{2} \cdot \frac{x^k_{+,2}-x^k_{-,2}}{2} \cdot \frac{x^l_{+,3}-x^l_{-,3}}{2}  } }^{\,r} , \label{eq:SahlmannThiemannLambda}
\end{align}
confer (4.6) in~\cite{Towards2}. Hence, a variable substitution in which  six charges represent these differences $x_{Jj}^-\coloneqq \frac{x^j_{+,J}-x^j_{-,J}}{2}$ and the remaining six the respective sums $x_{Jj}^+\coloneqq \frac{x^j_{+,J}+x^j_{-,J}}{2}$
lets the latter integrals be of normal Gaussian type as they will not appear in the root / fractional power of the absolute value above anymore\footnote{Note that \cite{Towards2} actually substitutes $\nicefrac{1}{2}$ times the difference and the sum in order to get rid off those denominators. This 8 is furthermore one part of the numerical prefactor $\nicefrac{1}{48}$ of the definition of the volume operator, cf. \eqref{eq:defQ}, whose other part $\nicefrac{1}{3!}$ is compensated by fixing the order of the edges in the vector fields, charges and now $x$.}. 
We realise that the restriction to cubic graphs allows to reduce the originally eighteen-dimensional integral --- coming from three U(1)-charges and six edges per vertex (three ingoing, three outgoing ones) --- to a nine-dimensional one. This is due to the eigenvalue  $\lambda^r_{J_k j_k}$ seeing only the differences of the ingoing and outgoing charges per edge in this scenario and we have 
\begin{align}
    \lambda^r\lr{\{x_{Jj}^-\}} = t^{\frac{3r}{4}} \sqrt{\det(x_{Jj}^-)}^{\,r}. \label{eq:Volx-}
\end{align}
In the following, we will drop the minus label at quantities such as $x^-_{Jj}$ and $p^-_{Jj}$ because we will focus on the computation of the remaining nine-dimensional integral involved in $\langle \prod_{k=1}^N {\hat q}^{j_k}_{J_k}(r) \rangle_{\cs}$ for a graph of cubic topology. 
Taking a look at equation (4.20) of~\cite{Towards2}, which is their equivalent of~\eqref{eq:Expectationvalue} above, and after considering the state's norm, we find
\begin{align}
    \frac{2^N}{t^N}\langle \prod_{k=1}^N {\hat q}^{j_k}_{J_k}(r) \rangle_{\cs} =  \lr{1\pm\mathcal{K}_t} \sqrt{\frac{2}{\pi}}^9 \infint {}^9x \es{-2\sum_{Jj}\lr{x^j_J}^2} \prod_{k=1}^{N}\lambda^r_{J_k j_k}\lr{ \left\{ x^j_J + \tfrac{p^j_J}{T} \right\} }, \label{eq:420Towards2}
\end{align}
wherein the eigenvalues $ \lambda^r_{J_k j_k} $ take the form
\begin{align}
    \lambda^r_{J_k j_k} \lr{ \left\{ x^j_J + \tfrac{p^j_J}{T} \right\} } \coloneqq 2 \frac{\lambda^r\lr{ \left\{ x^j_J + \frac{p^j_J}{T} \right\} } - \lambda^r\lr{ \left\{ x^j_J + \frac{p^j_J}{T}+ \frac{T}{2}\delta^{ii_k}_{JJ_k}} \right\} }{t}. \label{eq:cubiclambda}
\end{align}
Note that we used $\lr{1\pm\mathcal{K}_t}$ as a shorter equivalent for the prefactor used in~\cite{Towards2} --- with $\mathcal{K}_t$ being $\mathcal{O}\lr{t^{\infty}}$, the bracket is assumed to be 1 in the semiclassical limit $t\to0$ eventually. Also, we omitted their $\sigma_k$ as a subscript of $\lambda$, which they used to tell whether an edge is in- or outgoing. As they explain, this orientation becomes irrelevant for cubic graphs and U$(1)^3$. Lastly, the factor of $\frac{2}{t}$ in~\eqref{eq:cubiclambda} was introduced because on the one hand we obtain a factor of $2$ for the regularisation of the Poisson bracket as will be  shown in \cite{GTtoAppear} and on the other hand we replace the $\tau_j$ matrices by $\i$ when substituting ${\rm SU}(2)$ by ${\rm U}(1)^3$ and need to divide by $\frac{1}{\hbar}$ in the quantisation step\footnote{Note that \cite{Towards2} additionally absorbed this prefactor into a new definition of the $\qr$-operator, cf. (4.2) therein (where the 2 in the denominator is supposed to be in the nominator, whereas the usual former regularisation of the Poisson bracket without the additional factor $2$ was used, from our understanding).}.

The simplest scenario now is $N=1$, resulting in a reduction involving a difference of two charge matrices of only $3\times3$ charges each, with the subtrahend's charge matrix containing a shift in one of the entries via $+\frac{T}{2}\delta^{ii_k}_{JJ_k}$. We call this quantity the \textit{basic building block} of the kind of semiclassical expectation values  in \eqref{eq:Expectationvalue}. In the next subsection, we will present how  the computation thereof can performed by means of KCHFs and their asymptotics finally yielding the result shown in~\eqref{3edgesFinal}.

\subsubsection{Analytical computation of \texorpdfstring{$\langle \qr \rangle_{\cs}$}{q} for a graph of cubic topology} \label{subsubsec:rigorous3edges}

The aim of this subsection is to provide a way of analytically calculating  the semiclassical expectation values of the $\qr$-operators for graphs with cubic topology, as introduced before. Without loss of generality we will assign the shift involved in the volume operator's action to the $x^1_1$-component of the first edge and the first U(1)-copy (cf.~\eqref{eq:qr} and~\eqref{eq:startingExpectationvalue})\footnote{Starting with the shift in any other matrix element, one can always rearrange the matrix elements in such a way that the shift appears in this component, only picking up minus signs that will be annihilated by the absolute value; or just perform the same steps with respect to that matrix element.}. This is reflected in the super- and subscript of $\hat{q}^1_1(r)$. 
We will furthermore also state the general result at the end, which is deducible from this specific starting point as we will ultimately see. 

Our starting point right after the initial Poisson resummation then reads
\begin{align}
    &{\frac{2}{t}} \langle\hat{q}^1_1(r) \rangle_{\cs}  = \nonumber\\
    &  = \frac{2}{t}\frac{\lp{}^{6r}}{a^{6r}\csnorm}  \sumset{N} \lr{\tfrac{2\pi\sqrt{2}}{T}}^{\!9} \infint^{9}x^i_I \es{2\sum_{Ii}\lr{-\lr{x^i_I}^2+2\frac{p^i_I-\pi\i N^i_I}{T}x^i_I}}\frac{\lr{|\det{X}|^r-|\det{\widetilde{X}}|^r}}{T^{3r}}  \nonumber\\
    &  = \frac{2}{t}\frac{\lr{\frac{2\pi\sqrt{2}}{T}}^{\!9}T^{3r}}{\csnorm}  \isumset{N} \es{\sum_{i}2\lr{\frac{p_i-\pi\i N_i}{T}}^2} \infint^{9}x_i\es{-2\sum_{i}\lr{x_i-\frac{p_i-\pi\i N_i}{T}}^2} \lr{|\det{X}|^r-|\det{\widetilde{X}}|^r} ,\label{StartingPoint}
\end{align}
where
\begin{align}
   X \coloneqq \begin{pmatrix}
x^1_1 & x^2_1 & x^3_1 \\
x^1_2 & x^2_2 & x^3_2 \\
x^1_3 & x^2_3 & x^3_3
\end{pmatrix}
\eqqcolon \begin{pmatrix}
x_1 & x_2 & x_3 \\
x_4 & x_5 & x_6 \\
x_7 & x_8 & x_9
\end{pmatrix} \qquad\text{and}\qquad \widetilde{X}\coloneqq \begin{pmatrix}
x_1+\frac{T}{2} & x_2 & x_3 \\
x_4 & x_5 & x_6 \\
x_7 & x_8 & x_9
\end{pmatrix}. \label{3edgesmatrices}
\end{align}
By $\int_{-\infty}^{\infty}\mathrm{d}^{9}x^i_I$ we understand $\int_{-\infty}^{\infty}\cdots\int_{-\infty}^{\infty} \mathrm{d}x^1_1 \mathrm{d}x^2_1\cdots\mathrm{d}x^3_3 $ and $\sum_{Ii} $ stands for $\sum_{\stackrel{i=1,2,3}{I=1,2,3}}$, while from the second line onwards $i=1,\ldots,9$ is no longer labelling the U(1)-copy but instead running through the matrix elements $x_i$ and accordingly $p_i$ and $N_i$. Note that these quantities are precisely the $x^-_{Jj}$, $p^-_{Jj}$ and $n^-_{Jj}$ of \cite{Towards2} that correspond to the difference quantities in the course of the aforementioned substitution.
In contrast to the calculations before, we use now $r$ as the exponent of the determinants' absolute values. This is only due to clearer formulae. Accordingly, the exponents of $\lp{}^{3r}$ and  $a^{3r}$ changed to $\lp{}^{6r}$ and  $a^{6r}$, respectively.
As in the calculations before, we defined $x_i \coloneqq Tn_i $, with $T^2 \coloneqq t$, during the Poisson resummation.

To outline the procedure --- which will mainly be the same throughout these calculations ---, we first isolate the $x_1$-integration of the unshifted part:
\begin{align}
    & \infint x_1 \es{-2\lr{x_1-\frac{p_1-\pi\i N_1}{T}}^2} \lrabs{\det X}^r = \nonumber\\
    & = \infint x_1 \es{-2\lr{x_1-\frac{p_1-\pi\i N_1}{T}}^2} \lrabs{x_1x_5x_9+x_2x_6x_7+x_3x_4x_8-x_1x_6x_8-x_2x_4x_9-x_3x_5x_7}^r.
\end{align}
We then see that we can cast this expression into one of the form (\ref{KCHF}) via the substitution
\begin{align}
    x'_1 &\coloneqq \det{X} ,\label{x1subst}\\
    x'_{2,\ldots,9} &\coloneqq x_{2,\ldots,9},
\end{align}
with
\begin{align}
    \det{\lr{\frac{\d x'}{\d x}}} = \det \begin{pmatrix}
x_5x_9-x_6x_8 & x_6x_7-x_4x_9 & x_4x_8-x_5x_7 & \dots \\
0 & 1 & 0 & \dots \\
0 & 0 & 1 & \dots \\
\vdots & \vdots & \vdots & \ddots
\end{pmatrix} = x_5x_9-x_6x_8.
\end{align}
This allows us to rewrite the integral above as
\begin{align}
    \infint x_1 & \es{-2\lr{x_1-\frac{p_1-\pi\i N_1}{T}}^2} \lrabs{\det X}^r = \nonumber\\
    &=\infint x'_1 \es{-2\lr{\frac{x'_1}{x'_5x'_9-x'_6x'_8} + \frac{x'_3x'_5x'_7+x'_2x'_4x'_9-x'_2x'_6x'_7-x'_3x'_4x'_8}{x'_5x'_9-x'_6x'_8} - \frac{p_1-\pi\i N_1}{T}}^2} \frac{\lrabs{x'_1}^r}{\lrabs{x'_5x'_9-x'_6x'_8}} \nonumber\\
    & \eqqcolon \infint x'_1 \es{-2\lr{\frac{x'_1}{x'_5x'_9-x'_6x'_8} + x_0}^2} \frac{\lrabs{x'_1}^r}{\lrabs{x'_5x'_9-x'_6x'_8}}. \label{eq:9intFirstIntegral}
\end{align}
The long offset $x_0$ of the Gaussian stems from inserting the inversion $x_1 = x_1(x'_1,\ldots,x'_9)$ according to (\ref{x1subst}). Note that during the next integrations, shown in detail in Appendix~\ref{Appendix9integrals}, we directly abbreviate those by single letters like $a,b,\sigma,\tilde{\sigma},$ etc.

Note that the substitution gives rise to the denominator $\lrabs{x'_5x'_9-x'_6x'_8}$, which can for some combinations approach or equal 0. This is, however, not an issue with two terms of the Gaussian's argument being $\sim \nicefrac{1}{\lr{x'_5x'_9-x'_6x'_8}}$, too, hence suppressing those contributions. One could, furthermore, expect this denominator to cause problems for the following $x'_5$-, $x'_9$-, $x'_6$- \& $x'_8$-integrations, but it will neatly merge with the $x'_1$-integration's result and yield an in an analogous manner integrable expression.

Applying now the integration according to (\ref{KCHF}) and (\ref{KummerTrafo}) in the fashion of
\begin{align}
    \infint x \es{-2\lr{\frac{x}{s}-x_0}^2}\lrabs{x}^r = (\sqrt{2})^{-1-r}\lrabs{s}^{1+r} \Gsum \e{-2\lr{x_0}^2}\kchf{\opr,\frac{1}{2},2\lr{x_0}^2}
\end{align}
leads to 
\begin{align}
    \infint x_1 & \es{-2\lr{x_1-\frac{p_1-\pi\i N_1}{T}}^2} \lrabs{\det X}^r = \infint x'_1 \es{-2\lr{\frac{x'_1}{x'_5x'_9-x'_6x'_8} + x_0}^2} \frac{\lrabs{x'_1}^r}{\lrabs{x'_5x'_9-x'_6x'_8}} \nonumber\\ &=\frac{\lrabs{x'_5x'_9-x'_6x'_8}^{1+r}}{\lrabs{x'_5x'_9-x'_6x'_8}}(\sqrt{2})^{-1-r} \Gsum \e{-2\lr{x_0}^2} \kchf{\opr,\frac{1}{2},2\lr{x_0}^2}.\label{afterx1int}
\end{align}

Note that $x_0 = x_0\lr{x'_2,\ldots,x'_9,p_1}$ and we therefore still have a dependence on all the remaining $x_i$, additionally to the Gaussians in the $x_i$ that we just omitted for brevity. The $x_1$-dependence, however, was replaced by a $p_1$-dependence through the integration --- as expected.

Next, we isolate the $x'_5$-integration together with the result of (\ref{afterx1int}):
\begin{align}
    \infint x'_5 \es{-2\lr{x'_5-\frac{p_5-\pi\i N_5}{T}}^2} \lrabs{x'_5x'_9-x'_6x'_8}^r (\sqrt{2})^{-1-r}\Gsum \e{-2\lr{x_0}^2} \kchf{\opr,\frac{1}{2},2\lr{x_0}^2} \nonumber . 
\end{align}
Regarding the $r$-th power of the absolute value, we see that we can again perform a substitution to cast its argument into the desired form. However, the presence of the KCHF as well as the Gaussian in $x_0\equiv x_0(x'_5)$ make a straightforward analytical integration unfeasible. Taking a closer look at the KCHF's argument $x_0$ as defined during~(\ref{eq:9intFirstIntegral}), we see \begin{equation}x_0 = f\lr{\left\{x_{i\backslash 1}\right\}} - \frac{p_1-\pi\i N_1}{T}, \end{equation}
where $f$ is a function of all $x_i$ except $x_1$. As we want to consider semiclassical expectation values, we have in mind the limit of small $t$ and hence small $T = \sqrt{t}$. This allows us to perform an asymptotic expansion for large arguments of the KCHF, (\ref{expansion}), for the $x_1$-integration's result (\ref{afterx1int}), yielding
\begin{align}
\label{eq:Resx1Int}
    \infint x_1 & \es{-2\lr{x_1-\frac{p_1-\pi\i N_1}{T}}^2} \lrabs{\det X}^r = \nonumber\\
    & = \lrabs{x'_5x'_9-x'_6x'_8}^r(\sqrt{2})^{-1-r} \Gsum \e{-2\lr{x_0}^2} \kchf{\opr,\frac{1}{2},2\lr{x_0}^2} \nonumber\\
    &\!\!\stackrel{t\to 0}{\approx} \frac{1}{\sqrt{2}} \lrabs{x'_5x'_9-x'_6x'_8}^r \Ghalf \lr{\lr{x_0}^2}^\rhalf \lr{1-\rrfour\frac{\lr{x_0}^{-2}}{2}+\mathcal{O}\lr{\lr{x_0}^{-4}}},
\end{align}
where we cancelled the factor $\lr{\sqrt{2}}^r$ within $\lr{2(x_0)^2}^\rhalf$ with the according prefactor. Here, as in the former example in section~\ref{subsec:mtmoproots}, only that series of the asymptotic expansion was kept that cancelled the damping Gaussian in $x_0$. Note that the results of ~\cite{Brunnemann1,Brunnemann2} and~\cite{Towards2} also feature a series of integer powers in the momenta and $T$, coming from either estimates or a Taylor expansion applied before the integration. In our case here, the Kummer function that is obtained from the integration and its subsequent asymptotic expansion for large arguments lead to a series in integer powers as well. In contrast to ~\cite{Brunnemann1,Brunnemann2} and~\cite{Towards2}, here the terms still involve absolute values with the initial exponent $r$, thereby retaining information about this quantity.

Both appearances of $x_0$ in the result above have the effect that the expression cannot be integrated analytically with respect to $x'_5$ by means of KCHFs, but we can circumvent these last obstacles as well. First of all, we go back to (\ref{StartingPoint}) and note that there is a Gaussian in $\nicefrac{\lr{p_1-\pi\i N_1}}{T}$ due to completing the square of the Gaussian in $x_1$. Now, since we already performed the $x_1$-integration, there is no possibility of additional Gaussians in $\nicefrac{N_1}{T}$ to arise\footnote{Also, as will be seen later in (\ref{eq:statesnorm}), dividing by the norm $\csnorm $ will indeed give rise to Gaussians in $\nicefrac{p_i}{T}$, but none in $\nicefrac{N_1}{T}$.} and we can argue that all contributions with $N_1\neq 0$ are damped by that very Gaussian, with only the solution $N_1=0$ surviving. Note that applying this  has widely been used in the literature~\cite{GCS2,GCS3,Towards2,Brunnemann2}, where it was also assumed that only the $N=0$ terms contribute in the semiclassical limit, as it is an ultimate consequence of the Poisson resummation. This then causes $x_0\big|_{N_1=0}\in\mathbbm{R}$ and hence, using the same symbol $x_0$, we have
$$ \lr{\lr{x_0}^2}^\rhalf = \lrabs{x_0}^r .$$
Multiplying
\begin{align}
    \lrabs{x'_5x'_9-x'_6x'_8}^r \lrabs{x_0}^r &= \lrabs{x'_3x'_5x'_7+x'_2x'_4x'_9-x'_2x'_6x'_7-x'_3x'_4x'_8 - \lr{x'_5x'_9-x'_6x'_8}\frac{p_1}{T}}^r \nonumber\\
    &\eqqcolon \lrabs{ x'_5\lr{x'_3x'_7-x'_9\frac{p_1}{T}} + \tilde{x}_0 }^r, \label{determinantx5}
\end{align}
we see that we face in~\eqref{eq:Resx1Int} a similar situation as before the initial $x_1$-integration --- just with $x_1$ being replaced by $\frac{p_1}{T}$ within the determinant-like absolute value's argument. This expression itself can therefore likewise be integrated via a variable substitution, if we can also somehow handle the last term involving $\lr{x_0}^{-2}$. To clear this hurdle, remember first that this term arose via the asymptotic expansion of KCHFs, as $x_0\sim\nicefrac{1}{T}$. But given this, we can additionally perform a Taylor expansion in order to directly get a series in $T$, in place of  the intermediate variable $x_0$. This yields
\begin{equation}
    1-\rrfour\frac{\lr{x_0}^{-2}}{2} \approx 1-\rrfour\frac{T^2}{2p_1{}^2} + \mathcal{O}\lr{T^3} \eqqcolon \mathscr{S}. \label{eq:Taylor9int}
\end{equation}
Having no $x_i$-dependence up to the second order in $T$, we have achieved our goal and only need to integrate the determinant-like expression in (\ref{determinantx5}) against the Gaussian in $x'_5$ just by means of the $x_1$-integration's steps. Note that the series' $~\sim T^3$ contribution does indeed include the integration variables. For increasing the expansion's order, one has to proceed differently for this term, i.e. try a different substitution for example. However, as this is of higher order than we are currently interested in, we leave this to future investigations.

To continue, the starting point here then reads
\begin{equation}
    \Ghalf\mathscr{S}\infint x'_5 \es{-2\lr{x'_5-\frac{p_5-\pi\i N_5}{T}}^2} \lrabs{ x'_5\lr{x'_3x'_7-x'_9\frac{p_1}{T}} + \tilde{x}_0 }^r  \nonumber,
\end{equation}
where we also carried along the emerged factors $\Ghalf$ and $\mathscr{S}$.

We can then proceed by the next substitution
\begin{align}
    x''_5 &\coloneqq x'_5\lr{x'_3x'_7-x'_9\frac{p_1}{T}} + \tilde{x}_0 ,\\
    x''_{2,3,4,6,7,8,9} &\coloneqq x'_{2,3,4,6,7,8,9},\\
    \det{\lr{\frac{\d x''}{\d x'}}} &\,= x'_3x'_7-x'_9\frac{p_1}{T} ,
\end{align}
which then requires us to integrate the Gaussians against $$\frac{|x''_5|^r}{|x'_3x'_7-x'_9\frac{p_1}{T}|}.$$

The general procedure is exactly the one we just conducted and summarise within the next paragraph. The remaining integrations can be found in Appendix~\ref{Appendix9integrals}, where we then directly start in a more concise manner of performing the two just mentioned substitutions at once.

Let us summarise the method we used. We started with substituting the determinant argument within the absolute value as a new integration variable, causing a further offset in the respective Gaussian. For this expression of the integral, we can then use (\ref{KCHF}) and (\ref{KummerTrafo}) to obtain a result containing a KCHF. With the KCHF's argument being dependent on all remaining $x_i$, we can not proceed directly with further integrations. But as the KCHF's argument contains also the term $\nicefrac{p_1}{T}$, we can perform the asymptotic expansion for large arguments of the KCHF, (\ref{expansion}), on the result of the first integration, obtaining a power series in the argument's inverse. This series was then transformed into a power series in $T$ itself via a standard Taylor expansion. Via the asymptotic expansion, however, an additional factor arose --- the KCHF's argument to the power of $\nicefrac{r}{2}$. With the argument of the KCHF itself being the square of a complex number, we were able to transform this term into the absolute value to the power of $r$ by getting rid of the imaginary part, which was proportional to $N_1$ and thus of higher order than the $N_1=0$ contribution: The involved Gaussian in $\nicefrac{\lr{p_1-\pi\i N_1}}{T}$ in (\ref{StartingPoint})  damps all contributions $N_1\neq 0$ to zero, except for the $N_1=0$ case. This allowed us to solely work with the real part of the argument of the KCHF  by only considering the $N_1=0$-contribution and hence obtaining the aforementioned absolute value. From that point on, we face a similar situation as in the beginning just with the $x_1$-integration being already performed, which causes $x_1$ being replaced by $\nicefrac{p_1}{T}$ in the absolute value's argument. Therefore, we can go through the same steps for the remaining eight integrals\footnote{Two integrals will in fact be somewhat different, but the main method remains the same.}.

For the shifted contribution, the procedure remains the same, too. By considering the substitution $\tilde{x}_1 \coloneqq x_1 +\frac{T}{2} $ and hence
\begin{equation}
    \infint x_1 |\det \widetilde{X}|^r \es{-2\lr{x_1 - \frac{p_1-\pi\i N_1}{T}}^2} = \infint \tilde{x}_1 |\det X|^r \es{-2\lr{\tilde{x}_1 - \frac{p_1+\frac{T^2}{2}-\pi\i N_1}{T}}^2},
\end{equation}
we see that the only change is $p_1 \mapsto p_1+\frac{T}{2}^2$. This, of course, allows for an analogous treatment of the integrations. Moreover, it is important that also the series in $T$ that arise due to the asymptotic expansion of the KCHFs does not change: The reason for this is that we were able to perform this expansion with $\nicefrac{p_1}{T}$ being large. As we now have $\nicefrac{p_1}{T}+\frac{T}{2}$, the latter part has no significant contribution in that expansion's power series since we assume small $t$ and thus also small $T$ . However, the shift still leaves a trace in the expansion's absolute value factor --- which is the important part that survives in the commutator's difference.

Since we are interested in the classical limit of the operator $\frac{2}{t}\hat{q}^1_1(r)$, which should yield the classical cotriad for $r=\frac{1}{2}$, we need to consider the involved multiplication by $\frac{2}{t}$ to obtain the lowest order in $t$. After going through the remaining integrals, which are presented in detail in Appendix~\ref{Appendix9integrals}, we end up with the final result for the semiclassical expectation value of the basic building block for cubic graphs:
\begin{equation}
\label{3edgesFinal}
\boxed{
\begin{aligned}
    \frac{2\langle \hat{q}^1_1(r) \rangle_{\cs}}{t} & \approx - r \frac{|\det p|^r\Delta^1_1\lr{p}}{\det p}  +  \mathscr{F}(\{p_i\})\;\!t + \mathcal{O}\lr{t^{\frac{3}{2}}},
\end{aligned}
}
\end{equation}
where we chose the classicality parameter $t=T^2=\nicefrac{{\lp}^2}{a^2}$ in a final Taylor expansion in $T$ in order to get the correct limit for the lowest order in $T$. Recall that we replaced the initial exponent $\rhalf$ by $r$. In the result above, $\det p$ is the determinant of the matrix of the momenta $p^i_I$, i.e. $\det X\vert_{x^i_I\mapsto p^i_I}$, and $\Delta^1_1\lr{p} \coloneqq p^2_2 p^3_3 - p^2_3 p^3_2 \equiv p_5 p_9 - p_6 p_8$ is the minor of that matrix with respect to $p^1_1 \equiv p_1$, where the holonomy's shift occurred. Lastly, $\mathscr{F}(\{p_i\})$ is a function containing all the $\{p_i\}$, defined as in \eqref{eq:T4-term}\footnote{Note that this term contributes with $T^4$ (before dividing by $t$) despite having considered $\mathscr{S}$ only up to $T^2$. This is due to the final result's multiplicative structure, making the series' terms $\mathcal{O}\lr{T^3}$ at least $\mathcal{O}\lr{T^5}$ --- cf. \eqref{puttingtogether}.} in Appendix~\ref{Appendix9integrals}. The structure of the result's lowest order contribution is indeed what one would expect when comparing it with the classical Poisson bracket: Starting with the absolute value of a determinant to the power of $r$, by differentiation we get that determinant to the power of $r-1$ and an additional factor via the chain rule that is just the minor of the matrix with respect to the momentum with respect to which we differentiated the determinant.

We may now deduce the general result from the one above as follows:
\begin{equation}
\label{3edgesFinalGeneral}
\boxed{
\begin{aligned}
    \frac{2\langle \hat{q}^{i_0}_{I_0}(r) \rangle_{\cs}}{t} & \approx -  r \frac{|\det p|^r\Delta^{i_0}_{I_0}\lr{p}}{\det p}  +  \widetilde{\mathscr{F}}(\{p_i\})\;\!t + \mathcal{O}\lr{t^{\frac{3}{2}}}.
\end{aligned}
}
\end{equation}
Therein, $\widetilde{\mathscr{F}}(\{p_i\})$ is of similar structure as the $\mathscr{F}(\{p_i\})$ before, but the specific arrangement of the $p_i$ within the expression depends on the chosen $i_0$ and $I_0$. For the zeroth order, however, the generalisation was possible in this straightforward manner as all the steps leading to this term can be performed for any other starting choice of $i_0$ and $I_0$ in just the same way and all upcoming factors of $p_i$ added up to $\det p$.

Note that we did not need to use estimates for obtaining this result. While~\cite{Brunnemann1,Brunnemann2} considered general graphs and were thus forced to use estimates, we, in turn, only considered cubic graphs so far. And in contrast to~\cite{Towards2}, where they used a Taylor expansion to replace fractional powers of the momentum operators by integer ones, we used them at intermediate stages that, however, did not affect the integration variables up to the respective order in $T$ (cf.~\eqref{eq:Taylor9int} or later and also see the notes after \eqref{xsquareintegration} and \eqref{beforeLastTaylorWithVariable}).

If we further compare our result in \eqref{3edgesFinalGeneral} with the one of Sahlmann and Thiemann in \cite{Towards2} in Theorem 4.2 for the special case $N=1$ (eq. (4.45) therein) then the structures of the final results look slightly different. The reason for this is that in \eqref{3edgesFinalGeneral} the final result is still expressed in terms of fluxes, whereas in Theorem 4.2 the dependence of the fluxes on the lattice parameter and thus lattice fluctuations have already been considered explicitly. We will discuss the latter in more detail in subsection \ref{subsec:SemClassCubic} as well as in subsection \ref{subsec:highervalent}, below \eqref{eq:BBBSanThiemann}, in which for the case of cubic graphs the semiclassical expectation $\langle \prod_{k=1}^N {\hat q}^{j_k}_{J_k}(r) \rangle_{\cs}$  is presented using the method introduced in \cite{Towards2}. In order to investigate the cosmological singularity we can, as in \cite{Brunnemann1,Brunnemann2}, work with the result expressed in terms of the fluxes, which we will discuss in the next subsection.

\subsubsection{Cosmological singularity for the graph of cubic topology}
\label{sec:CosmoSing}
As far as the cosmological singularity is concerned, we need to look at the limit $p_i=0$. This limit cannot be taken in the result~\eqref{3edgesFinal} or~\eqref{3edgesFinalGeneral} and we have to proceed differently for this case. The main reason for this is that we cannot use the asymptotic expansion for large arguments of the Kummer functions anymore: The arguments of the KCHFs that we came across before were always $\sim \nicefrac{p_i}{T}$ for some $p_i$ or functions thereof. Hence, it was justified to assume that the argument is large  as $T^2=t\to0$. This kind of argument, however, is not applicable for $p_i=0$ and as we always consider only contributions from  $N_i=0$ in the end, also terms in the argument that are proportional to $\nicefrac{N_i}{T}$ cannot be used instead. This forces us to use estimates in this case, as we are left with integrals still including KCHFs when we cannot apply their asymptotic expansion. So now, the starting point  is \eqref{StartingPoint}$\big|_{p_i=0}$ and we estimate the difference in the determinants the following way:
\begin{align}
    \lrabs{ \det X}^r - \lrabs{\det \widetilde{X}}^r & = \lrabs{ \det X}^r - \lrabs{ \det X + \frac{T}{2}\Delta^1_1 \lr{X}}^r \nonumber\\
    & \!\!\!\!\; \stackrel{\eqref{approxdiff}}{\leq} \frac{T^r}{2^r} \lrabs{\Delta^1_1 \lr{X}}^r = \frac{T^r}{2^r} \lrabs{x_5x_9-x_6x_8}^r \nonumber\\
    & \!\!\!\!\; \stackrel{\eqref{approxsum}}{\leq} \frac{T^r}{2^r} \lr{ \lrabs{x_5x_9}^r+\lrabs{x_6x_8}^r} \nonumber\\
    & \!\!\!\!\; \stackrel{\eqref{2ab}}{\leq}  \frac{T^r}{2^{2r}} \lr{ \lr{\lrabs{x_5}^2+\lrabs{x_9}^2}^r + \lr{\lrabs{x_6}^2+\lrabs{x_8}^2}^r } \nonumber\\
    & \!\!\!\!\; \stackrel{\eqref{approxsum}}{\leq} \frac{T^r}{2^{2r}}\lr{ \lrabs{x_5}^{2r} + \lrabs{x_9}^{2r} + \lrabs{x_6}^{2r} + \lrabs{x_8}^{2r} } , \label{eq:Estimates9intP=0}
\end{align}
where used several estimates that can be found in Appendix~\ref{sec:Estimates} and which will be explained in more detail in section~\ref{sec:wEstimates}, which covers approaches that necessitate the application of estimates also for the $p_i \neq 0$ case. 
This expression above can now clearly be integrated resulting in KCHFs: All contained integrals are of the form
\begin{align}
    \infint x \es{-2x^2} \lrabs{x}^{2r} = \frac{\Gamma\lr{r+\tfrac{1}{2}}}{2^{r+\frac{1}{2}}}, \label{eq:Integral9IntP=0}
\end{align}
where the five of standard Gaussian type follow for $r=0$.
As before, we are interested in the semiclassical limit of the operator $\hat{q}^1_1(r)$ and thus divide the result by $t$ due to the involved commutator. With only considering the $N_i=0$ contributions\footnote{Note that the reason for only including these parts remains the same as before and is not affected by the choice $p_i=0$.}, we finally obtain
\begin{equation}
 \label{eq:resultRigorousP=0}
 \boxed{
\begin{aligned}
    \frac{2\langle \hat{q}^1_1(r) \rangle_{\cs}}{t} \ & \stackrel{p_i=0}{=} \ \frac{2}{t}\frac{\lr{\frac{2\pi\sqrt{2}}{T}}^{\!9}T^{3r}}{\csnorm} \infint^{9}x_i\es{-2\sum_{i}\lr{x_i}^2} \lr{|\det{X}|^r-|\det{\widetilde{X}}|^r} \\
    &\ \;\! \leq \frac{2}{t}\frac{\lr{\frac{2\pi\sqrt{2}}{T}}^{\!9}T^{4r}}{\csnorm \, 2^{2r}} \infint^{9}x_i\es{-2\sum_{i}\lr{x_i}^2} \lr{ \lrabs{x_5}^{2r} + \lrabs{x_9}^{2r} + \lrabs{x_6}^{2r} + \lrabs{x_8}^{2r} } \\
    &\ \;\! =  \frac{8}{\sqrt{\pi}\;\!2^{3r}}\Gamma\lr{r+\tfrac{1}{2}}t^{2r-1},
\end{aligned}
}
\end{equation}
where we used again $t=\nicefrac{\lp{}^2}{a^2}=T^2$. We see that we even obtain a $t^0$-result for the commutator divided by $t$ when choosing $r=\tfrac{1}{2}$, which corresponds to considering exactly $\hat V$ itself --- see \eqref{StartingPoint} for the choice of $r$ and recall that we replaced the initial exponent $\rhalf$ by $r$.
We want to point out that we were not expecting to obtain a result $\sim t^0$ for the $p_i=0$ case, as we already did not get one for the quantum mechanical scenario in our companion paper~\cite{paper1} (cf. (3.13) therein). However, while $\hbar$ carried a positive exponent there, the exponent of $t$ can become negative in the formula above if we, for example, consider the important case of $\sqrt{\hat V}$. In those cases, the limit in which $t$ is send to zero is not well-defined. If we compare our result with the one of Brunnemann and Thiemann in \cite{Brunnemann1,Brunnemann2}, then the result in (4.7) on page 25 in \cite{Brunnemann1}\footnote{Note that one has to divide by $t$.} demonstrates that for their estimation a finite $t^0$ result is obtained for choosing $r=\frac{4}{3}>1$, which in their notation corresponds to $V^{\frac{4}{3}}$. Given this, we realise that the final result for the upper bound and its implications do depend on the way how the fractional power is handled in the estimate, which is not too surprising but rather expected.

We can again deduce the general result from the one above,
\begin{equation}
 \label{eq:resultRigorousP=0General}
 \boxed{
\begin{aligned}
    \frac{2\langle \hat{q}^{i_0}_{I_0}(r) \rangle_{\cs}}{t} \ & \stackrel{p_i=0}{\leq}  \frac{8}{\sqrt{\pi}\;\!2^{3r}}\Gamma\lr{r+\tfrac{1}{2}}t^{2r-1},
\end{aligned}
}
\end{equation}
as we now face no difference at all: The initial choice of $i_0$ and $I_0$ only changes the resulting minor of the determinant during the chain of estimates \eqref{eq:Estimates9intP=0} and likewise the four integration variables that appear there afterwards. However, the integration of all those four will still be of the kind of \eqref{eq:Integral9IntP=0}, i.e. independent of the specific $x_i$. Hence, the same holds for the final result.

Before we address in the next subsection the scenario of higher valent vertices than the six valent ones in the cubic case and also briefly summarise our results of section~\ref{sec:woEstimates} at the end, we want to discuss the semiclassical continuum limit for the cubic graph and in this context the corresponding lattice fluctuations in higher than linear order to compare our results with the ones in \cite{Towards2}.

\subsubsection{Semiclassical continuum limit for graphs of cubic topology}
\label{subsec:SemClassCubic}
In subsection \ref{subsubsec:rigorous3edges}, we computed the semiclassical limit of  $\frac{2}{t}\langle \hat{q}^{i_0}_{I_0}(r) \rangle_{\cs}$ for graphs of cubic topology, which in the lowest order we expect to agree with  $\frac{1}{a^{6r}}\{\int_{e_{I_0}}A^{i_0}(x),V^{2r}_{R_x}\}$, where $V_{R_x}$ denotes the volume of a region $R_x$ around the point $x$. In this subsection, we again consider the limit in which we send the classicality parameter $t$ to zero but in addition combine this with the limit in which the regularisation parameter $a$ associated with the cubic graph is sent to zero.  To send both parameters, the semiclassicality as well as the lattice regularisation parameter to zero simultaneously works for the leading order only. In higher order due to the involved lattice corrections in general these two limits are in conflict meaning that in general higher order corrections can become huge if the regularisation parameter is sent to zero. To deal with this situation in \cite{Towards2} a lattice regularisation parameter $\epsilon$ of the form $\epsilon=\lp{}^\alpha a^{1-\alpha}$ with $0<\alpha<\frac{1}{2}$ was introduced. This allows to combine the contributions in terms of powers of the semiclassicality parameter $t=\lp{}^2/a^2$ with those from the lattice corrections and judge whether the higher order corrections remain small compared to the leading order for a small but non-vanishing semiclassicality parameter $t$. This depends of course on the choice of $\alpha$ and in \cite{Towards2} a choice of $\alpha=\frac{1}{6}$ was favoured for the operators considered in \cite{Towards2}.  In this subsection, we will focus on the details of the leading order only in order to show that the method to compute the semiclasscial limit of the cubic graphs presented in this article has the correct classical limit. A more detailed discussion about the lattice fluctuations and the necessity to adjust the power counting accordingly can be found in subsection \ref{subsec:highervalent} after the result in \eqref{eq:BBBSanThiemann}. With such an analysis of the leading order contribution we can check whether the result obtained in \ref{subsubsec:rigorous3edges} that is still discretised on a graph of cubic topology yields the correct expression for the continuum limit. For this purpose, it is sufficient to consider only the leading order of \eqref{3edgesFinal} because all higher order terms vanish in this limit anyway. In order to discuss the computation of this limit in detail, we reintroduce the minus label of the variables of the cubic graph that we introduced below \eqref{eq:SahlmannThiemannLambda} and \eqref{eq:Volx-} and dropped from thereon. Denoting the regulator by $\epsilon$, we show in Appendix \ref{sec:AppSemLimitCubic}  that
\begin{align}
\label{eq:ContLimit}
  \lim\limits_{t\to 0}   \frac{2\langle \hat{q}^{i_0}_{I_0}(r) \rangle_{\cs}}{t} &= - r \frac{\lrabs{\det p^-}^r\Delta^{i_0}_{I_0}\lr{p^-}}{\det p^-}=
    2\i \:\! h^{i_0}_{I_0}\left\{\left(h^{i_0}_{I_0}\right)^{-1},V^{2r}\lr{R_{\Box_\epsilon}}\right\}
\end{align}
and
\begin{align}
\lim\limits_{\epsilon\to 0} \lim\limits_{t\to 0}   \frac{2\langle \hat{q}^{i_0}_{I_0}(r) \rangle_{\cs}}{t} &=
\lim\limits_{\epsilon\to 0}\left(2\i \:\! h^{i_0}_{I_0}\left\{\left(h^{i_0}_{I_0}\right)^{-1},V^{2r}\lr{R_{\Box_\epsilon}}\right\}\right) = \frac{1}{a^{6r}}\left\{\int_{e_{I_0}}A^{i_0}, V^{2r}\lr{R_x}\right\} .
\end{align}
Therefore, in the case of cubic topology using the techniques of Kummer functions and their asymptotic expansion, we have shown that we obtain the correct classical limit in lowest order in the classicality parameter. As one can see from the detailed derivation in Appendix \ref{sec:AppSemLimitCubic}, there are two crucial ingredients for obtaining the correct classical limit. One is the regularisation constant $Z\coloneqq C_{\rm reg}=\frac{1}{48}$ that was found in \cite{VolAshtekar} and through an independent consistency check in \cite{VolGiesel,VolGiesel2}. The second one is an additional factor of $2$ that needs to be considered when using the Thiemann identity and regularising the involved classical Poisson bracket. The occurrence of this additional factor of $2$ will be discussed in detail in \cite{GTtoAppear}.  Note that if we perform a similar analysis for graphs of other than cubic topology, which have a higher valence, then one does not obtain the correct classical continuum limit due to the fact that the final result still involves the regularisation constant $Z=C_{\rm reg}$ to some rational power $r$. This can for instance be seen in the next subsection and the result \eqref{eq:BBBSanThiemann}, where we discuss such a scenario. 

\subsection{Higher valent vertices} \label{subsec:highervalent}

\subsubsection{General setup}

The calculation for cubic graphs before was feasible in that fashion since we were facing only one determinant as the root's argument --- stemming from having to consider only the basic building block of the $\hat q$-operator's eigenvalues. If we now go back from those to the full expression in~(\ref{eq:Expectationvalue}), we have to deal with the root of a sum of determinants in the generic case. Taking a look at the integrand of the expectation value we want to calculate, which is determined by~(\ref{eq:Expectationvalue}), and extracting the non-Gaussian part
\begin{align}
    \left| \sum_{IJK} \epsilon(IJK)\epsilon_{ijk} x^i_I x^j_J x^k_K \right|^{\frac{r}{2}} - \left| \sum_{IJK} \epsilon(IJK)\epsilon_{ijk}\lr{x^i_I+T\delta^{ii_0}_{II_0}}\lr{x^j_J+T\delta^{ji_0}_{JI_0}}\lr{x^k_K+T\delta^{ki_0}_{KI_0}}\right|^{\frac{r}{2}} \nonumber
\end{align}
from it, we can divide these sums of determinants into different parts via an application of Laplace's formula on the terms including $x^{i_0}_{I_0}$. This leads to 
\begin{align}
    &\left| \sum_{IJK} \epsilon(IJK)\epsilon_{ijk} x^i_I x^j_J x^k_K \right|^{\frac{r}{2}} = \nonumber\\
    & \quad= \left| x^{i_0}_{I_0} \sum_{JK} \Delta^{i_0}_{I_0}\lr{x_{JK}} - x^{i_0+1}_{I_0} \sum_{JK} \Delta^{i_0+1}_{I_0}\lr{x_{JK}} + x^{i_0+2}_{I_0} \sum_{JK} \Delta^{i_0+2}_{I_0}\lr{x_{JK}} + \det X_{\setminus i_0,I_0} \right|^{\frac{r}{2}}
\end{align}
and
\begin{align}
    &\left| \sum_{IJK} \epsilon(IJK)\epsilon_{ijk} \lr{x^i_I+T\delta^{ii_0}_{II_0}}\lr{x^j_J+T\delta^{ji_0}_{JI_0}}\lr{x^k_K+T\delta^{ki_0}_{KI_0}}\right|^{\frac{r}{2}} = \nonumber\\
    & \quad = \left| \lr{x^{i_0}_{I_0}+T} \sum_{JK} \Delta^{i_0}_{I_0}\lr{x_{JK}} - x^{i_0+1}_{I_0} \sum_{JK} \Delta^{i_0+1}_{I_0}\lr{x_{JK}} + x^{i_0+2}_{I_0} \sum_{JK} \Delta^{i_0+2}_{I_0}\lr{x_{JK}} + \right.\nonumber\\
    & \qquad\ \left.\vphantom{\sum_{JK}} +\det X_{\setminus i_0,I_0} \right|^{\frac{r}{2}} .
\end{align}
In the above equations, we abbreviated $ \sum_{IJK} \ \widehat{=}\ \sum_{I,J,K \colon e_I\cap e_J\cap e_K = v}$ as the sum over all edges $e_I, e_J, e_K$ such that $ e_I \cap e_J \cap e_K = v $ and without two or three edges being the same within a given  combination (so far also guaranteed by $\epsilon\lr{IJK}$). Similarly, $\sum_{JK}$ stands for the sum over all edges $e_J, e_K$ such that $e_{I_0}\cap e_J \cap e_K = v \land J,K \neq I_0 \land J\neq K$.
Looking at the above formulae, the respective first summands on the right hand sides are the ones where the action of the commutator's inverse holonomy leaves its trace: acting on edge $e_{I_0}$ and U(1)-copy $i_0$, we get $x^{i_0}_{I_0}+T$ in the commutator's second eigenvalue, while there is no shift within the first one. This (shifted) $x^{i_0}_{I_0}$ is then multiplied by the sum of all $\Delta^{i_0}_{I_0}\lr{x_{JK}}$, by which we denote the minors of the matrices $x_{JK}$ with respect to $x^{i_0}_{I_0}$, where the sum over $J,K$ is understood to be over all edges $e_J, e_K$ that join $e_{I_0}$ at the vertex $v$, hence $x_{JK}$ consists of the 9 charges (divided by $T$) of these edges. The following two summands then are the remaining two terms of Laplace's formula. Note that we use periodicity in the superscripts $i_0+1$, $i_0+2$ --- i.e., $4\mapsto1$ and $5\mapsto2$. Lastly, $\det X_{\setminus i_0,I_0}$ collects all determinants that do not contain $x^{i_0}_{I_0}$.

By this decomposition, we see that we do not only face the root of a sum, but also have multiple terms at hand that contain the (shifted) $x^{i_0}_{I_0}$ --- and multiple terms that do not. This renders the strategy of the previous, non-estimate calculation for the cubic scenario in section~\ref{subsubsec:rigorous3edges}, namely substituting the determinant by $x^{i_0}_{I_0}$ or $x^{i_0}_{I_0}+T$ respectively, unpractical: We saw in the most fundamental case of three edges that we need to perform all nine integrations by hand. This clearly is all the more correct when advancing to the many edges scenario where we would need to substitute the whole sum --- all parts that do and do not contain $x^{i_0}_{I_0}$, $x^{i_0}_{I_0}+T$ --- by the (shifted) $x^{i_0}_{I_0}$. Hence, not only the number of integrals increases as $3M$, for $M$ many edges, but also the insertion of the substitution gets more complicated. These are the reasons why we will ultimately turn to estimates for the many edges scenario. First of all, however, the next subsection tackles a scenario where we do not yet need estimates --- and not even Kummer's functions ---, with the cost of being unable to cover the $p=0$ case.

\subsubsection{Computation of \texorpdfstring{$\langle \prod_{k=1}^N {\hat q}^{j_k}_{J_k}(r) \rangle_{\cs}$}{N q}  for general graphs with \texorpdfstring{$M$}{M} edges via Sahlmann and Thiemann} \label{subsubsec:SahlmannThiemann}

We saw that KCHFs can be used to calculate semiclassical expectation values of $\qr$ for cubic graphs. The question then arises whether one can, the other way around, apply the procedure of~\cite{Towards2} also on graphs of not necessarily cubic topology.

For understanding how their procedure works, we first of all consider again only the basic building block of one single determinant that was the basis of the cubic graphs' treatment.  With using again $r$ instead of $\rhalf$, we obtain
\begin{align}
    \langle \prod_{k=1}^N {\hat q}^{j_k}_{J_k}(r) \rangle_{\cs} &= \frac{{\lp}^{6rN}\lrabs{Z_{\gamma}}^{rN}}{a^{6rN}\csnorm } \sumset{N}\lr{\frac{2\pi}{T}}^9 \infint^9 x^i_I \frac{1}{T^{3rN}} \es{\sum_{Ii}\lr{\frac{p^i_I-\pi\i N^i_I}{T}}^2} \e{-\sum_{Ii}\lr{x^i_I}^2} \nonumber \\
    & \qquad \cdot \prod_{k=1}^N \lr{ \lrabs{\det\lr{X+P}}^r - \lrabs{ \det\lr{{\tilde X}_k + P}}^r }, \label{eq:BeginningTowards2}
\end{align}
where we use the notation $X+P = \lr{\lr{x+p}^i{}_I}$ with matrix elements $\lr{x+p}^i{}_I = x^i_I+\frac{p^i_I-\pi\i N^i_I}{T}$ and the shifted matrix ${\tilde X}_k = X + \lr{\delta^{j_kj}\delta_{J_kJ}\frac{T}{m_{\gamma}}}$ with the latter being a 3x3 matrix consisting of 8 zeroes and $\frac{T}{m_{\gamma}}$ in entry $\lr{j_k, J_k}$. We introduce $m_{\gamma}$ here to cover simultaneously both the scenario of the cubic graph (for which $m_{\gamma}=2$) and the general case of the next subsection, in which the introduction of $x^{\pm}$ is not possible and hence the shift remains to be $T$ (hence, $m_{\gamma}=1$ there).
Lastly, we now used $ Z_{\gamma} = Z_{\text{comb}}\lr{\gamma} \cdot C_{\text{reg}} = Z_{\text{comb}}\lr{\gamma} \cdot \frac{1}{48} $ instead of the plain $Z$. This is by far not a necessary redefinition, but is rather used to have formulae that compare more easily with results from the literature and also, in the end, with the findings for cubic graphs of subsection \ref{subsubsec:rigorous3edges}. It always holds that $C_{\text{reg}}=\frac{1}{48}$ and for arbitrary graphs there is not much more to achieve. So one can in general set $ Z_{\gamma} = Z = C_{\text{reg}} $ --- or $Z_{\text{comb}}\lr{\gamma}=1$. For cubic graphs, however, the combinatorics are such that one can also factor out another $48 \coloneqq Z_{\text{comb}}\lr{\gamma}\big\vert_{\gamma\text{ is cubic}}$ and end up without any $Z$ or numerical prefactors --- or $Z_{\gamma}\big\vert_{\gamma\text{ is cubic}}=1$. This was needed to recover the correct semiclassical continuum limit (cf. subsection \ref{subsec:SemClassCubic}).

We proceed in the spirit of equation (4.21) in~\cite{Towards2}, which reads adopted to our present notation 
\begin{align}
    \lrabs{\det\lr{X+P}}^r - \lrabs{\det\lr{{\tilde X}_k + P}}^r = \lrabs{\det P}^r \lr{ \lrabs{\det\lr{P\inv X+1}}^r - \lrabs{\det\lr{P\inv {\tilde X}_k+1}}^r } . \label{eq:BeforeTaylorDet}
\end{align}
Therein, the inverse matrix $P\inv$ is of the form
\begin{align}
    \lr{P\inv}^i{}_I = \frac{1}{\det P} \, \Delta^i_I \lr{P\transp} ,
\end{align}
where we denote with $\Delta^i_I \lr{P\transp}$ the minor of $P\transp$ with respect to its entry $(i,I)$:
\begin{equation}
    \Delta^i_I \lr{P\transp} \coloneqq \lr{P\transp}^{i+1}{}_{I+1}\cdot \lr{P\transp}^{i+2}{}_{I+2} - \lr{P\transp}^{i+2}{}_{I+1}\cdot \lr{P\transp}^{i+1}{}_{I+2},
\end{equation}
having the periodicity in the indices in mind. Note that the necessity to involve $P^{-1}$ directly shows why this method does not allow to consider the limit of the cosmological singularity where all $p^i_I$ are sent to zero. The occurrence of the minor at this stage is in fact already the reason why we will ultimately end up with a result that resembles the differentiation of a determinant with respect to one matrix element, where this minor arises due to the chain rule.

Let us check the $T$-dependence of the quantities we just introduced. As we have $\lr{P}^i{}_I = \frac{p^i_I-\pi\i N^i_I}{T}$, we can infer $\det P \sim T^{-3}$ and $\Delta^i_I \lr{P\transp} \sim T^{-2}$. Hence, via the formula above, $\lr{P\inv}^i{}_I \sim T$ as expected.

With the above $T$-dependence, we can perform an expansion of the determinants on the right hand side of~(\ref{eq:BeforeTaylorDet}) around 1. Using the result of~\cite{Towards2}, we may already now neglect the contributions with  $N^i_I\neq0$, as they will contribute with $\mathcal{O}\lr{T^\infty}$ due to the remaining Gaussians in $\nicefrac{\pi^2 \lr{N^i_I}^2}{T^2}$ in~\eqref{eq:BeginningTowards2}. With this, all contained expressions are now real (note that there were terms $\sim\i N^i_I$ included so far) and we proceed along the lines of ~\cite{Towards2} via the well-known decomposition
\begin{align}
    \det\lr{1+A} = 1 + \tr A + \frac{1}{2}\lr{\lr{\tr A}^2-\tr A^2} + \det A. \label{eq:DetDecomposition}
\end{align}
We apply~\eqref{eq:DetDecomposition} to the difference of the absolute values of the two determinants of~(\ref{eq:BeforeTaylorDet}) leading to
\begin{align}
    \lr{\det\lr{P\inv X+1}^2}^\rhalf - \lr{\det\lr{P\inv {\tilde X}_k+1}^2}^\rhalf . \nonumber
\end{align}
This means we need to consider the square of the decomposition in~\eqref{eq:DetDecomposition} and with the $T$-dependencies stated before, we obtain up to fourth order in $T$
\begin{align}
    \det\lr{1+A}^2 &= 1 + 2 \tr A + 2\lr{\tr A}^2 - \tr A^2 +  \lr{\tr A}^3 - \tr A \cdot \tr A^2 + \nonumber\\
    &\quad + \det A + \tr A \cdot \det A + \frac{1}{4} \lr{\tr A}^4 - \frac{1}{2}\lr{\tr A}^2\cdot\tr A^2 + \mathcal{O}\lr{T^5} \nonumber\\
    &\eqqcolon 1+z_A + \mathcal{O}\lr{T^5}\\
\Rightarrow    \lr{\det\lr{1+A}^2}^\rhalf &\approx \sum_{k=0}^5 \binom{\rhalf}{k} \lr{z_A}^k + \mathcal{O}\lr{T^5} . \label{eq:TaylorRootDetDecomp}
\end{align}

Note that $A$ is not solely $\sim T$ for the contribution of the shift (i.e. when $A=P\inv {\tilde X}_k$) but also contains a term $\sim T^2$ via the shift within ${\tilde X}_k$. Therefore, and due to the multiplication of $z_A$ with itself, ~\eqref{eq:TaylorRootDetDecomp}'s terms also include higher order terms in $T$ as well and we will discard those only later for reasons of clearer formulae.

With all this, we can cast~(\ref{eq:BeginningTowards2}) into the form

\begin{align}
    &\langle \prod_{k=1}^N {\hat q}^{j_k}_{J_k}(r) \rangle_{\cs} = \frac{T^{6rN}\lrabs{Z_{\gamma}}^{rN}}{\csnorm } \lr{\frac{2\pi}{T}}^9 \infint^9 x^i_I \frac{1}{T^{3rN}} \es{\sum_{Ii}\lr{\frac{p^i_I}{T}}^2} \e{-\sum_{Ii}\lr{x^i_I}^2} \lrabs{\det P}^{rN} \cdot \nonumber \\
    &  \cdot \prod_{k=1}^N \Bigg [ r\tr \lr{P\inv X} - r\tr \lr{P\inv {\tilde X}_k} + \rhalf \lr{r\lr{\tr \lr{P\inv X}}^2- \tr \lr{P\inv X}^2 } - \nonumber\\
    & - \rhalf \lr{r\lr{\tr \lr{P\inv {\tilde X}_k}}^2- \tr \lr{P\inv {\tilde X}_k}^2 } +r(r-1) \tr\lr{P\inv X} \tr\lr{P\inv X P\inv \lr{\delta^{j_kj}\delta_{J_kJ}T}} - \nonumber\\
    & - \frac{r^2(r-1)}{2}\lr{\tr\lr{P\inv X}}^2\cdot \tr\lr{P\inv \lr{\delta^{j_kj}\delta_{J_kJ}T}} + \mathcal{O}\lr{T^5} \Bigg ].
\end{align}

Using now that ${\tilde X}_k$ differs from $X$ just by addition of a matrix of zeros except for one entry\footnote{Note that we used this feature and similar identities like the following ones to already state simplified versions of the last two contributions above. This is again only due to having more concise formulae.}, ${\tilde X}_k = X + \lr{\delta^{j_kj}\delta_{J_kJ}\frac{T}{m_{\gamma}}}$, we can simplify the above expression with the help of
\begin{align}
    \tr\lr{P\inv X} - \tr\lr{P\inv {\tilde X}_k} &= -\tr\lr{P\inv \lr{\delta^{j_kj}\delta_{J_kJ}\tfrac{T}{m_{\gamma}}}} = -\tfrac{T}{m_{\gamma}} \lr{\lr{P\inv}\transp}^{j_k}{}_{J_k}
\end{align}
and
\begin{align}
    \tr\lr{P\inv X}^2 - &\tr\lr{P\inv {\tilde X}_k}^2 = \nonumber\\
    & = -2\tr\lr{\lr{P\inv X}\lr{P\inv \lr{\delta^{j_kj}\delta_{J_kJ}\tfrac{T}{m_{\gamma}}}}} - \tr\lr{P\inv\lr{\delta^{j_kj}\delta_{J_kJ}\tfrac{T}{m_{\gamma}}}}^2 \nonumber\\
    & = -2\tr\lr{\lr{P\inv X} \lr{P\inv \lr{\delta^{j_kj}\delta_{J_kJ}\tfrac{T}{m_{\gamma}}}}} - \frac{T^2}{m_{\gamma}^2} \lr{\lr{\lr{P\inv}\transp}^{j_k}{}_{J_k}}^2 ,
\end{align}
where we also used
\begin{align}
    \tr\lr{P\inv \lr{\delta^{j_kj}\delta_{J_kJ}\tfrac{T}{m_{\gamma}}}} = \tfrac{T}{m_{\gamma}} \lr{P\inv}^a{}_b \, \delta^{bj_k}\delta_{aJ_k} = \tfrac{T}{m_{\gamma}} \lr{P\inv}_{J_k}{}^{j_k} = \tfrac{T}{m_{\gamma}} \lr{\lr{P\inv}\transp}^{j_k}{}_{J_k} .
\end{align}
With this and reinserting the state's norm, we now have 
\begin{align}
    \langle \prod_{k=1}^N {\hat q}^{j_k}_{J_k}(r) \rangle_{\cs} &= \frac{T^{6rN}\lrabs{Z_{\gamma}}^{rN}}{\sqrt{\pi}^9 T^{3rN}}\lrabs{\det P}^{rN}\infint^9 x^i_I \es{-\sum_{Ii}\lr{x^i_I}^2} \cdot \nonumber\\
    & \hspace{-1.8cm} \cdot \prod_{k=1}^N \Bigg [ - r \frac{T}{m_{\gamma}} \lr{\lr{P\inv}\transp}^{j_k}{}_{J_k} + \frac{r(1-r)}{2} \frac{T^2}{m_{\gamma}^2} \lr{\lr{\lr{P\inv}\transp}^{j_k}{}_{J_k}}^2 - \nonumber\\
    & \hspace{-.7cm} - r^2 \frac{T}{m_{\gamma}} \lr{\lr{P\inv}\transp}^{j_k}{}_{J_k} \tr\lr{P\inv X} + r\tr\lr{\lr{P\inv X}\lr{P\inv \lr{\delta^{j_kj}\delta_{J_kJ}\tfrac{T}{m_{\gamma}}}}}+ \nonumber\\
    & \hspace{-.7cm}+r(r-1) \frac{T}{m_{\gamma}}\sum_{j,J} \lr{\lr{P\inv}\transp}^{j_k}{}_{J} \lr{\lr{P\inv}\transp}^{j}{}_{J} \lr{\lr{P\inv}\transp}^{j}{}_{J_k} \lr{X^j_J}^2 -\nonumber\\
    & \hspace{-.7cm}- \frac{r^2(r-1)}{2} \frac{T}{m_{\gamma}} \lr{\lr{P\inv}\transp}^{j_k}{}_{J_k} \sum_{j,J} \lr{\lr{\lr{P\inv}\transp}^{j}{}_{J}}^2 \lr{X^j_J}^2 +\mathcal{O}\lr{T^{5}} \Bigg ].
\end{align}
First of all, we can argue that the second line's terms will vanish via integration as all their contributions to the integrand are linear in one $x^i_I$.\footnote{We already dropped linear contributions from the terms in the third and fourth line, too.} Integrated with respect to the Gaussian in that respective $x^i_I$ then amounts to zero.
All remaining integrations are then either solely Gaussian ones, yielding $\sqrt{\pi}^9$, which cancels the corresponding factor in the norm of the state. Or they are of the form 
\begin{align}
    \infint x \es{-x^2} x^{2} = \frac{\sqrt{\pi}}{2} \nonumber
\end{align}
and therefore also cancel the $\pi$-factors. As we now face $N$-times the commutator, we have to divide our final result by $t^N$ to be able to compare it with the classical result of the respective Poisson bracket. Furthermore, as before, we choose the classicality parameter $t$ to be $t=\nicefrac{\lp{}^2}{a^2}$. 
Lastly, we reinsert also
\begin{align}
    P^i{}_I = \frac{p^i_I}{T} \Rightarrow \det P = \frac{\det p}{T^3} \Rightarrow \lr{\lr{P\inv}\transp}^{j_k}{}_{J_k} = T \frac{\Delta^{j_k}_{J_k} \lr{p}}{\det p}
\end{align}
and obtain for the final result (including again $\frac{2^N}{t^N}$):
\begin{equation}
\label{eq:BBBSanThiemann} 
\boxed{
\begin{aligned}
    &\frac{2^N\langle \prod_{k=1}^N  {\hat q}^{j_k}_{J_k}(r) \rangle_{\cs}}{t^N} =  \\
    & \ = \frac{2^N}{t^N}\lrabs{Z_{\gamma}}^{rN} \lrabs{\det p}^{rN} \prod_{k=1}^N \left[ -r\frac{t}{m_{\gamma}}\frac{\Delta^{j_k}_{J_k} \lr{p}}{\det p} + \frac{r(1-r)}{2}\frac{t^2}{m_{\gamma}^2} \frac{\lr{\Delta^{j_k}_{J_k} \lr{p}}^2}{\lr{\det p}^2}  + \right.\\
    &\ \left. + \frac{r(r-1)}{2} \frac{t^2}{m_{\gamma}}\sum_{j,J} \frac{\Delta^{j_k}_{J}\lr{p} \Delta^{j}_{J}\lr{p} \Delta^{j}_{J_k}\lr{p}}{\lr{\det p}^3} - \frac{r^2(r-1)}{4} \frac{t^2}{m_{\gamma}} \frac{\Delta^{j_k}_{J_k}\lr{p}}{\lr{\det p}^3} \sum_{j,J} \lr{\Delta^{j}_{J}\lr{p}}^2 + \mathcal{O}\lr{t^{\frac{5}{2}}} \right]   \\
    & \ = \lr{-2r}^N \lrabs{Z_{\gamma}}^{rN}\frac{1}{m_{\gamma}^N}\lr{\frac{\lrabs{\det p}^{r}}{\det p}}^{\!N} \prod_{k=1}^N \left[\Delta^{j_k}_{J_k} \lr{p} \right] + \mathcal{O}\lr{t^{\frac{1}{2}}} .
\end{aligned}
}
\end{equation}

If we consider $m_{\gamma}=2$ as well as $Z_{\gamma}=1$, the above equation's lowest order contribution now looks like the result of performing the nine integrations in the case of the cubic graph shown in~(\ref{3edgesFinal}), just for general $N$ now. As discussed in subsection \ref{subsec:SemClassCubic}, this demonstrates again that the semiclassical limit of the lowest order agrees only with the classical expression for cubic graphs, otherwise it differs by a factor caused by the regularisation constant $Z_{\gamma}$ that depends on the graph under consideration.

Comparing the above result with Theorem 4.2 in \cite{Towards2} (eq. (4.45) therein), we notice that the results are expressed in a slightly different manner. This comes from the fact that \cite{Towards2} expressed the final results explicitly in terms of lattice fluctuations which are, in the result presented here, still encoded in the fluxes involved in \eqref{eq:BBBSanThiemann}. To extract the lattice fluctuations in \cite{Towards2} and rewrite the flux dependent corrections in terms of quantities of order unity, the quantity  $s=t^{\frac{1}{2}-\alpha}$, with $0<\alpha<\frac{1}{2}$, is introduced that involves a specific choice for the range of $\alpha$. This led to the power series of (4.39) in \cite{Towards2} to end with the contribution $\sim s^3T$ and omit the already higher-order contribution $\sim s^2T^2$. 
As a result, the first correction in the result of Theorem 4.2 in \cite{Towards2} only includes the terms corresponding to the ones in the second line of \eqref{eq:BBBSanThiemann} above, while the remaining correction term $\sim t^2$ of the first line corresponds to the $s^2T^2$-term \cite{Towards2} neglected. Because these kind of lattice fluctuations will be present for any given choice of $\alpha$, the power counting has to be adapted and in this sense differs from the more simple situation in the U(1) and quantum mechanical case of our companion paper \cite{paper1}, where these kind of lattice fluctuations are absent. Nevertheless, we kept those higher order terms here in order to discuss the differences and similarities to the U(1) and quantum mechanical cases. 

In the U(1) case or the quantum mechanical case, these two kind of correction terms contribute to the same order in $t$ but can be seen to take a different role:  as  equation (5.8) of \cite{paper1} shows  the expectation value of $\lrabs{\hat p}^r$ is not just $\lrabs{p}^r$ for coherent states but, instead, fluctuations arise (note that this is already true for ${\hat p}^3$, e.g.). Then, equation (5.10) in \cite{paper1} also features multiple correction terms $\sim t^2$ and one can see that the respective second one is the derivative of the correction term $\sim t$ in (5.8), while the other one corresponds to the second derivative just like the term $\sim t$ corresponds to the first derivative.\footnote{Note that this applies also to the quantum mechanical scenario. For this, compare eq. (3.4) and (3.9) or (3.12) in \cite{paper1}.} Taking a look at Theorem 4.3 of \cite{Towards2} (eq. (4.48) therein), one can observe the same situation: Differentiating the fluctuation correction of the volume operator's expectation value as stated there yields terms $\sim q^{\frac{3r}{2}-3}$ --- just like all corrections $\sim s^2$ of (4.45) in \cite{Towards2} (for $N=1$).

We want to note that when going to a higher order in the power series (4.39) of \cite{Towards2}, there \textit{can} be additional terms between the $\lr{sx}^2$ and $\mathcal{O}\lr{sT}$ contribution depending on the value $\alpha$ takes. While it of course always holds that $sT \sim t^{1-\alpha}$ is of higher order than $\lr{sx}^2 \sim t^{1-2\alpha}$ (for $0 < \alpha < \frac{1}{2}$), one has to be careful about terms $\lr{sx}^n$. Those contribute with $\sim t^{\frac{n}{2}-n\alpha}$ and therefore, for a fixed value of $\alpha$, all $n < \frac{1-\alpha}{\nicefrac{1}{2}-\alpha}$ have to be considered. For the choice of $\alpha\approx\frac{1}{6}$, which \cite{Towards2} motivates below (3.12) therein, this yields $n<2.5$ and everything is fine. But as $\alpha$ approaches $\frac{1}{2}$, one would have to include all $n\in 2\mathbbm{N}$ (note that all contributions with odd $n$ vanish via the integration against the Gaussian).

We want to apply this procedure now to the general case, i.e. not just one determinant within the eigenvalue of $\hat q$ and not necessarily graphs of cubic topology. Hence, for this part, we can set $m_{\gamma}=1$ and therefore drop it. The difference of the commutator's two eigenvalues as the delicate part of the integration then reads
\begin{align}
    \prod_{k=1}^N \lr{ \lrabs{\sum_i \det\lr{X_i+P_i}}^r - \lrabs{ \sum_i \det\lr{{\tilde X}_{i,k} + P_i}}^r }, \nonumber
\end{align}
as we have to collect the contributions from all triples of edges, cf.~(\ref{eq:voleigen}), which we label by the subscript $i$. In order to cast the argument of the roots into a form that we can apply a Taylor expansion on, we proceed by factoring out $\det P_1$ and a subsequent decomposition according to~(\ref{eq:DetDecomposition}):
\begin{align}
    &\lrabs{\sum_i \det\lr{X_i+P_i}}^r = \lrabs{\det P_1}^r \cdot \Bigg| 1+\tr\lr{P_1\inv X_1} + \frac{1}{2}\lr{ \lr{\tr\lr{P_1\inv X_1}}^2 - \tr\lr{P_1\inv X_1}^2 } +  \nonumber\\
    & \qquad + \det \lr{P_1\inv X_1} +\sum_{i\neq1} \frac{\det P_i}{\det P_1} \bigg( 1 + \tr\lr{P_i\inv X_i} + \frac{1}{2}\lr{ \lr{\tr\lr{P_i\inv X_i}}^2 - \tr\lr{P_i\inv X_i}^2 } + \nonumber\\
    & \qquad +  \det \lr{P_i\inv X_i} \bigg) \Bigg|^r. \label{eq:FactorOutP1}
\end{align}
Thereby, we obtained a result that now allows for a Taylor expansion, here not around 1 but instead around $1 + \sum_{i\neq1} \frac{\det P_i}{\det P_1}$, which is the collection of all terms $ \mathcal{O}\lr{T^0}$. After performing this Taylor expansion, those $\mathcal{O}\lr{T^0}$-terms vanish via the commutators difference and only (some of) the higher order terms remain. Collecting all terms of one order in $T$ leads to more and more evolved formulae as the order in $T$ grows, so we will only consider the lowest order here, where there is only one kind of term contributing. This is the $\tr A$ part of the decomposition in \eqref{eq:DetDecomposition}, resulting in $r \tr A$ after the root's Taylor expansion  in~(\ref{eq:TaylorRootDetDecomp}). Collecting this contribution from all combinations of three edges, we obtain
\begin{align}
    &\prod_{k=1}^N \lr{ \lrabs{\sum_i \det\lr{X_i+P_i}}^r - \lrabs{ \sum_i \det\lr{{\tilde X}_{i,k} + P_i}}^r } = \lrabs{\det P_1}^{rN} \prod_{k=1}^N \Bigg[ r\tr\lr{P_1\inv X_1} - \nonumber\\
    & \quad -r\tr\lr{P_1\inv {\tilde X}_{1,k}} + r\sum_{i\neq1} \frac{\det P_i}{\det P_1}\tr\lr{P_i\inv X_i} - r\sum_{i\neq1} \frac{\det P_i}{\det P_1}\tr\lr{P_i\inv {\tilde X}_{i,k}} + \mathcal{O}\lr{T^{3}} \Bigg]  \nonumber\\
    & = \lr{-r}^N T^{-3rN+2N} \prod_{k=1}^N\lr{\sum_i \frac{\lrabs{\det p_i}^r}{\det p_i} \Delta^{j_k}_{J_k}\lr{p_i} } + \mathcal{O}\lr{T^{-3rN+2N+1}}.
\end{align}
Using this and, as before, in order to compare the final result to the classical Poisson bracket multiplying the result by $\frac{2^N}{t^N}$, choosing again the classicality parameter $t\sim \lp{}^2$ and taking into account that we used $r$ as exponent, instead of the initial $\rhalf$, our final result is given by
\begin{equation}
\label{eq:FinResSTManyEdes}
\boxed{
\begin{aligned}
   \frac{2^N\langle \prod_{k=1}^N {\hat q}^{j_k}_{J_k}(r) \rangle_{\cs}}{t^N} &= \lr{-2r}^N \lrabs{Z_{\gamma}}^{rN} \prod_{k=1}^N \lr{\sum_i \frac{\lrabs{\det p_i}^r}{\det p_i} \Delta^{j_k}_{J_k}\lr{p_i} } + \mathcal{O}\lr{t^{\frac{1}{2}}}.
\end{aligned}
}
\end{equation}
Now as far as the lowest order is considered, we do not get the exact classical result here for other than cubic graphs, as expected. This is caused by the fact that in contrast to the case of the cubic graph, here the regularisation constant $Z_{\gamma}=\frac{1}{48}$ does not exactly cancel all remaining factors in the final result.  As already mentioned above, this method cannot be applied to the case $p\to 0$ because it requires the existence of the inverse of the matrix of the classical triad labels and, as a consequence, the case of the cosmological singularity cannot be analysed with this method.
~\\
~\\
This finishes our treatment of volume operator computations without the use of estimates. We showed that KCHFs allow for an analytical calculation of the basic building block of semiclassical expectation values of the class of operators $\qr$ as well as expectation values with respect to states of cubic topology. We furthermore extended the Sahlmann and Thiemann procedure \cite{Towards2} to more general scenarios. In both cases, in contrast to the work of Brunnemann and Thiemann
\cite{Brunnemann1,Brunnemann2}, we could perform our computations without using estimates and were able to consider the semiclassical limit of $t\to0$. A special role played the case of $p=0$ where neither the  Sahlmann and Thiemann procedure nor our method of using the asymptotics of the Kummer functions could be applied. Therefore, we also had to rely on estimates in our computation for this case and ---  depending on the value of $r$ --- obtained a divergence for generic $r$, too, if we in addition to $p\to 0$ consider the limit $t\to 0$ as well.
The next section, which discusses again semiclassical expectation values of $\qr$ but this time via estimates, in turn, follows closer the path Brunnemann and Thiemann chose.

\section{Semiclassical analysis of operators involving the volume operator via estimates
} \label{sec:wEstimates}

In this section, we address the computation of semiclassical expectation values $\langle \prod_{k=1}^N {\hat q}^{j_k}_{J_k}(r) \rangle_{\cs}$ for generic graphs that forces us to involve additional estimates in our computation. While the previous section's calculations did not need this, estimates become inevitable when we want to consider the realm of the initial singularity, i.e. $p^i_I \to 0$. Note that we did indeed also need estimates for the computation of the $p^i_I=0$ case at the end of subsection~\ref{subsubsec:rigorous3edges} as the strategy we used before did not apply anymore. The same is true for the previous path along Sahlmann and Thiemann: With $p^i_I=0$, the very first step of factoring out $\det P$ or $\det P_1$, \eqref{eq:BeforeTaylorDet} or \eqref{eq:FactorOutP1} respectively, is not applicable anymore since the inverse of $P$ does not exist then. First of all, in subsection~\ref{subsec:LikeBrunnemannThiemann}, we consider the case $N=1$ and start with a method that goes along the work of Brunnemann and Thiemann~\cite{Brunnemann1,Brunnemann2} but is already adopted to the usage of KCHFs. Afterwards, we generalise this to the case of generic integer $N$ covering the most general scenario presented in our work here. 
Following this, in subsection~\ref{subsec:newEstimates}, we discuss possibilities how estimates can be found that are better suited to preserve the power of the classicality parameter or the momenta and what issues can arise in this context.

\subsection{Estimates along the lines of Brunnemann and Thiemann}

\subsubsection{Semiclassical expectation value for \texorpdfstring{$\qr$}{q} }
\label{subsec:LikeBrunnemannThiemann}
In order to discuss the details of the estimate for $\langle \qr \rangle_{\cs}$, we consider as our starting equation \eqref{eq:startingExpectationvalue}, which is still formulated at the level of the charges:\footnote{We now use again $Z$ for the regularisation constant as we do not consider special graphs anymore and hence $Z_{\text{comb}}\lr{\gamma}=1 \Rightarrow Z_{\gamma}=C_{\text{reg}}=\frac{1}{48} \eqqcolon Z$.}
\begin{align}
    \langle \qr \rangle_{\cs} &= \frac{1}{\csnorm a^{3r}}  \sum_{\left\{n^i_I\right\}\in\mathbbm{Z}} \e{\sum_{i,I}\lr{-t\lr{n^i_I}^2+2p^i_In^i_I}} \lpz \lr{\left| \sum_{IJK} \epsilon(IJK)\epsilon_{ijk}n^i_In^j_Jn^k_K \right|^{\frac{r}{2}} - \right. \nonumber\\
    & \qquad\qquad \left. -\left| \sum_{IJK} \epsilon(IJK)\epsilon_{ijk}\lr{n^i_I+\delta^{ii_0}\delta_{II_0}}\lr{n^j_J+\delta^{ji_0}\delta_{JI_0}}\lr{n^k_K+\delta^{ki_0}\delta_{KI_0}}\right|^{\frac{r}{2}}}. \tag{\ref{eq:startingExpectationvalue}}
\end{align}

This was important for the work of \cite{Brunnemann2} as their crucial estimate only applies to integer numbers:
\begin{align}
|a|^r-|b|^r \leq ||a|-|b|| \tag{\ref{eq:approxBrunn}},
\end{align}
where $a,b\in\mathbbm{Z}$ and $r\in\mathbbm{Q}_{[0,1]}$.
This allowed them to transform integrals including rational powers of (eigenvalues of) the momentum operator into one with integer powers. However, with the methods introduced in our companion paper \cite{paper1}, we know how to compute integrals containing rational powers of this kind with the help of KCHFs, which is the reason why we will use slightly different estimates that still contain information on the initial exponents and hope to improve the estimate used in \cite{Brunnemann2} accordingly. The estimates we use are given by
\begin{align}
    &|a|^r-|b|^r \leq |a-b|^r \tag{\ref{approxdiff}}, \\
   & |a+b|^r \leq|a|^r+|b|^r  \tag{\ref{approxsum}}, \\
   & 2|ab| \leq |a|^2+|b|^2 \quad\text{and} \label{2ab}\\
   & |a|^2+|b|^2 \leq \lr{|a|+|b|}^2 \label{a2b2},
\end{align}
where $ a,b,r\in\mathbbm{R}$ and $ 0\leq r\leq 1$.

To stick to the notation in the literature, we follow~\cite{Brunnemann1,Brunnemann2} and start with the difference of the volume eigenvalues involved in the commutator of $\qr$. Recall that the action of $\lr{h^{i_0}_{I_0}}^{-1}$ within the second term of $\qr$ leads to a shift in that volume eigenvalue:
\begin{align} \label{eq:startAlaBrunnemann}
\Delta\lambda^r & \coloneqq \lambda^r\lr{\left\{n^i_I\right\}} - \lambda^r\lr{\left\{n^i_I+\delta^{ii_0}\delta_{II_0}\right\}}\nonumber\\
& = \lpz \left( \left| \sum_{IJK} \epsilon(IJK)\epsilon_{ijk}n^i_In^j_Jn^k_K \right|^{\frac{r}{2}} - \right. \nonumber\\
& \hphantom{ \lpz}\qquad - \left. \left| \sum_{IJK} \epsilon(IJK)\epsilon_{ijk}n^i_In^j_Jn^k_K + 3\sum_{JK}\epsilon(I_0JK)\epsilon_{i_0jk}n^j_Jn^k_K \right|^{\frac{r}{2}} \right) ,
\end{align}
where we abbreviated again $\sum_{JK}$ as the sum over all edges $e_J, e_K$ such that $e_{I_0}\cap e_J \cap e_K = v \land J,K \neq I_0 \land J\neq K$ with $I_0$ being fixed.

Integrating this expression now with respect to the first charge against the coherent states, i.e. Gaussians, would yield a KCHF with the other charges in its argument. As mentioned before, the resulting integral cannot be computed  analytically by means of KCHFs and thus forces us to perform estimates in order to remove the sums of determinants under the fractional power:
\begin{align}
    \Delta\lambda^r & \overset{\text{(\ref{approxdiff})}}{\leq} \lpz 3^\rhalf \lrabs{\sum_{JK}\epsilon(I_0JK)\epsilon_{i_0jk}n^j_Jn^k_K}^\rhalf \nonumber\\
    & \overset{\text{(\ref{approxsum})}}{\leq} \lpz 3^\rhalf \lr{\sum_{JKjk}\lrabs{\epsilon(I_0JK)\epsilon_{i_0jk}n^j_Jn^k_K}}^\rhalf \nonumber\\
    &\:\, \leq \lpz 3^\rhalf \lr{\sum_{JKjk}\lrabs{n^j_Jn^k_K}}^\rhalf \nonumber\\
    & \overset{(\ref{2ab})}{\leq} \lpz 3^\rhalf \lr{\frac{1}{2}\sum_{JKjk}\lr{\lrabs{n^j_J}^2+\lrabs{n^k_K}^2}}^\rhalf \nonumber\\
    &\:\, \leq \lpz 3^\rhalf \lr{\frac{1}{2}3M\lr{\sum_{Jj}\lrabs{n^j_J}^2+\sum_{Kk}\lrabs{n^k_K}^2}}^\rhalf \nonumber\\
    & \overset{(\ref{a2b2})}{\leq} \lpz (9M)^\rhalf \lr{\sum_{Jj}\lrabs{n^j_J}}^r \nonumber\\
    & \overset{\text{(\ref{approxsum})}}{\leq} \lpz (9M)^\rhalf \sum_{Jj}\lrabs{n^j_J}^r . \label{voleigenestimate}
\end{align}
In the second line, there's no summation over $j,k$ inside the absolute value anymore since we applied formula~(\ref{approxsum}) to pull the sum out of the absolute value. In the next step, we estimated all $\epsilon$ by $+1$ from above. Formula~(\ref{a2b2}) was applied in the generalisation of more than two terms being summed up. The prefactor $3M$ stems from the empty sum over the U(1)-copies $k$ and the edges $K$ for the first term and likewise over $j,J$ for the second one. In the fashion of~\cite{Brunnemann2}, we estimated this by $3M$ albeit not all edges necessarily meet $e_{I_0}$ at $v$ and $j,k$ allowing only for two contributions, as $i_0$ is fixed and occupies the third U(1)-copy.

If we now compare this result with the one Brunnemann and Thiemann obtained in~\cite[eqn. (C.39) therein]{Brunnemann2},
\be
\lrabs{\Delta\lambda^r} \overset{\text{B.--T.}}{\leq} \lpz 9M \sum_{Jj}\lrabs{n^j_J}^2 ,
\ee
we notice that it is quite similar at first sight, but~(\ref{voleigenestimate}) still contains information in the charges about what root of the volume operator we are considering --- i.e. $r$. This also includes that the new value is smaller as $r<2$ and $M,\lrabs{n^j_J}\in\mathbbm{N}_0$.

However, looking only at the dimensions of the charges, we started with $\sim|n|^{\frac{3r}{2}}$ and reduced it to $\sim|n|^r$ whereas Brunnemann and Thiemann end up with $\sim|n|^2$. This will become important later on when we discuss possible improvements of estimates in subsection~\ref{subsec:newEstimates}.

With the result above, we can now tackle the expectation value of interest:

\begin{align}
    \langle\qr\rangle_{\cs} & \leq  \frac{\lpz (9M)^\rhalf}{a^{3r}\csnorm}\sumset{n} \e{\sum_{Ii}\lr{-t\lr{n^i_I}^2+2p^i_In^i_I}}\sum_{Jj}\lrabs{n^j_J}^r \nonumber\\
    & = \frac{\xi}{\csnorm} \sumset{N} \infint^{3M} x^i_I \lr{\frac{2\pi}{T}}^{3M}\e{\sum_{Ii}\lr{-\lr{x^i_I}^2+\frac{2p^i_I}{T}x^i_I-\frac{2\pi\i N^i_I}{T}x^i_I}}\sum_{Jj}\frac{\lrabs{x^j_J}^r}{T^r} \nonumber\\
    & = \frac{\xi}{\csnorm}\frac{(2\pi)^{3M} \sqrt{\pi}^{3M-1}}{T^{3M+r}} \sumset{N}\sum_{Jj}\e{\sum_{Ii\backslash Jj}\lr{\frac{p^i_I}{T}-\frac{\pi\i N^i_I}{T}}^2}\Gamma\lr{\tfrac{r+1}{2}} \cdot \nonumber\\
    &\quad \cdot \kchf{\frac{r+1}{2},\frac{1}{2},\lr{\frac{p^j_J}{T}-\frac{\pi\i N^j_J}{T}}^2} . \label{eq:BrunnemannKCHF}
\end{align}
From the first to the second line, we defined $\xi\coloneqq T^{3r}\lrabs{Z}^\rhalf(9M)^\rhalf$. We then notice that out of the $3M$-dimensional integration of the sum of $3M$ many terms $\sim\lrabs{x^j_J}^r$, only one $x^j_J$-integration per addend yields a KCHF,
\be
\infint x^j_J \, \e{\lr{x^j_J}^2+\lr{\frac{2p^j_J}{T}-\frac{2\pi\i N^j_J}{T}}x^j_J}\lrabs{x^j_J}^r = \Gamma\lr{\tfrac{r+1}{2}}\kchf{\frac{r+1}{2},\frac{1}{2},\lr{\frac{p^j_J}{T}-\frac{\pi\i N^j_J}{T}}^2} ,
\ee
whereas the other $3M-1$ integrals are just of the form
\be
\infint x^i_I \,\e{-\lr{x^i_I}^2+\lr{\frac{2p^i_I}{T}-\frac{2\pi\i N^i_I}{T}}x^i_I} = \sqrt{\pi}\es{\lr{\frac{p^i_I}{T}-\frac{\pi\i N^i_I}{T}}^2} .
\ee
Therefore, we denoted by $\sum_{Ii\backslash Jj}$ in the Gaussian the sum over all combinations of edges and U(1)-copies that do not correspond to the particular configuration of the $\sum_{Jj}$-sum: We sum over all $(I,i)$ with $I=1,2,\dots,M$ and $i=1,2,3$ leaving out the combination $(J,j)$ --- and $(i_0, I_0)$, of course, which we lost right at the beginning of \eqref{voleigenestimate}.

Including the states' norm~\eqref{eq:norm}
\be
\csnorm = \lr{\frac{2\pi\sqrt{\pi}}{T}}^{3M} \e{\sum_{Ii}\lr{\frac{p^i_I}{T}}^2}\prod_{Ii}\lr{1+K^i_{t(I)}},
\ee
applying formula~(\ref{expansion}) for the asymptotic expansion for large arguments of the KCHF and performing all the intermediate steps as before (that is  using that only the contribution for $N^i_I=0\ \forall\ i,I$ has to be taken into account and dividing by $t$ to compare it with the classical Poisson bracket), we finally obtain
\begin{equation}
\label{eq:ResultALaBrunnemann}
\boxed{
\begin{aligned}
    \frac{\langle\qr\rangle_{\cs}}{t} & \leq \eqref{eq:BrunnemannKCHF} \overset{t\to0}{=} T^{3r}\lrabs{Z}^{\rhalf}(9M)^\rhalf \sum_{Jj}\zeroinfsum{n} \frac{\lr{-\frac{r}{2}}_n\lr{\frac{1-r}{2}}_n}{n!} \lrabs{p^j_J}^{r-2n} t^{n-r-1} . 
\end{aligned}
}
\end{equation}

As the configuration where the volume operator enters with linear power is of importance, we may also directly state the corresponding formula for $r=\frac{1}{2}$ and after dividing by $t$:
\begin{equation}
\label{eq:r1/2ResultALaBrunnemann}
\boxed{
\begin{aligned}
    \frac{\langle{\hat q^{i_0}_{I_0}}\lr{\frac{1}{2}}\rangle_{\cs}}{t} & \overset{t\to0}{\lesssim} T^{\frac{3}{2}} |Z|^{\frac{1}{4}}(9M)^{\frac{1}{4}} \sum_{Jj}\lr{ \frac{\sqrt{\lrabs{p^j_J}}}{{\sqrt{t}}^3} - \frac{1}{16}\frac{1}{{\sqrt{t}\sqrt{\lrabs{p^j_J}}}^3} - \frac{15}{256}\frac{\sqrt{t}}{{\sqrt{\lrabs{p^j_J}}}^7}+\mathcal{O}\lr{t^{\frac{3}{2}}} } ,
\end{aligned}
}
\end{equation}
\begin{equation}
\label{eq:r1/2ResultALaBrunnemannlp}
\boxed{
\begin{aligned}
    \frac{\langle{\hat q^{i_0}_{I_0}}\lr{\frac{1}{2}}\rangle_{\cs}}{t}  \overset{t\to0}{\lesssim} 
     |Z|^{\frac{1}{4}}(9M)^{\frac{1}{4}} \sum_{Jj} \frac{\sqrt{\lrabs{p^j_J}}}{t^{\frac{3}{4}}} + \mathcal{O}\lr{t^{\frac{1}{4}}}.
\end{aligned}
}
\end{equation}

If interested in the limit $p=0$, however, one has to proceed differently from~(\ref{eq:BrunnemannKCHF}) onwards.  What we did before was to use the KCHF's argument's part $\nicefrac{p^i_I}{T}$ in order to say that it is large as $T\to 0$. This clearly is not valid anymore if $p^i_I = 0$. Instead, we might proceed the following way by using the very definition of the KCHF,~(\ref{eq:defKCHF1}), with a previous Kummer transformation,~(\ref{KummerTrafo}):

\begin{align}
    \frac{\langle\qr\rangle_{\cs}}{t} & \:\!\stackrel{p^i_I=0}{\leq} \frac{\lrabs{Z}^\rhalf \lr{9M}^\rhalf}{\sqrt{\pi}} \Gamma\lr{\tfrac{r+1}{2}} t^{r-1} \sumset{N}\sum_{Jj}\es{-\sum_{Ii\backslash Jj}\frac{\pi^2 }{t}\lr{N^i_I}^2} \cdot \nonumber\\
    & \qquad \cdot \kchf{\frac{r+1}{2},\frac{1}{2},-\frac{\pi^2 }{t}\lr{N^j_J}^2} \nonumber\\
    & \stackrel{(\ref{KummerTrafo})}{=} \frac{\lrabs{Z}^\rhalf \lr{9M}^\rhalf}{\sqrt{\pi}} \Gamma\lr{\tfrac{r+1}{2}} t^{r-1} \sumset{N}\sum_{Jj}\es{-\sum_{Ii}\frac{\pi^2 }{t}\lr{N^i_I}^2} \cdot \nonumber\\
    & \qquad\cdot \kchf{-\frac{r}{2},\frac{1}{2},\frac{\pi^2 }{t}\lr{N^j_J}^2} \nonumber\\
    & \stackrel{(\ref{eq:defKCHF1})}{=} \frac{\lrabs{Z}^\rhalf \lr{9M}^\rhalf}{\sqrt{\pi}} \Gamma\lr{\tfrac{r+1}{2}} t^{r-1} \sumset{N}\sum_{Jj}\es{-\sum_{Ii}\frac{\pi^2 }{t}\lr{N^i_I}^2} \cdot \nonumber\\
    & \qquad \cdot \sum_{n=0}^\infty \frac{\lr{-\frac{r}{2}}_n}{\lr{\frac{1}{2}}_n n!}\lr{\frac{\pi^2 }{t}\lr{N^j_J}^2}^n \nonumber\\
    & \ \: = \frac{\lrabs{Z}^\rhalf \lr{9M}^\rhalf}{\sqrt{\pi}} \Gamma\lr{\tfrac{r+1}{2}} t^{r-1} \sumset{N}\sum_{Jj}\es{-\sum_{Ii}\frac{\pi^2 }{t}\lr{N^i_I}^2} \cdot \nonumber\\
    & \qquad \cdot \left( 1-r\frac{\pi^2}{t}\lr{N^j_J}^2 + \frac{2-r}{6}\frac{\pi^4}{t^2}\lr{N^j_J}^4 + \mathcal{O}\lr{t^{-3}} \right).
\end{align}
Hence, for $p^i_I = 0$, we now have a series in $\nicefrac{\lr{N^i_I}^2}{t}$. Concerning the limit of small $t$, this does however not cause problems due to the overall damping Gaussians that allow us to consider only the $N^i_I=0$ solutions again. Therefore, we altogether obtain

\begin{equation}
\boxed{
    \frac{\langle\qr\rangle_{\cs}}{t} \stackrel{p^i_I=0}{\leq} \frac{\lrabs{Z}^\rhalf \lr{9M}^{1+\rhalf}}{\sqrt{\pi}} \Gamma\lr{\tfrac{r+1}{2}} t^{r-1} \label{eq:ResultALaBrunnemannP=0}
}
\end{equation}

as an upper bound for the case of $p^i_I=0$ and $t\to 0$ when using the estimate like the one of Brunnemann and Thiemann. So we see that we even obtain a $t^0$-result when choosing $r=1$ --- which corresponds to considering $\hat V$, cf. \eqref{eq:qr}. However, note again that we did not expect the lowest order to be $\sim t^0$ anyway, since not even the quantum mechanical example in our companion paper~\cite{paper1} did feature this behaviour.

Taking a look at the respective results~\eqref{eq:ResultALaBrunnemann},~\eqref{eq:r1/2ResultALaBrunnemann} and~\eqref{eq:ResultALaBrunnemannP=0}, we see that exponent of the momenta was not preserved by the estimates. We started with the volume to the power of $r$, which translates to $p^{\nicefrac{3r}{2}}$. Via the estimate~\eqref{voleigenestimate}, however, we altered the exponent of the charges from $n^{\nicefrac{3r}{2}}$ to $n^r$, leading ultimately also to $p^r$ --- as we see in~\eqref{eq:ResultALaBrunnemann} and~\eqref{eq:r1/2ResultALaBrunnemann}. Regarding the $t$-dependency, we obtained $t^{\frac{r}{2}-1}$ for the commutator over $t$ in lowest order, confer~\eqref{eq:ResultALaBrunnemann}. For the case of $p^i_I=0$, this changed to $t^{r-1}$ in~\eqref{eq:ResultALaBrunnemannP=0}. Note that the exponent changed albeit using no additional estimates in either case, they are both based on \eqref{eq:BrunnemannKCHF}. While the $p\neq 0$ approach proceeded with the asymptotics for large arguments of the KCHFs, for the $p=0$ case we continued with the very definition of $\kchf{a,b,z}$ that can be found in \eqref{eq:defKCHF1}.

\subsubsection{Semiclassical expectation value for \texorpdfstring{$\prod_{k=1}^N {\hat q}^{i_k}_{I_k}$}{q to the power of N} } \label{subsubsec:Nq}

We now generalise the previous procedure to general products of the operator $\qr$, i.e. $\prod_{k=1}^N {\hat q}^{i_k}_{I_k}$. Therein, the shift is allowed to act on a different edge and a different U(1)-copy for the respective operators of the product, labelled by $k$.

Having~\eqref{voleigenestimate} as the estimate for the eigenvalue of one of the ${\hat q}^{i_k}_{I_k}$ operators, we can straightforwardly state our analogue of~\cite{Brunnemann2}'s (5.2):
\begin{align}
    & \langle \prod_{k=1}^N {\hat q}^{i_k}_{I_k}(r) \rangle_{\cs} \leq \frac{\lp{}^{3rN}\lrabs{Z}^{\frac{rN}{2}} (9M)^{\frac{rN}{2}}}{a^{3rN}\csnorm} \sumset{n} \e{ \sum_{Ii}\lr{-t\lr{n^i_I}^2 +2p^i_I n^i_I} } \lr{ \sum_{Jj} \lrabs{n^j_J}^r }^N \nonumber\\
    &  \leq \frac{T^{3rN}\lrabs{Z}^{\frac{rN}{2}} (9M)^{\frac{rN}{2}}}{\sqrt{\pi}^{3M}} \sumset{N} \!\! \e{-\sum_{Ii}\frac{\pi^2 \lr{N^i_I}^2}{t}} \infint^{3M} x^i_I \es{ -\sum_{Ii} \lr{ x^i_I -\frac{p^i_I-\pi\i N^i_I}{T} }^2 } \lr{ \sum_{Ii} \frac{\lrabs{x^i_I}^r}{T^r}}^N. \label{eq:startBrunnemannPath}
\end{align}

As we can not integrate this expression as it is, due to the sum to the power of $N$, we have to further manipulate this part:
\begin{align}
    \lr{ \sum_{Ii} \frac{\lrabs{x^i_I}^r}{T^r}}^N &= \frac{1}{T^{rN}} \sum_{\{ \nk \}} c_{\nk} \prod_{Ii} \lrabs{x^i_I}^{r\nik}.
\end{align}
This means that we sum over all the $\mathscr{k}$-many distributions of $N$ into non-negative integers $\nik$ such that $\sum_{Ii} \nik = N$. The single products of the sum's addends are weighted by the combinatorical coefficients
\begin{align}
c_{\nk} = \frac{N!}{\prod_{Ii}\nik !} = \binom{N}{\mathscr{n}^1_{1,\mathscr{k}}, \mathscr{n}^2_{1,\mathscr{k}}, \ldots , \mathscr{n}^3_{M,\mathscr{k}}} ,
\end{align}
which are just the multinomial coefficients. We then realise that we face integrals of Gaussians against products of the single integration variables $x^i_I$ to the power of $r\nik$. As these decouple, we can indeed apply the integration by means of KCHFs. This leads to 
\begin{align}
    \mathcal{I}_{\nik} \coloneqq & \infint{x^i_I} \es{-\lr{x^i_I - \frac{p^i_I-\pi\i N^i_I}{T}}^2} \lrabs{x^i_I}^{r\nik} \nonumber\\
     =& \, \Gamma\lr{\tfrac{1+r\nik}{2}} \es{-\lr{\frac{p^i_I-\pi\i N^i_I}{T}}^2} \kchf{\tfrac{1+r\nik}{2},\tfrac{1}{2},\lr{\tfrac{p^i_I-\pi\i N^i_I}{T}}^2} . \label{eq:Inik}
\end{align}
We can now proceed in two ways: First, we consider the general case and use the asymptotic expansion for large arguments of the KCHF and then, secondly, investigate the $p=0$ scenario like we did at the end of subsection~\ref{subsec:LikeBrunnemannThiemann}. For the first approach, we get
\begin{align}
    \mathcal{I}_{\nik} \overset{t\to 0}{=} \Gamma\lr{\tfrac{1}{2}} \frac{\lrabs{p^i_I}^{r\nik}}{T^{r\nik}} \lr{ 1 - \tfrac{r\nik\lr{1-r\nik}}{4} \tfrac{t}{\lr{p^i_I}^2} + \mathcal{O}\lr{t^2} } ,
\end{align}
including all the intermediate steps just as before --- i.e. only one part of the asymptotic expansion contributes due to the overall Gaussian and we can consider the $N^i_I=0$ contribution only. Inserting this into the formula for the expectation value, we obtain
\begin{align}
    \langle \prod_{k=1}^N & {\hat q}^{i_k}_{I_k}(r) \rangle_{\cs}  \overset{t\to 0}{\leq} \nonumber\\
    &\overset{t\to 0}{\leq} T^{3rN}\lrabs{Z}^{\frac{rN}{2}}\lr{9M}^{\frac{rN}{2}} \sum_{\{\nk\}} \frac{c_{\nk}}{T^{rN}} \prod_{Ii} \frac{\lrabs{p^i_I}^{r\nik}}{T^{r\nik}} \lr{ 1- \tfrac{r\nik\lr{1-r\nik}}{4} \tfrac{t}{\lr{p^i_I}^2} +\mathcal{O}\lr{t^2} } \nonumber\\
    &\ \leq t^{\frac{rN}{2}} \lrabs{Z}^{\tfrac{rN}{2}} \lr{9M}^{\frac{rN}{2}} \lr{3M}^N  \lrabs{p_{\text{max}}}^{rN} + \mathcal{O}\lr{t^{\frac{rN}{2}+1}}.
\end{align}
We introduced $p_{\text{max}} \coloneqq \max_{Ii} \{p^i_I\}$ as an upper bound for all the $p^i_I$ in order to be able to obtain this concise result. With the help of $p_{\text{max}}$, we could combine the product of all the different momenta and their respective exponents $r\nik$ via $\sum_{Ii}\nik = N$ to $p_{\text{max}}^{rN}$. Lastly, as all expressions are then independent of the specific $\nk$, we used
\begin{align}
    \sum_{\{\nk\}} c_{\nk} = \lr{3M}^N .
\end{align}

Including the additional division by $t^N$ in order to be able to compare it the the classical Poisson bracket's result, we finally arrive at
\begin{equation} \label{eq:ResultALaBrunnemannThiemannGeneral}
\boxed{
    \frac{\langle \prod_{k=1}^N {\hat q}^{i_k}_{I_k}(r) \rangle_{\cs}}{t^N}  \overset{t\to 0}{\leq} t^{\lr{\frac{r}{2}-1}N} \lrabs{Z}^{\tfrac{rN}{2}} \lr{9M}^{\frac{rN}{2}} \lr{3M}^N  \lrabs{p_{\text{max}}}^{rN} + \mathcal{O}\lr{t^{\lr{\frac{r}{2}-1}N+1}}
}
\end{equation}

Continuing directly with the $p=0$ scenario, we have to change the procedure from \eqref{eq:Inik} onwards. We do this similar to the calculation at the end of subsection \ref{subsec:LikeBrunnemannThiemann} and first of all insert $p=0$ and perform a Kummer transformation (cf. \eqref{KummerTrafo}):
\begin{align}
    \mathcal{I}_{\nik} & \overset{p = 0}{=} \e{\frac{\pi^2\lr{N^i_I}^2}{t}}\Gamma\lr{\tfrac{1+r\nik}{2}} \kchf{\tfrac{1+r\nik}{2},\tfrac{1}{2},-\tfrac{\pi^2\lr{N^i_I}^2}{t}} \nonumber\\
    & \; = \Gamma\lr{\tfrac{1+\nik}{2}} \kchf{-\tfrac{r\nik}{2},\tfrac{1}{2},\tfrac{\pi^2\lr{N^i_I}^2}{t}}.
\end{align}
The Gaussian in $\nicefrac{-\pi^2\lr{N^i_I}^2}{t}$ from \eqref{eq:startBrunnemannPath} then allows us to consider again only $N^i_I=0$, as we find via using the very definition \eqref{eq:defKCHF1} of the KCHF
\begin{align}
    \e{-\frac{\pi^2\lr{N^i_I}^2}{t}} \cdot  \mathcal{I}_{\nik} \overset{p= 0}{=} \Gamma\lr{\tfrac{1+r\nik}{2}} \e{-\frac{\pi^2\lr{N^i_I}^2}{t}} \lr{ 1 - r\nik \frac{\pi^2\lr{N^i_I}^2}{t} +\mathcal{O}\lr{t^{-2}} } = \Gamma\lr{\tfrac{1+r\nik}{2}} .
\end{align}
With this, we can state
\begin{align}
        \langle \prod_{k=1}^N {\hat q}^{i_k}_{I_k}(r) \rangle_{\cs} \overset{p=0}{\leq}  t^{rN} \lr{9M}^{\frac{rN}{2}} \lrabs{Z}^{\frac{rN}{2}} \sum_{\{\nk\}} \frac{c_{\mathscr{n}_k}}{\sqrt{\pi}^{3M}} \prod_{Ii} \Gamma\lr{\tfrac{1+r\nik}{2}},
\end{align}
and including the division by $t^N$
\begin{equation} \label{eq:ResultALaBrunnemannThiemannGeneralP=0}
\boxed{
\langle \frac{\prod_{k=1}^N {\hat q}^{i_k}_{I_k}(r)}{t^N} \rangle_{\cs} \overset{p=0}{\leq}  t^{(r-1)N} \lr{9M}^{\frac{rN}{2}} \lrabs{Z}^{\frac{rN}{2}} \sum_{\{\nk\}} \frac{c_{\mathscr{n}_k}}{\sqrt{\pi}^{3M}} \prod_{Ii} \Gamma\lr{\tfrac{1+r\nik}{2}}.
}
\end{equation}
Note that the $\sqrt{\pi}^{3M}$ in the denominator gets reduced by all those $\Gamma\lr{\nicefrac{\lr{1+r\nik}}{2}}$ for which $\nik=0$, leaving $\sqrt{\pi}^{\:\!\sharp(\nik)}$ with $\sharp(\nik)$ standing for the number of non-zero $\nik$ within the decomposition $\nk$, i.e. the number of integrals that resulted in $\Gamma\lr{\nicefrac{\lr{1+r\nik}}{2}}$ instead of $\Gamma\lr{\tfrac{1}{2}}=\sqrt{\pi}$. This feature of a fraction of a gamma function and a remaining $\sqrt{\pi}$ from the normalisation when considering $p=0$ can also be seen in \eqref{eq:resultRigorousP=0} and even in the quantum mechanical case: confer equation (3.13) of our companion paper~\cite{paper1}.

We therefore showed that we can also consider general $N$-fold products of the operator $\hat{q}^{i_k}_{I_k}$ when following the Brunnemann and Thiemann path but using KCHFs in order to preserve fractional powers. The results, however, feature the same divergence for $t\to 0$, but we were able to retain some information about the initial, fractional power of the volume operator that we started with.

The considerations in the next subsections are now about the question of how to improve the existing estimates in such a way that they do conserve the power in the momenta and the classicality parameter.

\subsection{Towards an improved estimate} \label{subsec:towards}

What we can deduce so far about whether an estimate is potentially conserving the correct exponents of $p$ and $t$ is that we need to have a result that still is a difference in two KCHFs, reflecting the commutator's two expectation values: estimating the difference by a single term, and hence yielding only one single KCHF after the integration, causes the lowest order term in the asymptotic expansion to survive, which thereby modifies the initial powers in $p$ and $t$. Furthermore, we need to try conserving the exponent of the charges during our estimates.

However, we also know that the approach via KCHFs does not work if we have too complex expressions within the roots such as products of different charges $n^i_I$ and non-constant additions to the charges. This forces us to find an estimate that allows to further remodel the expressions by, e.g., factoring out one charge after the other to enable the integration.

We first of all recapitulate what we are facing:
\begin{align}
    \lambda^r\lr{\left\{n^i_I\right\}} -& \lambda^r\lr{\left\{n^i_I+\delta^{ii_0}\delta_{II_0}\right\}}  = \lpz \left( \lrabs{\sum_{IJK}\det\lr{n^i_In^j_Jn^k_K}}^\rhalf - \right. \nonumber\\
    & - \left. \lrabs{\sum_{IJK}\det\lr{\lr{n^i_I+\delta^{ii_0}\delta_{II_0}}\lr{n^j_J+\delta^{ji_0}\delta_{JI_0}}\lr{n^k_K+\delta^{ki_0}\delta_{KI_0}}}}^\rhalf \right) . \label{rawdifference}
\end{align}

As indicated before, we might achieve progress by factoring out one charge. So as a first step, we apply Laplace's rule on an expression mimicking one representative of the charge matrices' differences:
\be
\left| \det
\begin{pmatrix}
a & b & c \\
d & e & f \\
g & h & i
\end{pmatrix} \right|^\rhalf - \left| \det
\begin{pmatrix}
(a-1) & b & c \\
d & e & f \\
g & h & i
\end{pmatrix} \right|^\rhalf = \lrabs{a\det(a_-)+{\widetilde C}}^\rhalf - \lrabs{(a-1)\det(a_-)+{\widetilde C}}^\rhalf ,
\ee
where $\det(a_-)$ denotes the minor of the matrix with respect to entry $a$ and ${\widetilde C}\coloneqq- b\det(b_-)+c\det(c_-)$, with $\det(b_-), \det(c_-)$ being the minors of the matrix with respect to $b, c$ respectively.

Obviously, the most progress would be achieved by just dropping the ${\widetilde C}$ as we would not only avoid many integrals, but especially cast the fractional power into an expression that can be analytically integrated and expressed by means of KCHFs, since we could then factor out $\det(a_-)$ and hence face the familiar expression giving rise to a difference in two KCHFs.

Proceeding with finding an appropriate approximation, we reshape the difference as
\begin{equation}
\label{eq:DiffC}
    \lrabs{a\det(a_-)+{\widetilde C}}^\rhalf - \lrabs{(a-1)\det(a_-)+{\widetilde C}}^\rhalf \eqqcolon \lr{\det(a_-)}^\rhalf\lr{\lrabs{a+C}^\rhalf - \lrabs{(a-1)+C}^\rhalf},
\end{equation}
on which we can then apply the new estimate~(\ref{approxfinal})
\be
    |a+C|^\rhalf-|(a-1)+C|^\rhalf \leq |a|^\rhalf - |a-1|^\rhalf +2 \ , \tag{\ref{approxfinal}}
\ee
where $a,C,r\in\mathbbm{R}$ and $ 0\leq r\leq 2$.
As we see when considering $a=0, C=1$, getting rid of $C$ without any cost was in fact not possible and we needed to include the offset $+2$ instead.

Applying this now on the difference in the eigenvalues~(\ref{rawdifference}), with $a$ in~\eqref{eq:DiffC} corresponding to the charge $n^{i_0}_{I_0}$ that gets shifted, we obtain
\begin{align}
    \Delta\lambda^r & \leq \lpz \lrabs{\sum_{JK} \epsilon \lr{I_0JK} \epsilon_{i_0jk} n^j_J n^k_K }^\rhalf \left[ \lrabs{n^{i_0}_{I_0}}^\rhalf - \lrabs{n^{i_0}_{I_0}+1}^\rhalf+ 2 \right] . \label{estimatetowards}
\end{align}

We again denoted by $\sum_{JK}$ the abbreviation for the sum over all edges $e_J, e_K$ such that $e_{I_0}\cap e_J \cap e_K = v \land J,K \neq I_0 \land J\neq K$, which just reflects the expression to be the sum of all minors with respect to $n^{i_0}_{I_0}$ of matrices containing this charge. Thus, the term $\det(a_-)$ from before is now a sum. The remaining two terms of the Laplace expansion as well as all determinants of charge matrices not containing $n^{i_0}_{I_0}$ are absorbed into $C$ and therefore lost after applying the approximation.

This expression can now be integrated straightforwardly with respect to $n^{i_0}_{I_0}$ by means of KCHFs, while the fractional power of the sum over all minors of the initial charge matrix before could not.

Inserting the above estimate into the expectation value of $\qr$ in \eqref{eq:Expectationvalue}, applying the Poisson resummation formula and integrating over $n^{i_0}_{I_0}$, we get
\begin{align}
    \langle\qr&\rangle_{\cs}  \leq \nonumber\\
    & \leq \frac{T^{3r}\lrabs{Z}^\rhalf}{\sqrt{\pi}^{3M}}\sumset{N} \e{-\sum_{Ii}\lr{\frac{2\pi\i p^i_IN^i_I}{t}+\frac{\pi^2\lr{N^i_I}^2}{t}}}\infint^{3M-1} x^{i\backslash i_0}_{I\backslash I_0} \es{-\sum_{Ii\backslash I_0i_0} \lr{x^i_I-\frac{p^i_I-\pi\i N^i_I}{T}}^2} \cdot \nonumber\\
    &\quad \cdot\lrabs{\sum_{JK}\epsilon\lr{I_0JK}\epsilon_{i_0jk} x^j_J x^k_K }^\rhalf 
     T^{-r} \left[ T^{-\rhalf}\Gamma\lr{\tfrac{r+2}{4}} \kchf{-\frac{r}{4},\frac{1}{2},-\lr{\frac{\pzero-\pi\i N^{i_0}_{I_0}}{T}}^2}- \right. \nonumber\\
   & \qquad\qquad \left. -T^{-\rhalf}\Gamma\lr{\tfrac{r+2}{4}} \kchf{-\frac{r}{4},\frac{1}{2},-\lr{\frac{\pzero+T^2-\pi\i N^{i_0}_{I_0}}{T}}^2}  +2\sqrt{\pi} \right] .
\end{align}

For not making it unnecessarily confusing, the structure of the equation before is kept and the square bracket's single terms correspond one to one to each other. All additional factors before the sum over all minors stem from going over to the expectation value of $\qr$ and applying the Poisson resummation formula.
The problem now is that the integral involving  the Gaussians as well as the fractional power of the sum over all determinants of the minors of the charge matrices can not be solved analytically. Hence, we have to apply further estimates.
We do this in the fashion of the approximation~(\ref{approxsum}),\be
    |a+b|^r\leq|a|^r+|b|^r \quad (\text{where} \ a,b,r\in\mathbbm{R} \ \text{and } 0\leq r\leq 1)\tag{\ref{approxsum}},
\ee
which applied to the present scenario then reads
\begin{align}
    \lrabs{\sum_{JK}\epsilon\lr{I_0JK}\epsilon_{i_0jk} x^j_J x^k_K }^\rhalf & \leq \sum_{JKjk} \lrabs{\epsilon\lr{I_0JK}\epsilon_{i_0jk}x^j_J x^k_K}^\rhalf \nonumber\\
    & = \sum_{JKjk} \lrabs{x^j_J}^\rhalf\lrabs{x^k_K}^\rhalf .
\end{align}
Therein, the sum over $j,k$ is such that $j,k\neq i_0$ and in the first line, the $j,k$ under the sum means there is no further summation over double indices inside the absolute value. The same is true for $J,K$ and $I_0$.

Therefore, all remaining integrals are now
\begin{align}
    & \infint {}^{3M-1}  x^{i\backslash i_0}_{I\backslash I_0}  \es{-\sum_{Ii\backslash I_0i_0} \lr{x^i_I-\frac{p^i_I-\pi\i N^i_I}{T}}^2} \sum_{{JKjk}} \lrabs{x^j_J}^\rhalf\lrabs{x^k_K}^\rhalf = \nonumber \\ 
    & = \sqrt{\pi}^{3M-3}\,\Gamma^2\lr{\tfrac{r+2}{4}} \sum_{{JKjk}} \kchf{-\frac{r}{4},\frac{1}{2},-\lr{\frac{p^j_J-\pi\i N^j_J}{T}}^2} \kchf{-\frac{r}{4},\frac{1}{2},-\lr{\frac{p^k_K-\pi\i N^k_K}{T}}^2},
\end{align}
since we have $3M-3$ many integrals over Gaussians for each summand and only two of them are multiplied by the roots, giving rise to the two KCHFs. Note that the sum over $J,K,j,k$ is still considered to include only those edges $e_J \neq e_K$ that meet $e_{I_0}$ at $v$ and, likewise, only those U(1)-copies $j\neq k$ with $j,k\neq i_0$ .

Putting everything together, we obtain
\begin{align}
    &\langle\qr\rangle_{\cs}  \leq \nonumber\\
    & \leq \frac{T^{3r}\lrabs{Z}^\rhalf T^{-\frac{3r}{2}}}{\sqrt{\pi}^{3}} \!\!\! \sumset{N} \e{-\sum_{Ii}\lr{\frac{2\pi\i p^i_IN^i_I}{t}+\frac{\pi^2\lr{N^i_I}^2}{t}}} \Gamma^3\lr{\tfrac{r+2}{4}} \left[ \kchf{-\tfrac{r}{4},\tfrac{1}{2}, -\lr{\tfrac{\pzero -\pi\i N^{i_0}_{I_0}}{T}}^2} \right.-
    \nonumber\\
   & \qquad - \left. \kchf{-\tfrac{r}{4},\tfrac{1}{2}, -\lr{\tfrac{\pzero +T^2-\pi\i N^{i_0}_{I_0}}{T}}^2} +2\Gamma^{-1}\lr{\tfrac{r+2}{4}}T^{\frac{r}{2}}\sqrt{\pi}\right]\cdot \nonumber\\
    & \quad \cdot \sum_{{JKjk}} \kchf{-\tfrac{r}{4},\tfrac{1}{2}, -\lr{\tfrac{p^j_J-\pi\i N^j_J}{T}}^2} \kchf{-\tfrac{r}{4},\tfrac{1}{2}, -\lr{\tfrac{p^k_K-\pi\i N^k_K}{T}}^2} \ . \label{eq:+2EstimateKCHFs}
\end{align}

Applying the asymptotic expansion for large arguments, we notice that again only the expansion's sum without the Gaussian prefactor in $\nicefrac{p^i_I}{T}$ does not vanish for $T\to 0$ and similarly, only the summand $\{N^i_I\} = 0$ contributes out of the sum over all $\{N^i_I\}$. 
As before, for our final upper bound, we divide the entire expression by $t$ in order to be able to compare it with the classical Poisson bracket. Our final result for the upper bound for a generic graph with $M$ edges reads in full detail 
\begin{equation}
\boxed{
\begin{aligned}
    & \frac{\langle\qr\rangle_{\cs}}{t} \lesssim T^{3r}\lrabs{Z}^\rhalf T^{-\frac{3r}{2}-2} \left[ \lrabs{\frac{\pzero }{T}}^{\frac{r}{2}} \sum_{s=0}^{\infty} \frac{\lr{-\frac{r}{4}}_s\lr{\frac{2-r}{4}}_s}{s!}\lr{\frac{\pzero }{T}}^{-2s} - \right. \\
    & \qquad\qquad \left. - \lrabs{\frac{\pzero +T^2}{T}}^{\frac{r}{2}} \sum_{s=0}^{\infty} \frac{\lr{-\frac{r}{4}}_s\lr{\frac{2-r}{4}}_s}{s!}\lr{\frac{\pzero+T^2}{T}}^{-2s} +2T^{\frac{r}{2}}\right] \cdot \\
    & \quad\quad \cdot \sum_{{JKjk}} \lr{ \lrabs{\frac{p^j_J}{T}}^{\frac{r}{2}} \sum_{s=0}^{\infty} \frac{\lr{-\frac{r}{4}}_s\lr{\frac{2-r}{4}}_s}{s!}\lr{\frac{p^j_J}{T}}^{-2s} } \lr{ \lrabs{\frac{p^k_K}{T}}^{\frac{r}{2}} \sum_{s=0}^{\infty} \frac{\lr{-\frac{r}{4}}_s\lr{\frac{2-r}{4}}_s}{s!}\lr{\frac{p^k_K}{T}}^{-2s} } .
\end{aligned}
}
\end{equation}
Considering now all terms up to $s=1$, we see first of all that the zeroth order terms in the square bracket's sums cancel each other, but the offset $2T^{\frac{r}{2}}$ remains. Finally, we obtain
\begin{equation}
\label{finaltowards}
\boxed{
\begin{aligned}
    & \frac{\langle\qr\rangle_{\cs}}{t} \lesssim  |Z|^\frac{r}{2}T^{\frac{3r}{2}-2} \left[ 2T^{\frac{r}{2}} - \frac{r}{2}\sgn \pzero\, \lrabs{\pzero}^{\rhalf-1}T^{2-\rhalf} +\mathcal{O}\lr{T^{4-\rhalf}} \right] \cdot \\ 
    & \qquad \cdot \sum_{{JKjk}} \lr{ \lrabs{p^j_Jp^k_K}^\rhalf T^{-r} - r(2-r)\frac{\lr{p^j_J}^2+\lr{p^k_K}^2}{16\lr{p^j_J}^2\lr{p^k_K}^2} \lrabs{p^j_Jp^k_K}^\rhalf T^{2-r} +\mathcal{O}\lr{T^{4-r}} } \\
    &\lesssim |Z|^\frac{r}{2}2M \lr{ 2\lrabs{\pmax}^r t^{\frac{r}{2}-1} - \frac{r}{2}\lrabs{\pmax}^r \sgn \pzero\,  \lrabs{\pzero}^{\frac{r}{2}-1}t^0 - \frac{r(2-r)}{4} \lrabs{\pmax}^{r-2} t^{\rhalf}}  + \mathcal{O}\lr{t^1}, 
\end{aligned}
}
\end{equation}
where we introduced $\lrabs{\pmax} = \max_{i\neq i_0, I\neq I_0}\lr{\left\{ \lrabs{p^i_I} \right\}}$ --- with $e_I$ being an edge meeting with $e_{I_0}$ at $v$. This was possible under the assumption that $\lrabs{\pmax}$ is not increasing faster than $t$ approaches 0 such that the last term in the second line is still contributing less than the leading term before. Also, we estimated the sum over all edges joining $e_{I_0}$ at $v$ and the remaining two U(1)--copies by $2M$.

We realise our obtained result also diverges for $t\to 0$. We can directly see that this is due to the result not entirely being a difference in two KCHFs: The diverging term proportional to $t^\rhalf$ stems from multiplying the offset's term $2T^\rhalf$ in the square bracket with the term proportional to $T^{-r}$ of the sum over all contributions of the minors. Without the offset in~\eqref{eq:DiffC}, the leading order term would be the final result's second one, stemming from the leading order term of the difference of the two KCHFs with the shifted momentum in their argument. This is the square bracket's second term multiplied by the lowest order term $\sim T^{-r}$ of the sum over all contributions of the minors. This expression not only resembles the expected $t$--dependency, but also the exponents of the momenta are correct. Comparing it with what one obtains via differentiating, which would be the case for the classical Poisson bracket, we find accordance when it comes to the exponents of the momenta: the two unshifted momenta remain unchanged --- each one contributing with $ p^\rhalf$ yielding  an overall $p^r$ ---, while the momentum $\pzero$'s exponent is decreased by 1, also giving rise to the prefactor $\rhalf$.

Being interested in the limit of $p\to 0$, we have to proceed differently from \eqref{eq:+2EstimateKCHFs} onwards, as the asymptotic expansion for large arguments of the Kummer functions is not applicable anymore. Continuing in the same fashion as we already did before, we set $p=0$ in \eqref{eq:+2EstimateKCHFs} and see that the overall Gaussian in $\nicefrac{\pi^2 \lr{N^i_I}^2}{t}$ allows us again to only consider the $N^i_I = 0$ contribution. This implies for most KCHFs $\kchf{a,b,0} = 1$, with only the one including the shift resulting in a power series in $t$. Using $\sum_{JKjk}1 \leq \lr{3M}^2$, the classicality parameter $t=\nicefrac{\lp{}^2}{a^2}$ and diving the final result again by $t$, we get for the upper bound in the $p\to 0$ limit 
\begin{equation} \label{finalTowardsp=0}
\boxed{
    \begin{aligned}
  \frac{\langle \qr\rangle_{\cs}}{t} \overset{p^i_I=0}{\leq} \frac{\lrabs{Z}^{\rhalf}\lr{3M}^2}{\sqrt{\pi}^3} \Gamma^3\lr{\tfrac{r+2}{4}} \lr{\frac{2\sqrt{\pi}}{\Gamma\lr{\tfrac{r+2}{4}}} t^{r-1} + \rhalf t^{\tfrac{3r}{4}} + \mathcal{O}\lr{t^{1+\tfrac{3r}{4}}}}.
\end{aligned}
}
\end{equation}
We can again link the lowest order contributing term with the estimate's offset +2 and without it, the lowest order term had been the non-diverging $\sim t^{\nicefrac{3r}{4}}$ one. We also realise that despite having tried to construct an improved estimate, the offset's influence caused the same $t$-dependencies for both the general and the $p=0$ scenario that we got in subsection \ref{subsec:LikeBrunnemannThiemann} and hence it is again diverging if we take in addition the limit $t\to 0$.

\subsection{On new estimates} 
\label{subsec:newEstimates}
So far, we used estimates that in the end allowed us to integrate Gaussians in $x^i_I$ against functions of the form $\lrabs{x^i_I}^r$. However, this caused problems concerning either the powers in $p$ or $t$ --- or both. Having in mind that we now know how to analytically integrate roots of determinants, we can combine estimates to remove roots \textit{of sums} of determinants, but then integrate the actual roots of determinants without the need of further estimations.

Assigning again the shift to the $n^1_1$-charge w.l.o.g., we assume there are $N$ determinants that contain $n^1_1$ and hence, write the difference in the roots as
\begin{align}
\lrabs{\sum_{\mathfrak{n}=1}^N \det\mathcal{N}_\mathfrak{n} + C}^\rhalf - \lrabs{\sum_{\mathfrak{n}=1}^N \det\widetilde{\mathcal{N}}_\mathfrak{n} + C}^\rhalf \leq \lrabs{ \sum_{\mathfrak{n}=1}^N \lr{ \det\mathcal{N}_\mathfrak{n} - \det\widetilde{\mathcal{N}}_\mathfrak{n} } }^\rhalf,
\end{align}
where we denoted the charge matrices that contain $n^1_1$ by $\mathcal{N}_\mathfrak{n}$ and, accordingly, those with the shifted $n^1_1+1$ by $\widetilde{\mathcal{N}}_\mathfrak{n}$. All remaining determinants of charge matrices not containing $n^1_1$ were put into $C$ and we see that those vanish by applying the estimate (\ref{approxdiff}).

Next, we perform the Laplace expansion with respect to the charges $n^i_1$ of edge 1 on the determinants:
\begin{align}
\lrabs{ \sum_{\mathfrak{n}=1}^N \lr{ \det\mathcal{N}_\mathfrak{n} - \det\widetilde{\mathcal{N}}_\mathfrak{n} } }^\rhalf & =  \left| \sum_{\mathfrak{n}=1}^N \lr{ n^1_1 \Delta^1_1 \left(\mathcal{N}_{\mathfrak{n}}} - n^2_1 \Delta^2_1 \lr{\mathcal{N}_{\mathfrak{n}}} + n^3_1 \Delta^3_1 \lr{\mathcal{N}_{\mathfrak{n}}} - \right. \right. \nonumber\\
& \qquad\quad\ \left. \left. -  \lr{n^1_1+1} \Delta^1_1 \lr{\mathcal{N}_{\mathfrak{n}}} + n^2_1 \Delta^2_1 \lr{\mathcal{N}_{\mathfrak{n}}} - n^3_1 \Delta^3_1 \lr{\mathcal{N}_{\mathfrak{n}}} \right) \vphantom{\sum_{\mathfrak{n}=1}^N} \right|^\rhalf \nonumber\\
& = \lrabs{ \sum_{\mathfrak{n}=1}^N \Delta^1_1 \lr{\mathcal{N}_{\mathfrak{n}}} }^\rhalf \equiv \lrabs{ \sum_{IJ} \det \begin{pmatrix}
n^2_I & n^3_I \\
n^2_J & n^3_J  
\end{pmatrix} }^\rhalf \leq \sum_{IJ} \lrabs{  \det \begin{pmatrix}
n^2_I & n^3_I \\
n^2_J & n^3_J  
\end{pmatrix} }^\rhalf,
\end{align}
where --- as before --- the sum over $I,J$ considers all edges $e_I\neq e_J$ that meet $e_1$ at the vertex $v$.
We furthermore denoted the minor of the determinant of the charge matrix $\mathcal{N}_\mathfrak{n}$ with respect to the charge $n^i_1$ by $\Delta^i_1 \lr{\mathcal{N}_{\mathfrak{n}}}$:
\begin{equation}
\det\mathcal{N}_\mathfrak{n} = \det \left(\begin{array}{*3{c}}
    n_1^1 & n^2_1 & n^3_1 \\[.1cm]
    n^1_I & \tikzmark{left}{$n^2_I$} & n^3_I  \\
    n^1_J & \tikzmark{lowerleft}{$n^2_J$}  & \tikzmark{right}{$n^3_J$}
  \end{array}\ \right)
    \Highlight[first]
    \begin{tikzpicture}[overlay, remember picture]
    \draw[<-,thick] (-0.4,-0.2) to[out=0,in=180] (0.6,0.1) node[right]{$\Delta^1_1 \lr{\mathcal{N}_{\mathfrak{n}}}$ .};
    \end{tikzpicture}
\end{equation}

First of all, we notice that due to our estimate of the difference of two fractional powers as the fractional power of the difference, also all minors of the determinants except for the one with respect to the shifted $n^1_1$ vanish. We are then left with the fractional power of the sum of all minors of the determinants over all edges that meet at $v$ with $e_1$. Since we put the shift into the first U(1)-charge, only the respective 2 and 3 components of the other edges' charges contribute. The last step then was to apply the estimate in (\ref{approxsum}) in order to cast the integral in a form that involves instead of a fractional power of a sum  the sum of fractional powers, which can be solved in terms of KCHFs.

What is left to compute is a sum over all fractional powers of the minors of the determinants integrated against all $3M$ Gaussians in the charges $n^i_I$. With each minor of the determinant depending only on four of those charges, we immediately have $3M-4$ many integrals to result in $\sqrt{\pi}$ via the standard Gaussian integral. Note that before, in the analytical treatment of the eigenvalue's basic building block, we had to perform all $3M=9$ integrations without some being just standard Gaussian ones.

After performing the Poisson resummation formula, we are left with the following involved integrals:
\begin{align}
\langle  \hat{q}^1_1(r) \rangle_{\cs} & \leq \lr{\frac{T}{2\pi\sqrt{\pi}}}^{3M} \e{-\sum_{iI} \lr{\frac{p^i_i}{T}}^2 } \frac{\lp{}^{6s} |Z|^s }{a^{6s}\prod_i\lr{1+K^i}} \lr{\frac{2\pi}{T}}^{3M} \sumset{N} \e{\sum_{iI} \lr{ \frac{p^i_I-\pi\i N^i_I}{T} }^2 } \cdot \nonumber\\
& \quad\ \cdot \infint^{3M} x^i_I \es{-\sum_{iI} \lr{ x^i_I - \frac{p^i_I-\pi\i N^i_I}{T} }^2 } \frac{1}{T^{2s}} \sum_{IJ} \lrabs{ x^2_Ix^3_J - x^3_Ix^2_J }^s \nonumber\\
& = \frac{T^{4s} |Z|^s }{\sqrt{\pi}^4\prod_i\lr{1+K^i}} \sumset{N} \e{-\sum_{iI} \lr{ \frac{2\pi\i p^i_I N^i_I}{T^2 } + \frac{\pi^2 \lr{N^i_I}^2}{T^2}  } } \cdot \nonumber\\
&\quad\ \cdot \sum_{IJ} \infint x^2_I \infint x^3_I \infint x^2_J \infint x^3_J \es{ -\sum_{\stackrel{k=2,3}{K=I,J}}\lr{ x^k_K - \frac{p^k_K-\pi\i N^k_K}{T} }^2 } \! \lrabs{ x^2_Ix^3_J-x^3_Ix^2_J }^s, \label{moreedgesfirst}
\end{align}
where we used $s\coloneqq \rhalf$. We see that we can substitute the determinant-like argument of the absolute value similarly to before (cf. subsection \ref{subsubsec:rigorous3edges}) via
\begin{align}
\tilde{x}^2_I & \coloneqq x^2_Ix^3_J-x^3_Ix^2_J \\
\d \tilde{x}^2_I &\hphantom{:} = \lrabs{x^3_J} \d x^2_I \ .
\end{align}
Then, isolating the $\tilde{x}^2_I$-integration, we have via (\ref{KCHF}) and (\ref{expansion})
\begin{align}
   & \infint \tilde{x}^2_I \es{-\lr{\frac{\tilde{x}^2_I}{x^3_J} + \frac{x^3_Ix^2_J}{x^3_J} - \frac{p^2_I-\pi\i N^2_I}{T} }^2} \frac{\lrabs{\tilde{x}^2_I}^s}{\lrabs{x^3_J}}  \nonumber\\
   & =  \Gsum \lrabs{x^3_J}^s \es{-\lr{\frac{x^3_Ix^2_J}{x^3_J} - \frac{p^2_I-\pi\i N^2_I}{T}}^2} \kchf{\frac{1+s}{2},\frac{1}{2},\lr{\frac{x^3_Ix^2_J}{x^3_J} - \frac{p^2_I-\pi\i N^2_I}{T}}^2} \nonumber\\
   & \approx \Ghalf \lrabs{x^3_J}^s \lrabs{\frac{x^3_Ix^2_J}{x^3_J} - \frac{p^2_I}{T}}^s \lr{1-\frac{s(1-s)}{4}\lr{\frac{x^3_Ix^2_J}{x^3_J} - \frac{p^2_I}{T}}^{-2}+\mathcal{O}\lr{T^3}} \nonumber\\
   & \approx \Ghalf  \lrabs{x^3_Ix^2_J - \frac{p^2_I}{T}x^3_J}^s \underbrace{\lr{1-\frac{s(1-s)}{4} \frac{T^2}{\lr{p^2_I}^2} +\mathcal{O}\lr{T^3}}}_{\eqqcolon \mathscr{T}},
\end{align}
where we also considered only the $N^2_I=0$ contribution for the same reasoning as before.

Continuing with the (isolated) $x^3_J$-integration, we substitute
\begin{align}
\tilde{x}^3_J & \coloneqq \frac{p^2_I}{T}x^3_J - x^3_Ix^2_J  \\
\d \tilde{x}^3_J &\hphantom{:} = \lrabs{\frac{p^2_I}{T}} \d x^3_J \ .
\end{align}
and hence have
\begin{align}
     \Ghalf \mathscr{T} \!\!& \infint \tilde{x}^3_J \es{-\lr{ \frac{\tilde{x}^3_J}{\nicefrac{p^2_I}{T}} + \frac{x^3_Ix^2_J}{\nicefrac{p^2_I}{T}} - \frac{p^3_J-\pi\i N^3_J}{T} }^2} \lrabs{\frac{T}{p^2_I}} \lrabs{\tilde{x}^3_J}^s = \nonumber\\
    & = \Ghalf \mathscr{T} \Gsum \lrabs{\frac{p^2_I}{T}}^s \es{-\lr{\frac{x^3_Ix^2_J}{\nicefrac{p^2_I}{T}}-\frac{p^3_J-\pi\i N^3_J}{T}}^2}\kchf{\frac{1+s}{2},\frac{1}{2},\lr{\frac{x^3_Ix^2_J}{\nicefrac{p^2_I}{T}}-\frac{p^3_J-\pi\i N^3_J}{T}}^2} \nonumber\\
    & \approx \Ghalf^2\mathscr{T} \lrabs{\frac{p^2_I}{T}}^s \lrabs{\frac{x^3_Ix^2_J}{\nicefrac{p^2_I}{T}}-\frac{p^3_J}{T}}^s \lr{1-\frac{s\lr{1-s}}{4} \lr{\frac{x^3_Ix^2_J}{\nicefrac{p^2_I}{T}}-\frac{p^3_J}{T}}^{-2}+\mathcal{O}\lr{T^3}} \nonumber\\
    & \approx \Ghalf^2 \mathscr{T} \lrabs{ x^3_Ix^2_J - \frac{p^2_Ip^3_J}{T^2} }^s \underbrace{\lr{1-\frac{s\lr{1-s}}{4}\frac{T^2}{\lr{p^3_J}^2}+\mathcal{O}\lr{T^3}}}_{\eqqcolon \mathscr{T}'} \label{eq:LastNewApproachAsymptoticExpansion}
\end{align}
via the asymptotic expansion of the KCHF from the second to the third line --- considering only the $N^3_J=0$ contribution now --- and the subsequent Taylor expansion.

Consequently, we continue by substituting
\begin{align}
\tilde{x}^3_I & \coloneqq x^3_Ix^2_J - \frac{p^2_Ip^3_J}{T^2}  \\
\d \tilde{x}^3_I &\hphantom{:} = \lrabs{x^2_J} \d x^3_I 
\end{align}
in order to obtain the isolated $x^3_I$-integration
\begin{align}
     \Ghalf^2&\mathscr{T}\mathscr{T}' \infint \tilde{x}^3_I \es{-\lr{ \frac{\tilde{x}^3_I}{x^2_J} + \frac{p^2_Ip^3_J}{T^2x^2_J}-\frac{p^3_I-\pi\i N^3_I}{T} }^2} \frac{\lrabs{\tilde{x}^3_I}^s}{\lrabs{x^2_J}} = \nonumber\\
    & = \Ghalf^2\mathscr{T}\mathscr{T}' \Gsum \lrabs{x^2_J}^s \es{-\lr{\frac{p^2_Ip^3_J}{T^2x^2_J}-\frac{p^3_I-\pi\i N^3_I}{T}}^2}\kchf{\frac{1+s}{2},\frac{1}{2},\lr{\frac{p^2_Ip^3_J}{T^2x^2_J}-\frac{p^3_I-\pi\i N^3_I}{T}}^2} \nonumber\\
    & \approx \Ghalf^3\mathscr{T}\mathscr{T}' \lrabs{\frac{p^2_Ip^3_J}{T^2}-\frac{p^3_I}{T}x^2_J}^s \lr{1-\frac{s(1-s)}{4} \lr{\frac{p^2_Ip^3_J}{T^2x^2_J}-\frac{p^3_I}{T}}^{-2}+\mathcal{O}\lr{T^5}} \nonumber\\
    & \approx \Ghalf^3\mathscr{T}\mathscr{T}' \lrabs{\frac{p^2_Ip^3_J}{T^2}-\frac{p^3_I}{T}x^2_J}^s \lr{1-\frac{s(1-s)}{4} \frac{\lr{x^2_J}^2\,T^4}{\lr{p^2_Ip^3_J}^2} +\mathcal{O}\lr{T^5}}.
\end{align}
We again performed the usual steps, including neglecting all but the $N^3_I=0$ contribution. With the expansion series' first non-constant term in $T$ being $x^2_J$-dependent, the last integration differs from the three before. However, via the substitution
\begin{align}
\tilde{x}^2_J & \coloneqq \frac{p^3_I}{T}x^2_J - \frac{p^2_Ip^3_J}{T^2} \\
\d \tilde{x}^2_J &\hphantom{:} = \lrabs{\frac{p^3_I}{T}} \d x^2_J 
\end{align}
and (\ref{xsquareintegration}), it is feasible in the same manner --- just with three resulting KCHFs and their succeeding expansion:
\begin{align}
    & \Ghalf^3\mathscr{T}\mathscr{T}' \infint \tilde{x}^2_J \lrabs{\frac{T}{p^3_I}} \es{-\lr{\frac{\tilde{x}^2_J}{\nicefrac{p^3_I}{T}}+\frac{p^2_Ip^3_J}{T^2\nicefrac{p^3_I}{T}}-\frac{p^2_J-\pi\i N^2_J}{T}}^2} \lrabs{\tilde{x}^2_J}^s \cdot \nonumber\\
    & \quad \cdot \lr{ 1- \frac{s(1-s)}{4} \frac{T^4}{\lr{p^2_Ip^3_J}^2} \lr{\frac{\tilde{x}^2_J}{\nicefrac{p^3_I}{T}} + \frac{p^2_Ip^3_J}{T^2\nicefrac{p^3_I}{T}}}^2  } =  \nonumber\\
    & = \Ghalf^3\mathscr{T}\mathscr{T}' \Gsum \lrabs{\frac{p^3_I}{T}}^2 \, \frac{\e{-\lr{\frac{\Delta^1_1\lr{p}}{p^3_IT}}^2}}{8 \lr{p^2_Ip^3_Ip^3_J}^2} \cdot \nonumber\\
    & \qquad \cdot \left[ 2\lr{p^2_Ip^3_J}^2\lr{4\lr{p^3_I}^2-s(1-s)T^2} \kchf{\frac{1+s}{2},\frac{1}{2},\lr{\frac{\Delta^1_1\lr{p}}{p^3_IT}}^2} + \right.\nonumber\\
    & \qquad\quad\ + s\lr{1-s^2}T^2 \left( \lr{p^3_IT}^2 \kchf{\frac{3+s}{2},\frac{1}{2},\lr{\frac{\Delta^1_1\lr{p}}{p^3_IT}}^2} \right. \nonumber\\
    & \left. \left. \qquad\qquad\qquad\qquad\qquad\quad - 4 \Delta^1_1\lr{p} \, p^2_Ip^3_J\, \kchf{\frac{3+s}{2},\frac{3}{2},\lr{\frac{\Delta^1_1\lr{p}}{p^3_IT}}^2} \right) \right] \nonumber\\
    & \approx \Ghalf^4\mathscr{T}\mathscr{T}' \lrabs{\frac{\Delta^1_1\lr{p}}{T^2}}^s \lr{ 1 - \frac{s(1-s)}{4} \lr{\frac{p^3_I}{\Delta^1_1\lr{p}}}^2T^2 - \frac{s(1-s)}{4} \lr{\frac{p^2_J}{p^2_Ip^3_J}}^2 T^2 + \mathcal{O}\lr{T^3}}, \label{moreedgeslastintegralresult}
\end{align}
where we performed the asymptotic expansion for large arguments of the KCHFs and the follow-up Taylor expansion simultaneously from the third to the fourth line. We furthermore defined the minor with respect to the entry (1,1) of the $p$-matrix:
$$\Delta^1_1\lr{p} \coloneqq p^2_Ip^3_J-p^3_Ip^2_J.$$

The last step consists of combining this expression with the remaining parts of (\ref{moreedgesfirst}) as well as multiplying the last expansion's bracket with $\mathscr{T}$ and $\mathscr{T}'$ and keeping all terms up to $\sim T^2$:
\begin{align}
    \langle  \hat{q}^1_1(r) \rangle_{\cs} & \leq  \frac{T^{4s} |Z|^s }{ \prod_i\lr{1+K^i}} \sum_{IJ} \lrabs{\frac{\Delta^1_1\lr{p}}{T^2}}^s \lr{ 1 - \frac{s(1-s)}{4}\mathscr{P}_{IJ}T^2 + \mathcal{O}\lr{T^3} } \nonumber\\
    & = \frac{|Z|^s }{ \prod_i\lr{1+K^i}} \sum_{IJ} \lrabs{\Delta^1_1\lr{p}}^s T^{2s} \lr{ 1 - \frac{s(1-s)}{4}\mathscr{P}_{IJ}T^2 + \mathcal{O}\lr{T^3} }.
\end{align}
We thereby abbreviated the function of all $\left\{p^{2,3}_{I,J}\right\}$ that stems from adding up the terms $\sim T^2$ of the expansions' series by
\begin{equation}
\mathscr{P}_{IJ} \coloneqq \frac{\lr{p^2_Ip^3_Ip^3_J}^2+\lr{\Delta^1_1\lr{p}}^2\lr{\lr{p^2_I}^2+\lr{p^2_J}^2+\lr{p^3_J}^2}}{\lr{\Delta^1_1\lr{p} \, p^2_Ip^3_J}^2}.
\end{equation}

With that, we divide additionally by $t$ and have as equivalent for the Poisson bracket's lowest order contribution\footnote{Note that the sum over $IJ$ therein considers the terms of $\Delta^1_1\lr{p}$ that are labelled by the edges $e_I, e_J$.}
\begin{equation}
 \boxed{
\begin{aligned} \label{eq:resultImprovedEstimate}
    \frac{\langle  \hat{q}^1_1(r) \rangle_{\cs}}{t} \lesssim |Z|^{\rhalf} \sum_{IJ} \lrabs{\Delta^1_1\lr{p}}^{\rhalf} t^{\rhalf-1} + \mathcal{O}\lr{t^{\rhalf}},
\end{aligned}
}
\end{equation}
from which we can again deduce the general result:
\begin{equation}
 \label{eq:resultImprovedEstimateGeneral}
 \boxed{
\begin{aligned}
    \frac{\langle  \hat{q}^{i_0}_{I_0}(r) \rangle_{\cs}}{t} \lesssim |Z|^{\rhalf} \sum_{IJ} \lrabs{\Delta^{i_0}_{I_0}\lr{p}}^{\rhalf} t^{\rhalf-1} + \mathcal{O}\lr{t^{\rhalf}}.
\end{aligned}
}
\end{equation}
Having started with an expression $\sim p^{3s} = p^{\nicefrac{3r}{2}}$ and ending up with one that is $\sim \Delta^{i_0}_{I_0}\lr{p}^{\rhalf}$, i.e. $\sim p^r$, this clearly does not reflect the expected differentiation result $\sim p^{\nicefrac{3r}{2}-1}$ of the classical Poisson bracket. Also the order in $t$ is not as desired since even a negative exponent of $t$ is obtained due to the prefactors (note that we have to set $s=\frac{1}{4}$, i.e. $r=\frac{1}{2}$, for $\sqrt{\hat{V}}$).

Note that we cannot state a result for $p=0$ for this strategy, as the asymptotic expansion within \eqref{eq:LastNewApproachAsymptoticExpansion} is not feasible in this case --- just like during the analytical computation of subsection \ref{subsubsec:rigorous3edges}, whose procedure we used here for the determinant-like argument of \eqref{moreedgesfirst}. We would then need to apply estimates on this argument and thereby end up with a similar result as 
\eqref{eq:resultRigorousP=0General} in subsection \ref{subsubsec:rigorous3edges} or \eqref{eq:ResultALaBrunnemannP=0} in subsection \ref{subsec:LikeBrunnemannThiemann}, only with different numerical prefactor.

Comparing the result above, \eqref{eq:resultImprovedEstimate}, with the analytical calculation of the commutator's eigenvalue's basic building block before in section \ref{sec:wEstimates} as well as the other attempts for many edges estimates, we can deduce that due to
\begin{enumerate}
    \item losing the difference of fractional powers of the absolute values and
    \item reducing the initial exponent of the $n^i_I$ or $x^i_I$ respectively
\end{enumerate}
via the estimates, we get again a result with altered powers in $p$ and $t$. It seems that we can link the first point to the modified power in $t$ and the second one to the modification in the power in $p$. While the latter is probably indeed true, the reason for the altered power in $t$ might be more complex. Reducing the initial exponent of the $n^i_I$ via the estimates we carried out, we obviously simultaneously destroyed the final power in the $p^i_I$. For the issue of the modified power in $t$, however, both points might actually be of importance. First of all, we saw in the course of the analytical calculation of the basic building block that in order to obtain the expected order of $t$, it was important that the last expansion was not a Taylor series of the type $1 - f\lr{\left\{p^i_I\right\}} T^2 + \mathcal{O}\lr{T^3}$ of some KCHF, but in fact (still) that of a difference of two roots, where the argument of one was the other one's plus a correction term $\sim T^2$. As a consequence, the $t^2$ contribution became the lowest order that contributed because all lower order terms cancelled within that difference. It seems that it is precisely this additional $T^2$ term that we lack when working with estimates that have no such difference in fractional powers of shifted arguments.

However, reducing the exponent of the $n^i_I$ causes less integrations to be performed, too, and those also change the orders in $T$ as the respective $x^i_I$ gets replaced, in some sense, by $\nicefrac{p^i_I}{T}$ once the integration has been performed. Therefore, point 2 might also be a reason for obtaining a modified power of $t$.\newline

To recap the new estimates we presented so far: We started in subsection~\ref{subsec:LikeBrunnemannThiemann} with a modified version of the sequence of estimates that Brunnemann and Thiemann~\cite{Brunnemann2} used. This allowed us to retrain information about the initial, fractional exponent of $\qr$ and also to use less estimates: after casting the eigenvalue's fractional power of a sum into a sum of fractional powers, we could directly perform the integration without the need of additional estimates.

In subsection~\ref{subsec:towards}, we tried to keep a difference in two absolute values, leading to a difference in two KCHFs --- one with and one without the shift. From what we learnt, this is necessary in order to have the asymptotic series' zeroth order terms cancel each other and thereby yielding the expected powers in both $p$ and $t$. However, the estimate had to take an offset $\sim t^0$ along that ultimately lead to a negative power in $t$.

Then, the first part of subsection~\ref{subsec:newEstimates} tried to find an estimate that was not going so far as to get rid of the integration of a determinant, since we saw during the analytic computation of the basic building blocks in subsection~\ref{subsubsec:rigorous3edges} that the integration of plain determinants are indeed feasible. However, this estimate then resulted again in a term with altered exponents of the charges and a single KCHF, leading to a result with different powers in $p$ and $t$ than desired.

We can therefore say that we look for an estimate that results in an expression that is still a difference of two terms $\sim n^{\nicefrac{3r}{2}}$ and with one of them additionally reflecting the shift.
So one could guess that an estimate in the fashion of (\ref{estimatetowards}) without the $+2$ would be a favourable estimate:
\begin{align}
\lrabs{\sum_{\mathfrak{n}=1}^N \det\mathcal{N}_\mathfrak{n} + C}^\rhalf - \lrabs{\sum_{\mathfrak{n}=1}^N \det\tilde{\mathcal{N}}_\mathfrak{n} + C}^\rhalf \stackrel{?}{\leq} \left[ \lrabs{n^1_1}^\rhalf - \lrabs{n^1_1+1}^\rhalf \right] \lrabs{ \sum_{\mathfrak{n}=1}^N \Delta^1_1 \lr{\mathcal{N}_{\mathfrak{n}}} }^\rhalf, \label{ideamoreedges}
\end{align}
which could then be further estimated by the analytically integrable expression
$$ (\ref{ideamoreedges}) \leq \left[ \lrabs{n^1_1}^\rhalf - \lrabs{n^1_1+1}^\rhalf \right] \sum_{\mathfrak{n}=1}^N\lrabs{  \Delta^1_1 \lr{\mathcal{N}_{\mathfrak{n}}} }^\rhalf.$$
However, one can easily convince oneself that (\ref{ideamoreedges}) is not a correct inequality and in order to fix it, one has to reintroduce the additional $+2$-term of~\eqref{estimatetowards}.
Interestingly enough, if we were blindly using the above false inequality as estimate, we would obtain a reasonable result. All one has to do is replacing the standard Gaussian $x^1_1$-integration in (\ref{moreedgesfirst}), which lead to $\sqrt{\pi}$, by
\begin{align}
\sqrt{\pi} & \mapsto\infint x^1_1\, \frac{1}{T^s} \es{-\lr{x^1_1 - \frac{p^1_1-\pi\i N^1_1}{T}}^2} \lr{ \lrabs{x^1_1}^s - \lrabs{x^1_1 + T}^s } \approx \nonumber \\
& \quad\ \approx \frac{\Ghalf}{T^{2s}} \lr{1-\frac{s(1-s)}{4}\frac{T^2}{\lr{p^1_1}^s}} \lr{ -s \frac{\lrabs{p^1_1}^s}{p^1_1}T^2 + \frac{2(1-s)}{2}\frac{\lrabs{p^1_1}^s}{\lr{p^1_1}^2}T^4 + \mathcal{O}\lr{T^6}} . \label{fakefirstintegration}
\end{align}
This would then lead to the new lowest order contribution
\begin{align}
\frac{\langle  \hat{q}^1_1(r) \rangle_{\cs}}{t} & \stackrel{\bigtimes}{\leq}   -\rhalf |Z|^{\rhalf} \sum_{IJ} \frac{\lrabs{p^1_1 \Delta^1_1\lr{p}}^{\rhalf}}{p^1_1} + \mathcal{O}\lr{t}, \label{ideaFinalResult}
\end{align}
whose structure is now quite intuitive: Before we started integrating, \eqref{ideamoreedges} estimated the difference in the roots of the sums of determinants in such a way that we lost all contributions that were not linked to the charge $n^1_1$ that we chose to be the one the shift acts on. What was left was the difference of that charge's unshifted and shifted root multiplied by the sum of all the roots of minors of the determinant with respect to $n^1_1$ --- transforming into $\Delta^1_1\lr{p}$ during the integrations. In the end, the result looks like an estimate that singles out the contributions of the shift only, while indeed preserving the exponent in $p$ and $t$. Note that the upper bound in \eqref{ideaFinalResult} features an overall exponent of $\nicefrac{3r}{2}-1$ of the momenta, which is precisely the $p$-dependency one would expect due to the Poisson bracket's differentiation.

\subsection{Comparison with the Brunnemann and Thiemann estimates} \label{subsec:comparison}

We can tell two main differences between the work of Brunnemann and Thiemann~\cite{Brunnemann1,Brunnemann2} and the one at hand. First, when performing calculations without estimates, we either only considered the case of $N=1$, i.e. no multiples of $\qr$-operators, during the analytic computation of subsection~\ref{subsubsec:rigorous3edges}. Or, while the treatment of subsection~\ref{subsubsec:SahlmannThiemann} did allow for it, we then couldn't tackle the scenario of $p=0$.
Second, considering the limit $t \to 0$, Brunnemann and Thiemann get diverging results, while at least the non-estimative computations of this work did not. Note that \cite{Brunnemann1}'s respective results $(4.6) \sim t^{(\frac{3r}{2}-2)N}$ (general case) and $(4.7) \sim t^{(\frac{3r}{2}-1)N}$ ($p=0$ scenario) share featuring a negative exponent of $t$ for, e.g., the important value $r=\frac{1}{2}$.

This leads to the observation: If $p=0$, then $t\to 0$ diverging. If $t\to 0$, then $p=0$ problematic.

The reasons, as it seems, is that considering $p=0$ is not possible without estimates, which ultimately leads to a negative exponent of $t$. In order to understand the mechanism that causes this alteration of the exponent, we take a look at the essential dependencies. We start with
\begin{align}
    \langle \V^r \rangle_{\cs} \sim \lp{}^{3r} \sum \lrabs{n^3}^{\frac{r}{2}} 
\end{align}
as the expectation value of $\qr$'s first term.
As we see e.g. by the transition from \eqref{voleigenestimate} to \eqref{eq:ResultALaBrunnemann}, integrating $\sum \lrabs{n}^r$ against the coherent states leads to a $t$-dependency of $\sim t^{-r}$ after also performing the asymptotic expansion for large arguments of the KCHF. Hence, the above yields
\begin{align}
    \langle \V^r \rangle_{\cs} \sim \lp{}^{3r} \sum \lrabs{n^3}^{\frac{r}{2}} \mapsto \lp{}^{3r} t^{-\frac{3r}{2}} = a^{3r} \cdot t^0 ,
\end{align}
via $\nicefrac{\lp}{a}= T = \sqrt{t}$, as it should be. Now, estimates change this $t$-conversion, but only for the altered exponent of the charges and not also for the initial $\lp{}^{3r}$ that sets the correct dimension for the volume to the power of $r$. This means, if we were to apply an estimate on the expectation value before, we would obtain
\begin{align}
    \langle \V^r \rangle_{\cs} \sim \lp{}^{3r} \sum \lrabs{n^3}^{\frac{r}{2}} \leq \lp{}^{3r} \sum \lrabs{n}^{2r} \mapsto \lp{}^{3r} t^{-2r} = a^{3r} \cdot t^{-\frac{r}{2}} ,
\end{align}
since the volume prefactor $\lp{}^{3r}$ can not compensate the increasingly negative exponent of $t$, stemming from the integration over the charges. This feature manifests itself also when comparing the $t$-dependencies of the $p=0$ scenario: While~\cite{Brunnemann1}'s (4.7) results in an estimate $\sim t^{\frac{3r}{2}-2}$ for $N=1$ and after additionally dividing by $t$, we arrived at a result $\sim t^{r-1}$ in~\eqref{eq:ResultALaBrunnemannP=0}. So the estimate one applies really leaves a trace in the resulting order in $t$. However, note that Brunnemann and Thiemann used additional estimates for evaluating the integral which we could perform via KCHFs.

One apparent key quantity during \cite{Brunnemann2}'s evaluation of the aforementioned integrals is $A^i_I$, introduced between their (5.6) and (5.7), as its constant part leads to a non-vanishing expression when considering $p\to 0$ in the end. Hence, from first sight, it seems to be of great importance. Nevertheless, we often called the \textit{crucial} estimate of Brunnemann and Thiemann the one that allowed them to get rid off the exponent $r$ of the charges, i.e. all the work they did in the appendix that leads to their (C.39). We will now investigate the influence of $A^i_I$ by calculating the respective integrals right before and after its introduction --- i.e. \cite{Brunnemann2}'s (5.3) and (5.7) --- in order to motivate why the crucial estimate is still the one of replacing the exponent $r$. Adapting our notation to their (5.3), inserting the norm already and completing the square in the Gaussian, we obtain\footnote{Note that in \cite{Brunnemann2}'s (5.6) and the further equations of page 29, the Gaussian in $\nicefrac{\pi^2\lr{N^i_I}^2}{t}$ lacks the minus sign. However, that was clearly just a typo as it is back on page 30, but has of course to be considered when comparing that page's equations with the upcoming ones. Also, they did not include the length scale $a$ and hence kept $\lp$ and $t$ alongside.}
\begin{align}
    &\text{\cite[(5.3)]{Brunnemann2}} = \nonumber\\
    & =\frac{\lp{}^{3rN}\lr{9M}^N\lrabs{Z}^{\frac{rN}{2}}}{\sqrt{\pi}^{3M} t^N} \sumset{N} \e{-\sum_{Ii}\frac{\pi^2\lr{N^i_I}^2}{t}} \infint^{3M} x^i_I \es{-\sum_{Ii}\lr{x^i_I-\frac{p^i_I-\pi\i N^i_I}{T}}^2} \lr{\sum_{Ii}\lr{x^i_I}^2}^N . \label{eq:BrunnemannWayStart}
\end{align}
We may now proceed differently from here, knowing that we can integrate expressions like the one above by means of KCHFs. The following computations are along the path of the ones of subsection~\ref{subsubsec:Nq}, so we may keep it short here. Comparing~\eqref{eq:BrunnemannWayStart} with~\eqref{eq:startBrunnemannPath} regarding the respective integrals, we see that we can just set $r=2$ for the integration variable's exponent and arrive at the present scenario --- as $x^i_I \in \mathbbm{R} \Rightarrow \lrabs{x^i_I}^2 = \lr{x^i_I}^2$. Note that the prefactors' exponents $r$ are more delicate: The exponents of $\lp$ and $\lrabs{Z}$ stay the same --- they are not part of any estimates ---, while $(9M)^{\frac{rN}{2}}$ stems from the chain of estimates --- \eqref{voleigenestimate} in our case --- and we have to set $r=2$ for this part to resemble the estimate (C.39) of Brunnemann and Thiemann. Going through the computation then involves performing the same steps as in \ref{subsubsec:Nq} so we just state the final result here:
\begin{align}
    \text{\cite[(5.3)]{Brunnemann2}} \overset{t\to 0}{\leq} \lr{27M^2}^N \lrabs{Z}^{\tfrac{rN}{2}}  \lr{p_{\text{max}}}^{2N} t^{\lr{\tfrac{3r}{2}-2}N} + \mathcal{O}\lr{t^{\lr{\tfrac{3r}{2}-2}N+1}}. \label{eq:BeforeAGeneral}
\end{align}
We again introduced $p_{\text{max}} \coloneqq \max_{Ii} \{p^i_I\}$ as an upper bound for all the $p^i_I$ in order to be able to obtain this concise result. Likewise, we obtain analogously to the computations performed before
\begin{align}
    \text{\cite[(5.3)]{Brunnemann2}} \overset{p=0}{=} \lr{9M}^N \lrabs{Z}^{\frac{rN}{2}} t^{\lr{\frac{3r}{2}-1}N} \sum_{\{\nk\}} \frac{c_{\nk}}{\sqrt{\pi}^{3M}} \prod_{Ii} \Gamma\lr{\nik+\tfrac{1}{2}} \label{eq:BeforeAp=0}
\end{align}
for the scenario of $p=0$. Note again that the $\sqrt{\pi}^{3M}$ in the denominator gets reduced by all those $\Gamma(\nik+\tfrac{1}{2})$ for which $\nik=0$, leaving $\sqrt{\pi}^{\sharp(\nik)}$ with $\sharp(\nik)$ standing for the number of non-zero $\nik$ within the decomposition $\nk$, i.e. the number of integrals that resulted in $\Gamma(\nik+\tfrac{1}{2})$ instead of $\Gamma(\tfrac{1}{2})=\sqrt{\pi}$.

We have now stated the result of the integrals that Brunnemann and Thiemann obtained before the introduction of $A^i_I$. Consequently, we now compute the integrals Brunnemann and Thiemann obtained right after the first appearance of $A^i_I$:
\begin{align}
    \text{\cite[(5.7)]{Brunnemann2}} & = \frac{\lp{}^{3rN} \lr{9M}^N \lrabs{Z}^{\frac{rN}{2}} }{\sqrt{\pi}^{3M} t^N} \sumset{N} \e{-\sum_{Ii}\frac{\pi^2\lr{N^i_I}^2}{t}} \infint^{3M} X^i_I \es{-\sum_{Ii}\lr{X^i_I}^2} \cdot \nonumber\\
    & \quad \cdot \lr{ \sum_{Ii} \lr{ A^i_I \lr{X^i_I}^2 - \frac{\lrabs{p^i_I} A^i_I }{2T} + \frac{\pi^2 \lr{N^i_I}^2}{T^2} } }^N.
\end{align}
As the next step, we again expand the sum to the power of $N$. Knowing that this will result in a sum of products of the respective addends, we may already now narrow down the expression to containing $N^i_I=0 \, \forall\, i,I$ only due to the overall Gaussian in $\nicefrac{\pi^2\lr{N^i_I}^2}{t}$ in order to face shorter formulae.\footnote{Note that we have a pure Gaussian in $\lr{X^i_I}^2$ without further $N^i_I$ contributions.} Performing the expansion, we find
\begin{align}
    \lr{ \sum_{Ii} \lr{ A^i_I \lr{X^i_I}^2 - \frac{\lrabs{p^i_I} A^i_I }{2T} } }^N = \sum_{\{\nk\}} d_{\nk} \prod_{Ii} \lr{A^i_I \lr{X^i_I}^2}^{\nik} \lr{-\frac{\lrabs{p^i_I} A^i_I}{2T}}^{\mik},
\end{align}
where we now had to introduce a second set of non-negative integers, $\mik$, as the sum we started with consisted of two differently structured addends --- unlike the pure sum over the integration variables before. Hence, we now have $\sum_{Ii} \lr{\nik + \mik} = N$ and
\begin{align}
d_{\nk} = \frac{N!}{\prod_{Ii}\nik! \cdot \mik!} = \binom{N}{\mathscr{n}^1_{1,\mathscr{k}}, \mathscr{n}^2_{2,\mathscr{k}}, \ldots , \mathscr{n}^3_{M,\mathscr{k}},\mathscr{m}^1_{1,\mathscr{k}}, \mathscr{m}^2_{2,\mathscr{k}},\ldots,\mathscr{m}^3_{M,\mathscr{k}}} .
\end{align}
With this, we face again the integration of Gaussians against the integration variable to the power of an even and non-negative integer. But as we now have pure Gaussians without an offset $\sim p^i_I$ or $ \sim N^i_I$, the integration reads
\begin{align}
    \infint X^i_I \es{-\lr{X^i_I}^2} \lr{X^i_I}^{2\nik} = \Gamma\lr{\nik+\tfrac{1}{2}}
\end{align}
and we do not need to perform an additional asymptotic expansion. Hence, we ultimately obtain
\begin{align}
    \text{\cite[(5.7)]{Brunnemann2}} & = \frac{t^{(\frac{3r}{2}-1)N} \lr{9M}^N \lrabs{Z}^{\frac{rN}{2}} }{\sqrt{\pi}^{3M} } \sum_{\{\nk\}} d_{\nk} \prod_{Ii} \lr{A^i_I}^{\nik} \Gamma\lr{\nik+\tfrac{1}{2}} \sqrt{\pi}^{\mik} \lr{-\frac{\lrabs{p^i_I}A^i_I}{2T}}^{\mik}. \label{eq:AfterAGeneral}
\end{align}
This also allows for a straightforward handling of the $p=0$ case. For at least some of the $\nk$, we have $\mik = 0 \, \forall i,I$ which prohibits the overall result to be zero due to the $p^i_I$ in the last factor. We therefore obtain
\begin{align}
    \text{\cite[(5.7)]{Brunnemann2}} & \overset{p=0}{=} \frac{t^{\lr{\frac{3r}{2}-1}N} \lr{9M}^N \lrabs{Z}^{\frac{rN}{2}}}{\sqrt{\pi}^{3M}} \!\!\!\! \sum_{\overset{\{\nk\}\text{ s.t.}}{\mik =0 \,\forall\, i,I}} \!\!\!\!\!\! d_{\nk} \prod_{Ii} \Gamma\lr{\nik+\tfrac{1}{2}}. \label{eq:AfterAp=0}
\end{align}

This finishes the calculations and we can now compare the results we obtained when integrating the steps right before and after the introduction of $A^i_I$ in \cite{Brunnemann2}. First of all, we can say that both new results for $p=0$, \eqref{eq:BeforeAp=0} and \eqref{eq:AfterAp=0}, feature a $t$-dependence of $t^{\lr{\frac{3r}{2}-1}N}$ and this is also true for the final $p=0$ result of Brunnemann and Thiemann --- cf. \cite[(5.10) with $p=0$]{Brunnemann2}. However, for the general result of the computation without $A^i_I$, we got $\sim t^{\lr{\frac{3r}{2}-2}N}$ (cf. \eqref{eq:BeforeAGeneral}) and $\sim t^{\frac{3}{2}\lr{r-1}N}$ for the general result including $A^i_I$ (cf. \eqref{eq:AfterAGeneral} and use $\sum_{Ii}\mik = N$ to get the lowest possible order). We can therefore conclude that for investigating the realm of the initial singularity, $p=0$, the introduction of $A^i_I$ does not seem to be of importance --- at least concerning whether the result is non-zero and its $t$-dependence; combinatorical prefactors do of course change.

Furthermore, it is not clear to the authors how this introduction was achieved anyway since the binomial formula estimate they used from their (5.6) to their (5.7) seems not to be applicable at this stage with $X^j_J \in \mathbbm{C}$ and also the first line of this intermediate sequence of steps assumes the imaginary part of $X^j_J$ to be zero. Also, the integration over $X^i_I$ should be a complex one. However, it seems that their computations run through if one does not introduce the absolute values and argues already at the level of their (5.3) that one can exclusively consider the solution $N^i_I=0 \, \forall i,I$.

Lastly, we want to point out that the above procedure of factoring out the $N$-fold product of the sums that one obtains via the estimates is applicable to the calculations of section \ref{sec:wEstimates}, too, where we only considered $N=1$. Generalising to arbitrary $N$ is possible as just described but the computations will become more elaborate, while we wouldn't expect significant changes.

\section{Conclusion and outlook}
\label{sec:conclusion}
In this article, we extended a method to analytically compute semiclassical expectation values based on Kummer's confluent hypergeometric functions from quantum mechanics or quantum mechanics on a circle respectively, introduced in our former work \cite{paper1}, to the case of U$(1)^3$ coherent states and the dynamical operators relevant in loop quantum gravity (LQG) in order to discuss a new procedure for computing semiclassical expectation values in LQG in addition to the already existing ones. In particular, we investigated the  question of singularity avoidance in LQG and compared our method to results by Brunnemann and Thiemann \cite{Brunnemann1,Brunnemann2}.  The utilisation of Kummer's confluent hypergeometric function allows to analytically evaluate integrals involving products of roots and Gaussians. To begin with, we first reviewed this procedure for a fractional power $r>-\frac{1}{2}$ of the momentum operator in quantum mechanics. The article is divided into two main parts, where the first one in section~\ref{sec:woEstimates} involves semiclassical computations that can be performed without estimates, whereas for the second part in  section~\ref{sec:wEstimates} the calculations do rely on estimates. As a first scenario for LQG  in subsection \ref{subsubsec:rigorous3edges}, similar to \cite{Towards1,Towards2}, we considered graphs of cubic topology and aimed at computing semiclassical expectation values of the crucial dynamical operators $\qr$, products of which include for instance the analogue of the inverse scale factor in LQC. Moreover, these operators are also involved in more complicated dynamical operators such as matter Hamiltonians or the Hamiltonian constraint in LQG. We showed that for cubic graphs and linear power of $\qr$ our technique allows to compute the semiclassical expectation value of $\qr$ analytically without using estimates, as opposed to \cite{Brunnemann1,Brunnemann2}, thereby extending results from the literature in the sense that the final outcomes still contain a stronger fingerprint of the initially involved fractional power. The final result for all semiclassical expectation values considered here can be written as a power series in the classicality parameter $t$ and one expects to get the classical result in the limit where $t$ is sent to zero. In the case of a graph of cubic topology and for non-vanishing classical triad labels of the complexifier coherent states, we were able to show that we obtain the correct classical limit in zeroth order of the classicality parameter without using estimates and, moreover, could perform the continuum limit in which the regulator is removed as well. In the latter step, we could confirm in subsection~\ref{subsec:SemClassCubic} that the regularisation constant of the volume operator  for the $U(1)^3$ case  needs to be $\frac{1}{48}$ in order to obtain the correct classical limit as was already pointed out in \cite{Towards2,VolGiesel,VolGiesel2} and how this is related to the different result found in \cite{Yang:2019xms} where the ${\rm SU(2)}$ case is considered using the graphical calculus will be discussed elsewhere in \cite{GTtoAppear}. To analyse the singularity avoidance, we need to investigate the case in which the triad label of the coherent states vanishes, that is $p\to 0$. Then, the asymptotic expansion of Kummer's functions cannot be used in a similar manner as before; therefore, the computation of the semiclassical expectation value becomes more involved. As a consequence, we needed to introduce estimates in this specific case, which are however different to the ones used in \cite{Brunnemann1,Brunnemann2}. In accordance with their result, we also obtain a finite upper bound for the semiclassical value of $\qr$ for a graph of cubic topology and obtain singularity avoidance. However,  the way how the fractional power enters into the final result differs and, as discussed in subsection~\ref{sec:CosmoSing}, therefore also for which values of the fractional power a finite expression in the $t\to 0$ limit exists. In our results, this happens if $\qr$ involves the volume operator linearly in the commutator, whereas for \cite{Brunnemann1,Brunnemann2} this happens for a fractional power of the volume operator of $\frac{4}{3}$, showing, as rather expected, that such properties do depend on the kind of estimates used in the computations. The results discussed so far are restricted to linear powers of the operator $\qr$ and cubic graphs. The next, more involved case was considered in subsection \ref{subsubsec:SahlmannThiemann}, where we recapitulated the procedure introduced by Sahlmann and Thiemann in \cite{Towards1,Towards2}, which there was applied to cubic graphs. We extended this method to more general graphs and obtained, again up to some expected rescaling caused by the regularisation constant, the expected classical expression in the zeroth order of the classicality parameter. The case of $p=0$, however, was not treatable with this procedure because this method requires that the matrix built from the classical triad labels of the complexifier coherent states is invertible, which is no longer given in the limit  $p\to 0$. 

In the second part of the article, in section~\ref{sec:wEstimates}, we analysed whether our method based on Kummer's confluent hypergeometric functions can be used to  improve the results for the upper bound regarding the singularity avoidance, that is the case $p=0$. As discussed in the applications in section~\ref{sec:wEstimates}, introducing estimates usually has the consequence that one estimates the original fractional powers by different powers in the classical label $p$, the classicality parameter $t$, or both.  Compared to the estimates used in \cite{Brunnemann1,Brunnemann2}, in some steps of our work we could keep fractional powers and did not need to estimate those by integer powers. Therefore, we aimed at trying to understand in more detail how the aforementioned modification of the order in $t$ and $p$ respectively arises when one uses estimates. As a first step, we carried out a computation that followed the path of \cite{Brunnemann1,Brunnemann2}, where it was shown i.a. that there exists an upper bound for the semiclassical expectation value of the operator-analogue of the inverse scale factor even when approaching the initial singularity via $p=0$. We modified the approach of \cite{Brunnemann1,Brunnemann2} in two ways: First, we did not need to get rid of the non-integer exponent of the charges as we could rely on the KCHF procedure. And secondly, the integration by means of KCHFs also allowed us to refrain from using additional estimates in order to evaluate the resulting integral. We discussed the case  of one single $\qr$ and  $N$-multiple ones separately in order to better demonstrate the differences and similarities of the two methods. Our result then features the same property when we want to consider $t\to 0$ additionally to $p=0$ --- or vice versa ---, namely that the expression is not well defined if both limits are taken. We were able to find two aspects of estimates that cause this issue: One is changing the initial exponent of the charges via an estimate, causing ultimately a modified exponent of $t$ as well. The other one is to apply estimates such that the initial difference due to the commutator is replaced by only one expression, whose integration then gives rise to one single KCHF and its power series. We saw during the analytical computation for graphs of cubic topology (cf. \ref{subsubsec:rigorous3edges} and the last steps of Appendix \ref{Appendix9integrals}) that in the end, the zeroth order of the commutator's two KCHFs cancel each other --- and this is of course not possible anymore when having only one function in an estimate. Maybe one finds an estimate that changes the overall exponent in $t$, i.e. as a prefactor of the series, in such a way that this series' lowest order contribution turns out to carry the correct order in $t$, but the authors are not too positive about this possibility; and for such complicated operators this might also not be expected. Having these two reasons in mind, we continued to test new estimates that respect these rules of having a difference in KCHFs and not altering the U(1)-charges' exponents involved in the eigenvalue of the volume operator. However, for the analysed modified estimates in subsection~\ref{subsec:newEstimates} there was always some issue occurring such that we ultimately had to break with one of the conditions --- but our analysis gives a more detailed picture of where this exactly comes from in the application of the estimate. This can help to perform a future analysis on improved estimates in a more focused manner. For instance,  one could use the ansatz for a new estimate that we stated at the end of subsection \ref{subsec:newEstimates} --- where we showed that an intuitive, yet inapplicable estimate would yield the expected classical result --- in order to reverse engineer a similar and indeed applicable estimate. 

Another follow up question is whether there exists a link between the approaches via KCHFs and the Sahlmann and Thiemann one based on a Taylor expansion. Appendix \ref{appendix:ComparisonProcedures} shows such a connection for the U(1) case, where one can associate the asymptotic expansion of the KCHF with the power series expansion of \cite{Towards2}. For higher-dimensional scenarios, however, this is not deducible in a similar straightforward manner as the KCHF way then means to successively perform the interwoven integrals. In contrast to this, the procedure Sahlmann and Thiemann used in \cite{Towards2} allows for tackling all integrations simultaneously after a disentanglement via a power series expansion.

A further and interesting generalisation of the methods presented in our work would be to extend the KCHF procedure for computing semiclassical expectation values and matrix elements to the case of SU(2) complexifier coherent states and understand how the techniques and results are related to the ones that one obtains via the semiclassical perturbation theory introduced in \cite{AQG3}.  Besides that, there exists also work on matrix valued KCHFs \cite[and references therein]{ToMatrixKCHF, LaplaceMatrixKCHF} that one might use in order to evaluate the integrals of the determinants. The authors looked into this, but could not find a way to handle the calculations properly so far. However, the authors are also aware that this is still an active field of research and with investing more time, there might be ways to tackle it.

\tocless{\section*{Acknowledgements}}
The work of D.W. was supported by a stipend provided from the FAU Erlangen-Nürnberg. Further, D.W. thanks the Studienstiftung des deutschen Volkes (German Academic Scholarship Foundation) for financial support at an earlier stage of the project.

\newpage

\appendix

\section{The Poisson (re-)summation formula} \label{sec:Poisson}

    In the course of our calculations, the Poisson (re-)summation formula will frequently be used when calculating expectation values and norms: 
    \begin{equation}
    \infsum{n}f(nT)=\frac{2\pi}{T}\infsum{N}\tilde{f}\lr{\frac{2\pi N}{T}},    \label{PoissonResum}
    \end{equation}
    with $\tilde{f}(k) = \infint x \, \e{-\i kx}f(x)$ being the Fourier transform of $f(x)$. It is applicable to functions $f\in L_1\lr{\mathbbm{R},\d x}$ for which
    \begin{equation}
        F\lr{x} \coloneqq \sum_{n=-\infty}^\infty f\lr{x+nT} \ , T>0,
    \end{equation}
    is absolutely and uniformly convergent for $x\in [0,T]$.

\section{The 9 integrations} \label{Appendix9integrals}

Here, we show the calculations behind section~\ref{subsubsec:rigorous3edges} and perform all 9 integrations, including intermediate expansions, in more detail --- and with the first two substitutions being implemented right from the beginning. As starting point, we want to calculate

\begin{align}
    & \frac{2}{t} \langle \hat{q}^1_1(r) \rangle_{\cs} = \\
    & = \frac{2}{t} \frac{\lp{}^{6r}\lr{\frac{2\pi\sqrt{2}}{T}}^{\!9}}{a^{6r}\mcsnorm \, T^{3r}}  \isumset{N} \!\! \es{2\sum\limits_{i}\lr{\frac{p_i-\pi\i N_i}{T}}^2} \infint^{9}x_i\es{-2\sum\limits_{i}\lr{x_i-\frac{p_i-\pi\i N_i}{T}}^2} \lr{|\det{X}|^r-|\det{\widetilde{X}}|^r} , \label{eq:appendixStart}
\end{align}
where
\begin{align}
   X \coloneqq \begin{pmatrix}
x^1_1 & x^2_1 & x^3_1 \\
x^1_2 & x^2_2 & x^3_2 \\
x^1_3 & x^2_3 & x^3_3
\end{pmatrix}
\eqqcolon \begin{pmatrix}
x_1 & x_2 & x_3 \\
x_4 & x_5 & x_6 \\
x_7 & x_8 & x_9
\end{pmatrix} \qquad\text{and}\qquad \widetilde{X}\coloneqq \begin{pmatrix}
x_1+\frac{T}{2} & x_2 & x_3 \\
x_4 & x_5 & x_6 \\
x_7 & x_8 & x_9
\end{pmatrix},
\end{align}
i.e. we assigned the shift to the $x^1_1$-component of the first edge and the first U(1)-copy. As in the calculations before, we defined $x_i \coloneqq Tn_i $, with $T^2 \coloneqq t$. In contrast to the calculations before, however, we now use $r$ as the exponent of the determinants' absolute values. This is only due to clearer formulae. Accordingly, the exponents of $\lp{}^{3r}$ and $a^{3r}$ changed to $\lp{}^{6r}$ and $a^{6r}$, respectively.

An important note is that all the above $x_i$, $p_i$ and $N_i$ represent the cubic graph's differences in the respective quantities of the out- and ingoing edges, cf. subsection~\ref{sec:CubicGraphSetup} and the paragraph following (4.7) in~\cite{Towards2}. The integrals of the related sums reduced to standard Gaussian ones that we considered solved and being compensated by the part $\norm{\Psi}^2_+$ of the state's norm
\begin{align}
    \csnorm &= \underbrace{\lr{\frac{2\pi\sqrt{\pi}}{T}}^9 \es{2\sum\limits_i\lr{\frac{p_i^+}{T}}^2}}_{\pcsnorm} \underbrace{\prod_i\lr{1+K^{(i)}_{t}} \cdot \lr{\frac{2\pi\sqrt{\pi}}{T}}^9 \es{2\sum\limits_i\lr{\frac{p_i}{T}}^2} \prod_i\lr{1+K^{(i)}_{t}}}_{\mcsnorm} \\
    & =\pcsnorm \, \mcsnorm  .
\end{align}
Note that we put both expansion remainders $\prod_i\lr{1+K^{(i)}_{t}}$ into the definition of $\mcsnorm$ as this allows us to have a more concise description of our formulae, i.e. no such term in the prefactor's denominator in~\eqref{eq:appendixStart}.

The expression above is clearly not integrable in this form (at least not by means of KCHFs), but we can perform a substitution
\begin{align}
    x'_1 &\coloneqq \det{X} ,\\
    x'_{2,\ldots,9} &\coloneqq x_{2,\ldots,9},
\end{align}
with
\begin{align}
    \det{\lr{\frac{\d x'}{\d x}}} = \det \begin{pmatrix}
x_5x_9-x_6x_8 & x_6x_7-x_4x_9 & x_4x_8-x_5x_7 & \dots \\
0 & 1 & 0 & \dots \\
0 & 0 & 1 & \dots \\
\vdots & \vdots & \vdots & \ddots
\end{pmatrix} = x_5x_9-x_6x_8
\end{align}
and arrive at integrating the Gaussians against $$\frac{|x'_1|^r}{|x'_5x'_9-x'_6x'_8|}.$$
We realise this will hinder the integrations with respect to one of the $x'_i$ of the denominator and straightforwardly circumvent this obstacle by another substitution:
\begin{align}
    x''_5 &\coloneqq x'_5x'_9-x'_6x'_8 ,\\
    x''_{1,2,3,4,6,7,8,9} &\coloneqq x'_{1,2,3,4,6,7,8,9},\\
    \det{\lr{\frac{\d x''}{\d x'}}} &\,= x'_9 .
\end{align}
Therefore, we need to integrate the Gaussians against $$\frac{|x''_1|^r}{|x''_5||x''_9|}.$$

Starting with the $x''_1$-integration, we get
\begin{align}
    \infint x''_1 \lrabs{x''_1}^r \e{-2\lr{\nicefrac{x''_1}{x''_5}+a}^2} = \sqrt{2}^{-1-r} \e{-2a^2}\lrabs{x''_5}^{1+r} \Gsum \kchf{\opr,\frac{1}{2},2a^2},
\end{align}
where we abbreviated
\begin{equation} \label{integrationsa}
a \coloneqq \frac{x''_2x''_4x''_9+\frac{x''_3x''_7x''_6x''_8}{x''_9}-x''_2x''_6x''_7-x''_3x''_4x''_8}{x''_5} + \frac{x''_3x''_7}{x''_9}-\frac{p_1-\pi\i N_1}{T}
\end{equation}
stemming from inserting $x'_1=x'_1(x_{1},\ldots,x_{9})$ and then $x''_1=x''_1(x'_{1},\ldots,x'_{9})$.

For the $x''_5$-integration, we also need to consider the factor $|x''_5|^{1+r}$, of course, as well as the above KCHF that depends via $a$ on $x''_5$ just like the Gaussian in $-a^2$:
\begin{align}
    \infint x''_5 \es{-2\lr{\nicefrac{x''_5}{x''_9}+b}^2}\e{-2a^2}|x''_5|^r\kchf{\opr,\frac{1}{2},2a^2}, \nonumber
\end{align}
with
\begin{equation}
b\coloneqq \frac{x''_6x''_8}{x''_9}-\frac{p_5-\pi\i N_5}{T}.    
\end{equation}
Since we cannot integrate Gaussians against KCHFs, we need to find a away to express the latter differently. Looking at the form of $a$, we see $a\sim\frac{1}{T}$ and we can therefore use the asymptotic expansion for large arguments of KCHFs. As seen in (\ref{expansion}), the expansion yields two different series with one including the factor $\e{a^2}$. The integrand above, however, contains the factor $\e{-a^2}$, which damps the contribution to zero. Therefore, we can neglect the expansion's series without the $\e{a^2}$ prefactor and get
\begin{align}
    \e{-2a^2}\kchf{\opr,\frac{1}{2},2a^2} \approx \frac{\Ghalf}{\Gsum}\lr{2a^2}^\rhalf \lr{1-\rrfour\frac{a^{-2}}{2}+\mathcal{O}\lr{\lr{a^2}^{-2}}}.
\end{align}
With that, the $x''_5$-integration now reads
\begin{align*}
    \infint x''_5 \es{-2\lr{\nicefrac{x''_5}{x''_9}+b}^2} |x''_5|^r \lr{2\lr{\frac{c}{x''_5}+d}^2}^{\!\rhalf} \left[ 1- \frac{r\lr{1-r}}{8} \lr{\frac{c}{x''_5}+d}^{-2} \right],
\end{align*}
where we rewrote $ a $ as $ \nicefrac{c}{x''_5}+d $, i.e. $c$ is the numerator of the fraction with $x''_5$ as denominator in (\ref{integrationsa}) and $d$ are the remaining two fractions independent of $x''_5$.

In the hope of simplifications like the cancelling of the lowest order, we now address the shifted contribution. There, the initial $x_1$-integration reads
\begin{align}
    \infint x_1 |\det \widetilde{X}|^r \es{-2\lr{x_1 - \frac{p_1-\pi\i N_1}{T}}^2} = \infint \tilde{x}_1 |\det X|^r \es{-2\lr{\tilde{x}_1 - \frac{p_1+\frac{T^2}{2}-\pi\i N_1}{T}}^2},
\end{align}
where we substituted $\tilde{x}_1 \coloneqq x_1 + \frac{T}{2}$ and understand $X$ in this case as the one of the unshifted contribution but with $\tilde{x}_1$ as the first entry. Hence, we can proceed just as before and only need to consider $p_1 \mapsto p_1+\frac{T^2}{2}$ in the Gaussian with respect to $\tilde{x}_1$. Note that we must not replace $p_1$ within the prefactor $\ex{\nicefrac{(p_1-\pi\i N_1)^2}{T^2}}$ that stems from completing the square.

We can then proceed with the same steps as before, first of all applying the two substitutions --- only starting with $\tilde{x}_1$ instead of $x_1$:
\begin{align}
    \infint x''_1 \lrabs{x''_1}^r \e{-2\lr{\nicefrac{x''_1}{x''_5}+\tilde{a}}^2} = \sqrt{2}^{-1-r} \e{-2\tilde{a}^2}\lrabs{x''_5}^{1+r} \Gsum \kchf{\opr,\frac{1}{2},2\tilde{a}^2},
\end{align}
where we consequently denoted $\tilde{a} \coloneqq \left. a \right|_{p_1\mapsto p_1+\frac{T^2}{2}} $. Applying the asymptotic expansion for large arguments, we get
\begin{align}
    \infint x''_1 \lrabs{x''_1}^r \e{-2\lr{\nicefrac{x''_1}{x''_5}+\tilde{a}}^2} \approx \frac{1}{\sqrt{2}} \lrabs{x''_5}^{1+r} \Ghalf\lr{\tilde{a}^2}^\rhalf \lr{1-\rrfour\frac{\tilde{a}^{-2}}{2}+\mathcal{O}\lr{\lr{\tilde{a}^2}^{-2}}}.
\end{align}
Hence, the $x''_5$-integration within the shifted contribution reads
\begin{align*}
    \infint x''_5 \es{-2\lr{\frac{x''_5}{x''_9}+b}^2} \lrabs{x''_5}^r \lr{\lr{\frac{c}{x''_5}+d-\frac{T}{2}}^2}^\rhalf \left[ 1- \frac{r\lr{1-r}}{8} \lr{\frac{c}{x''_5}+d-\frac{T}{2}}^{-2} \right].
\end{align*}
We notice that due to the shift $-\frac{T}{2}$ being present outside the square brackets' series, the lowest orders of the $x''_5$-integration of the unshifted and shifted contribution do not cancel --- at least not at this early stage.

Since the $x''_5$-integrations are not feasible in this form, we need to apply a further expansion: Noticing that $d\sim\frac{1}{T}$, we perform a Taylor expansion of the square brackets for $T\to 0$. This yields for both cases
\begin{align}
    \mathscr{S} \coloneqq 1-\rreight\frac{T^2}{p_1{}^2}+\mathcal{O}(T^3),
\end{align}
which has no $x''_5$-dependence anymore and therefore enables to proceed --- for the missing $N_1$, see the next paragraph. Combining the two contributions, we now have
\begin{align}
\frac{2}{t}&\langle \hat{q}^1_1(r) \rangle_{\cs} = \nonumber \\
  & = \frac{2}{t} \frac{T^{3r}\lr{\frac{2\pi\sqrt{2}}{T}}^{\!9}\Ghalf\mathscr{S}}{\mcsnorm \sqrt{2}}  \isumset{N} \es{2\sum\limits_{i}\lr{\frac{p_i-\pi\i N_i}{T}}^2} \infint^{7}x''_{i\setminus 1,5}\es{-2\sum\limits_{\mathclap{i\setminus 1,5}}\ \lr{x''_i-\frac{p_i-\pi\i N_i}{T}}^2} \frac{1}{|x''_9|} \cdot \nonumber\\
  & \quad \cdot \infint x''_5 \es{-2\lr{\frac{x''_5}{x''_9}+b}^2} \left[ \left| c+dx''_5 \right|^r - \left| c+(d-\frac{T}{2})x''_5 \right|^r \right] \\
  & = \frac{2}{t} \frac{T^{3r}\lr{\frac{2\pi\sqrt{2}}{T}}^{\!9}\Ghalf\mathscr{S}}{\mcsnorm \sqrt{2}}  \isumset{N} \es{2\sum\limits_{i}\lr{\frac{p_i-\pi\i N_i}{T}}^2} \infint^{7}x''_{i\setminus 1,5}\es{-2\sum\limits_{\mathclap{i\setminus 1,5}}\ \lr{x''_i-\frac{p_i-\pi\i N_i}{T}}^2} \frac{1}{|x''_9|} \cdot \nonumber\\
   & \quad \cdot \sqrt{2}^{-1-r}\Gsum \left[ \frac{1}{|d|} \left| d x''_9 \right|^{1+r} \e{-2\lr{b-\frac{c}{dx''_9}}^2} \kchf{\opr,\frac{1}{2},2\lr{b-\frac{c}{dx''_9}}^2} - \right. \nonumber\\
   & \qquad\quad\ \left.-\frac{1}{|d-\frac{T}{2}|} \left| (d-T) x''_9 \right|^{1+r} \e{-2\lr{b-\frac{c}{(d-\frac{T}{2})x''_9}}^2} \kchf{\opr,\frac{1}{2},2\lr{b-\frac{c}{(d-\frac{T}{2})x''_9}}^2} \right]. \label{resultx5}
\end{align}
In order to cast the expressions $\lr{\lr{\nicefrac{c}{x''_5}+d}^2}^\rhalf$ and $\lr{\lr{\nicefrac{c}{x''_5}+d-\frac{T}{2}}^2}^\rhalf$ into absolute values, we set $N_1=0$. This causes the new $d\in\mathbbm{R}$ (we kept the same $d$ for reasons of brevity and the limited alphabet), and hence the whole argument being real, which then allows to replace $\lr{\lr{\ldots}^2}^\rhalf \mapsto |\ldots|^r$. Setting $N_1=0$ was possible due to the overall prefactor $\exp{\lr{-\nicefrac{(\pi^2N_1{}^2)}{T^2}}}$ --- so, in other words, we only kept the $N_1=0$ solution as all other contributions are exponentially damped. For not making the notation even more elongate, we keep the part $\isumset{N} \es{2\sum_{i}\lr{\frac{p_i-\pi\i N_i}{T}}^2}$ as it is, regardless of successively considering $N_i=0$ only.

The structure of the result above suggests to proceed with the $x''_9$-integration. Again, we see that we would need to integrate over KCHFs and therefore apply the expansion for large arguments --- allowed due to $b\sim\frac{1}{T}$ --- to circumvent this. We consider again only the expansion's part that cancels the damping exponential function prefactor. Noticing that the situation is basically as the one before --- just with $a$, $\tilde{a}$ being replaced by $b-\frac{c}{dx''_9}$, $b-\frac{c}{(d-\frac{T}{2})x''_9}$, respectively --- we get the same expansion, too, with according replacements. Also the subsequent Taylor expansion of the expansions' series results in a similar expression:
\begin{equation}
    \mathscr{S}' \coloneqq 1 - \rreight \frac{T^2}{p_5{}^2} + \mathcal{O}(T^3),
\end{equation}
where we used in advance that we only consider the $N_5=0$ contribution, as motivated within the next steps.

Having said this, we can in general say that the procedure will (mostly) stay the same for the remaining integrations: We integrate Gaussians against absolute values, obtain KCHFs and apply the asymptotic expansion for large arguments where we only keep the series that is not damped by a Gaussian prefactor $\exp{\lr{-\nicefrac{p_i{}^2}{T^2}}}$. Then, we perform a Taylor expansion on the asymptotic expansion's series. For most cases, this will lead to $x''_i$-independent prefactors $\mathscr{S}'{}^{\ldots}{}' $, enabling to apply the same procedure again after setting the corresponding $N_i = 0$ --- allowed due to the Gaussian prefactor $\exp{\lr{-\nicefrac{(\pi^2N_i{}^2)}{T^2}}}$. In some cases, an $x''_i$-independent prefactor will not be obtained and we will need to integrate a Gaussian against the absolute value and square of an $x''_i$, but luckily that's feasible.

Applying these steps to (\ref{resultx5}), we get
\begin{align}
    & \mathrm{(\ref{resultx5})} \approx \frac{2}{t} \frac{T^{3r}\lr{\frac{2\pi\sqrt{2}}{T}}^{\!9}\Ghalf^2\mathscr{S}\mathscr{S}'}{\mcsnorm \sqrt{2}^2} \! \isumset{N} \!\! \es{2\sum\limits_{i}\lr{\frac{p_i-\pi\i N_i}{T}}^2} \! \infint^{6}x''_{i\setminus 1,5,9}\es{-2\sum\limits_{\mathclap{i\setminus 1,5,9}}\ \: \lr{x''_i-\frac{p_i-\pi\i N_i}{T}}^2} \cdot \nonumber \\
    &\cdot \infint x''_9 \es{-2\lr{x''_9-\frac{p_9-\pi\i N_9}{T}}^2} \bigg[ \left| x''_9\lr{x''_2x''_4-\frac{p_1p_5}{T^2}} + \frac{p_5}{T}x''_3x''_7 + \frac{p_1}{T}x''_6x''_8 - x''_2x''_6x''_7-x''_3x''_4x''_8 \right|^r - \nonumber\\
    &\qquad \left. - \left| x''_9\lr{x''_2x''_4-\frac{(p_1+\frac{T^2}{2})p_5}{T^2}} + \frac{p_5}{T}x''_3x''_7 + \frac{p_1+\frac{T^2}{2}}{T}x''_6x''_8 - x''_2x''_6x''_7-x''_3x''_4x''_8 \right|^r \right] \label{pre_resultx9} \\
    & = \frac{2}{t} \frac{T^{3r}\lr{\frac{2\pi\sqrt{2}}{T}}^{\!9}\Ghalf^2\mathscr{S}\mathscr{S}'}{\mcsnorm \sqrt{2}^2}  \isumset{N} \es{2\sum\limits_{i}\lr{\frac{p_i-\pi\i N_i}{T}}^2} \infint^{6}x''_{i\setminus 1,5,9}\es{-2\sum\limits_{\mathclap{i\setminus 1,5,9}}\ \: \lr{x''_i-\frac{p_i-\pi\i N_i}{T}}^2} \cdot \nonumber\\
   & \quad \cdot\sqrt{2}^{-1-r}\Gsum \left[ \left| x''_2x''_4-\frac{p_1p_5}{T^2} \right|^r \es{-2\sigma^2}\kchf{\opr,\frac{1}{2},2\sigma^2} - \right. \nonumber\\
   & \qquad \qquad \qquad \qquad\ - \left.\left| x''_2x''_4-\frac{\lr{p_1+\frac{T^2}{2}}}{T^2} \right|^r \es{-2\tilde{\sigma}^2}\kchf{\opr,\frac{1}{2},2\tilde{\sigma}^2} \right], \label{resultx9}
\end{align}
where we abbreviated
\begin{align}
     \sigma & \coloneqq \frac{- \frac{p_5}{T}x''_3x''_7 - \frac{p_1}{T}x''_6x''_8 + x''_2x''_6x''_7 + x''_3x''_4x''_8}{x''_2x''_4-\frac{p_1p_5}{T^2}} - \frac{p_9-\pi\i N_9}{T},\\
     \tilde{\sigma} & \coloneqq \frac{- \frac{p_5}{T}x''_3x''_7 - \frac{p_1+\nicefrac{T^2}{2}}{T}x''_6x''_8 + x''_2x''_6x''_7 + x''_3x''_4x''_8}{x''_2x''_4-\frac{\lr{p_1+\nicefrac{T^2}{2}} p_5}{T^2}} - \frac{p_9-\pi\i N_9}{T},
\end{align}
which are just the peaks of the Gaussians after substituting $x''_9$ such that it is alone the absolute values' argument. In \eqref{pre_resultx9} and \eqref{resultx9} above, we can observe how the denominator's $\sqrt{2}^n$ builds up to compensate the respective factor $\sqrt{2}^9$ in the numerator. Their origin is the prefactor of 2 in the Gaussians, which stems from the initial substitution right after \eqref{eq:SahlmannThiemannLambda}. The overall prefactor in the numerator stems from performing this substitution, while the denominator's one builds up via the successive integrations, as \eqref{resultx9} illustrates.

With both $\sigma, \tilde{\sigma} \sim \frac{1}{T}$, we can perform the asymptotic expansion and get as the Taylor expansion of the non-damped series
\begin{equation}
    \mathscr{S}'' \coloneqq 1 - \rreight \frac{T^2}{p_9{}^2} + \mathcal{O}(T^3),
\end{equation}{}
where also only the $N_9=0$ contribution was considered. This leads to
\begin{align}
    & \mathrm{(\ref{resultx9})} \approx \nonumber\\
    & \approx \frac{2}{t} \frac{T^{3r}\lr{\frac{2\pi\sqrt{2}}{T}}^{\!9}\Ghalf^3\mathscr{S}\mathscr{S}'\mathscr{S}''}{\mcsnorm \sqrt{2}^3}   \isumset{N}  \es{2\sum\limits_{i}\lr{\frac{p_i-\pi\i N_i}{T}}^2} \infint^{5}x''_{i\setminus 1,3,5,9}\es{-2\sum\limits_{\mathclap{\,i\setminus 1,3,5,9}}\quad\! \lr{x''_i-\frac{p_i-\pi\i N_i}{T}}^2} \cdot \nonumber\\
    & \quad \cdot \infint x''_3 \es{-2\lr{x''_3-\frac{p_3-\pi\i N_3}{T}}^2} \left[ \left| x''_3\lr{x''_4x''_8 - \frac{p_5}{T}x''_7} +\tau \right|^r - \left| x''_3\lr{x''_4x''_8 - \frac{p_5}{T}x''_7} +\tilde{\tau} \right|^r \right]\nonumber\\
    & = \frac{2}{t} \frac{T^{3r}\lr{\frac{2\pi\sqrt{2}}{T}}^{\!9}\Ghalf^3\mathscr{S}\mathscr{S}'\mathscr{S}''}{\mcsnorm \sqrt{2}^3}  \isumset{N} \es{2\sum\limits_{i}\lr{\frac{p_i-\pi\i N_i}{T}}^2} \infint^{5}x''_{i\setminus 1,3,5,9}\es{-2\sum\limits_{\mathclap{i\setminus 1,3,5,9}}\quad \! \lr{x''_i-\frac{p_i-\pi\i N_i}{T}}^2} \cdot \nonumber\\
    & \cdot \frac{\Gsum}{\sqrt{2}^{1+r}} \left| x''_4x''_8 - \tfrac{p_5}{T}x''_7 \right|^r \! \left[ \e{-2\lr{\frac{\tau}{x''_4x''_8-\frac{p_5}{T}x''_7}-\frac{p_3-\pi\i N_3}{T}}^{\!2}} \!\! \kchf{\!\tfrac{1+r}{2},\tfrac{1}{2},2\lr{\tfrac{\tau}{x''_4x''_8-\tfrac{p_5}{T}x''_7}-\tfrac{p_3-\pi\i N_3}{T}}^{\!\!2}} - \right. \nonumber\\
    & \qquad\qquad\qquad\qquad - \left. \e{-2\lr{\frac{\tilde{\tau}}{x''_4x''_8-\frac{p_5}{T}x''_7}-\frac{p_3-\pi\i N_3}{T}}^2} \kchf{\tfrac{1+r}{2},\tfrac{1}{2},2\lr{\tfrac{\tilde{\tau}}{x''_4x''_8-\tfrac{p_5}{T}x''_7}-\tfrac{p_3-\pi\i N_3}{T}}^2} \right] , \label{resultx3}
\end{align}
with
\begin{align}
     \tau & \coloneqq x''_2x''_6x''_7 - \frac{p_1}{T}x''_6x''_8 - \frac{p_9}{T}x''_2x''_4 + \frac{p_1p_5p_9}{T^3},\\
     \tilde{\tau} & \coloneqq x''_2x''_6x''_7 - \frac{p_1+\frac{T^2}{2}}{T}x''_6x''_8 - \frac{p_9}{T}x''_2x''_4 + \frac{\lr{p_1+\frac{T^2}{2}} p_5p_9}{T^3}.
\end{align}
Again, due to the original peak of the Gaussian, $\nicefrac{\lr{p_3-\pi\i N_3}}{T}$, being present in the KCHFs' arguments, we can again apply the asymptotic expansion and the following Taylor expansion:
\begin{align}
    & \mathrm{(\ref{resultx3})} \approx \nonumber\\
    &\approx \frac{2}{t} \frac{T^{3r}\lr{\frac{2\pi\sqrt{2}}{T}}^{\!9}\Ghalf^4\mathscr{S}\mathscr{S}'\mathscr{S}''}{\mcsnorm\sqrt{2}^4}  \isumset{N} \es{2\sum\limits_{i}\lr{\frac{p_i-\pi\i N_i}{T}}^2} \infint^{4}x''_{i=2,4,6,8}\es{-2\sum\limits_{\mathclap{i=2,4,6,8}}\quad\lr{x''_i-\frac{p_i-\pi\i N_i}{T}}^2} \cdot \nonumber\\
    & \quad \cdot \infint x''_7 \es{-2\lr{x''_7 - \frac{p_7-\pi\i N_7}{T}}^2} \lr{1-\rreight\frac{x''_7{}^2\,T^4}{p_1{}^2p_9{}^2}} \left[ \left| x''_7\lr{x''_2x''_6-\frac{p_3p_5}{T^2}} + \omega \right|^r \right. - \nonumber\\
    & \qquad\qquad \left. - \left| x''_7\lr{x''_2x''_6-\frac{p_3p_5}{T^2}} + \tilde{\omega} \right|^r \right] \label{beforex7}
\end{align}
where only the $N_3=0$ contribution was considered and
\begin{align}
     \omega & \coloneqq x''_8\lr{\frac{p_3}{T}x''_4-\frac{p_1}{T}x''_6}-\frac{p_9}{T}x''_2x''_4 + \frac{p_1p_5p_9}{T^3} ,\\
     \tilde{\omega} & \coloneqq x''_8\lr{\frac{p_3}{T}x''_4-\frac{\lr{p_1+\frac{T^2}{2}}}{T}x''_6}-\frac{p_9}{T}x''_2x''_4 + \frac{\lr{p_1+\frac{T^2}{2}}p_5p_9}{T^3} .
\end{align}
Note that during the Taylor expansion here, the obtained prefactor $1-\rreight\ \frac{(x''_7{}^2\,T^4)}{(p_1{}^2\,p_9{}^2)}$ is not $x''_i$-independent and despite being $\sim T^4$ upon first sight, it turns out to contribute as $\sim T^2$ after the integration over $x''_7$. This will make the next integration with respect to $x''_7$ longer but at least still feasible via
\begin{align}
    \infint x \es{-\lr{\frac{x}{c}+b}^2}|x|^r\lr{x+d}^2 = |c|^{1+r}\es{-b^2}\Gsum\left[ d^2\, \kchf{\opr,\frac{1}{2},b^2} \right. + \nonumber\\
    +\left. \frac{1+r}{2}\,\kchf{\frac{3+r}{2},\frac{1}{2},b^2} - 4\frac{1+r}{2} bcd\, \kchf{\frac{3+r}{2},\frac{3}{2},b^2} \right]. \label{xsquareintegration}
\end{align}

We furthermore want to point out that the Taylor expansion of the previous step, leading to \eqref{beforex7}, --- despite its result containing $x''_7$ --- did again not alter the exponents of the integration variables up to the first order in $T$. As before, it is the Taylor expansion of the series that resulted from the KCHFs' asymptotic expansions for large arguments and goes with $z^{-n}$ for the respective KCHF's argument $z$. Looking at \eqref{resultx3}, we notice that the argument of both KCHFs is $\sim (x''_7)^{-2}$, leading to a $(x''_7)^2$ dependency in the asymptotic expansion's series --- which then, as we see, does not change during the subsequent Taylor expansion.

Applying \eqref{xsquareintegration} now to (\ref{beforex7}) and after the expansions, we get
\begin{align}
    & \mathrm{(\ref{beforex7})} \approx \nonumber\\ 
    & \approx \frac{2}{t} \frac{T^{3r}\lr{\frac{2\pi\sqrt{2}}{T}}^{\!9}\Ghalf^5\mathscr{S}\mathscr{S}'\mathscr{S}''}{\mcsnorm \sqrt{2}^5}  \isumset{N} \es{2\sum\limits_{i}\lr{\frac{p_i-\pi\i N_i}{T}}^2} \infint^{3}x''_{i=2,4,6}\es{-2\sum\limits_{\mathclap{i=2,4,6}}\ \ \lr{x''_i-\frac{p_i-\pi\i N_i}{T}}^2} \cdot \nonumber\\
    & \quad \cdot \underbrace{\lr{1-\rreight P T^2}}_{\eqqcolon\mathscr{S}'''} \infint x''_8 \es{-2\lr{x''_8-\frac{p_8 - \pi\i N_8}{T}}^2} \left[ \left| x''_8 \lr{\frac{p_3}{T}x''_4-\frac{p_1}{T}x''_6} + \chi \right|^r - \right. \nonumber\\
    & \qquad\qquad \left. - \left| x''_8 \lr{\frac{p_3}{T}x''_4-\frac{p_1+\frac{T^2}{2}}{T}x''_6} + \tilde{\chi} \right|^r \right],\label{beforex8}
\end{align}
where
\begin{equation}
    P \coloneqq \frac{p_7{}^2}{p_1{}^2\,p_9{}^2} + \frac{p_3{}^2}{\lr{p_1p_9-p_3p_7}^2} \label{eq:defP}
\end{equation}
was defined within the expansion's prefactor --- the expansion of all the two times three KCHFs indeed breaks down nicely and we included $N_7=0$ --- and
\begin{align}
     \chi & \coloneqq x''_2\lr{\frac{p_7}{T}x''_6-\frac{p_9}{T}x''_4} + \frac{p_1p_5p_9-p_3p_5p_7}{T^3} \eqqcolon x''_2\lr{\frac{p_7}{T}x''_6-\frac{p_9}{T}x''_4} + \frac{\mathcal{P}}{T^3} ,\\
     \tilde{\chi} & \coloneqq x''_2\lr{\frac{p_7}{T}x''_6-\frac{p_9}{T}x''_4} + \frac{\mathcal{P}}{T^3} + \frac{p_5p_9}{2T} .
\end{align}
Then,
\begin{align}
    & \mathrm{(\ref{beforex8})} \approx \nonumber\\
    & \approx \frac{2}{t} \frac{T^{3r}\lr{\frac{2\pi\sqrt{2}}{T}}^{\!9}\Ghalf^6\mathscr{S}\mathscr{S}'\mathscr{S}''\mathscr{S}'''}{\mcsnorm\sqrt{2}^6}\isumset{N} \es{2\sum\limits_{i}\lr{\frac{p_i-\pi\i N_i}{T}}^2} \infint^{2}x''_{i=2,4}\es{-2\sum\limits_{\mathclap{i=2,4}}\ \lr{x''_i-\frac{p_i-\pi\i N_i}{T}}^2} \cdot \nonumber\\
    & \cdot  \infint x''_6 \es{-2\lr{x''_6-\frac{p_6 - \pi\i N_6}{T}}^2} \lr{1-\rreight\frac{\lr{p_3x''_4-p_1x''_6}^2 T^4}{\mathcal{P}^2}} \left[ \left| x''_6\lr{\frac{p_7}{T}x''_2-\frac{p_1p_8}{T^2}} + \vartheta \right|^r - \right. \nonumber\\
    & \qquad\qquad - \left. \left| x''_6\lr{\frac{p_7}{T}x''_2-\frac{\lr{p_1+\frac{T^2}{2}}p_8}{T^2}} + \tilde{\vartheta} \right|^r \right], \label{beforex6}
\end{align}
with
\begin{align}
     \vartheta & \coloneqq x''_4\lr{\frac{p_3p_8}{T^2}-\frac{p_9}{T}x''_2}+\frac{\mathcal{P}}{T^3} ,\\
     \tilde{\vartheta} & \coloneqq x''_4\lr{\frac{p_3p_8}{T^2}-\frac{p_9}{T}x''_2}+\frac{\mathcal{P}}{T^3} + \frac{p_5p_9}{2T} , \label{beforeLastTaylorWithVariable}
\end{align}
and we have again an $x''_i$-dependent prefactor from the expansions --- this time even with two of the remaining integration variables included. Note that this Taylor expansion did again not alter the integration variables' exponents up to the first order in $t$ contributions. Nevertheless, we can still continue as usual and after a longish integration in the style of (\ref{xsquareintegration}), the application of the asymptotic expansion for large arguments of all the six accrued KCHFs and the proceeding Taylor expansion, we get
\begin{align}
    & \mathrm{(\ref{beforex6})} \approx \nonumber\\
    & \approx \frac{2}{t} \frac{T^{3r}\lr{\frac{2\pi\sqrt{2}}{T}}^{\!9}\Ghalf^7\mathscr{S}\mathscr{S}'\mathscr{S}''\mathscr{S}'''}{\mcsnorm\sqrt{2}^7}\isumset{N} \es{2\sum\limits_{i}\lr{\frac{p_i-\pi\i N_i}{T}}^2} \infint^{2}x''_{i=2,4}\es{-2\sum\limits_{\mathclap{i=2,4}}\ \lr{x''_i-\frac{p_i-\pi\i N_i}{T}}^2} \cdot \nonumber\\
    & \quad \cdot \underbrace{\lr{1-\rreight \frac{p_1{}^2 p_6{}^2 \tilde{\mathcal{P}}^2+p_1{}^2p_8{}^2\mathcal{P}^2}{\tilde{\mathcal{P}}^2\mathcal{P}^2} T^2}}_{\eqqcolon \mathscr{S}''''} \left[ \left| x''_4\lr{\frac{p_3p_8}{T^2}-\frac{p_9}{T}x''_2} + \frac{p_6p_7}{T^2}x''_2 + \frac{\tilde{\mathcal{P}}}{T^3} \right|^r - \right.\nonumber\\
    &\qquad\qquad- \left. \left| x''_4\lr{\frac{p_3p_8}{T^2}-\frac{p_9}{T}x''_2} + \frac{p_6p_7}{T^2}x''_2 + \frac{\tilde{\mathcal{P}}}{T^3} + \frac{\Delta^1_1\lr{p}}{2T} \right|^r \right], \label{afterx6}
\end{align}
with
\begin{equation}
    \tilde{\mathcal{P}} \coloneqq \mathcal{P} - p_1p_6p_8 = p_1p_5p_9-p_3p_5p_7-p_1p_6p_8
\end{equation}
continuing to build up the determinant of the $p_i$'s matrix and
\begin{equation}
    \Delta^1_1\lr{p} \coloneqq p_5p_9-p_6p_8
\end{equation}{}
as the minor of that determinant with respect to the shifted $p_1 \equiv p^1_1$
\begin{equation}
\left(\begin{array}{*3{c}}
    p_1+\frac{T^2}{2} & p_2 & p_3 \\
    p_4 & \tikzmark{left}{$p_5$} & p_6  \\
    p_7 & p_8 & \tikzmark{right}{$p_9$}
  \end{array}\ \right) \  .
    \Highlight[first]
\end{equation}

We faced again an $x''_i$-independent prefactor $\mathscr{S}''''$ and can straightforwardly integrate over $x''_4$ first, yielding
\begin{align}
    \mathrm{(\ref{afterx6})}& \approx \frac{2}{t} \frac{T^{3r}\lr{\frac{2\pi\sqrt{2}}{T}}^{\!9}\Ghalf^8\mathscr{S}\mathscr{S}'\mathscr{S}''\mathscr{S}'''\mathscr{S}''''\mathscr{S}'''''}{\mcsnorm\sqrt{2}^8} \!\!\isumset{N} \!\!\! \es{2\sum\limits_{i}\lr{\frac{p_i-\pi\i N_i}{T}}^2}\!\!\! \infint x''_2\es{-2\lr{x''_2-\frac{p_2-\pi\i N_2}{T}}^2} \cdot \nonumber\\
    & \quad \cdot  \left[ \left| x''_2\,\frac{p_6p_7-p_4p_9}{T^2}+\frac{\hat{\mathcal{P}}}{T^3} \right|^r -  \left| x''_2\,\frac{p_6p_7-p_4p_9}{T^2}+\frac{\hat{\mathcal{P}}}{T^3} + \frac{\Delta^1_1\lr{p}}{2T} \right|^r \right], \label{beforex2}
\end{align}
with
\begin{equation}
     \hat{\mathcal{P}} \coloneqq \tilde{\mathcal{P}} +p_3p_4p_8
\end{equation}
as a further step towards the $p_i$'s determinant and the expansions' $x''_i$-independent prefactor
\begin{equation}
    \mathscr{S}''''' \coloneqq 1-\rreight\frac{p_3{}^2p_8{}^2}{\hat{\mathcal{P}}^2}T^2.
\end{equation}
All that's left is the last $x''_2$-integration. We see already that the absolute values' arguments are plausible: $\hat{\mathcal{P}}$ is building up the desired determinant of the $p_i$ and the shifted contribution has an additional $\Delta^1_1\lr{p}$ that reflects the minor of the determinant with respect to $p_1$ --- the $p_i$ on which the shift acted, leading directly to this extra term. Also the fact that $x''_2$ is multiplied by $p_6p_7-p_4p_9$ looks promising as this is the minor of the determinant with respect to $p_2$.

The $x''_2$-integration is just of the familiar form and we obtain after performing the asymptotic as well as the Taylor expansion and with considering only the $N_2=0$ contribution
\begin{align}
    (\mathrm{\ref{beforex2}}) & \approx  \frac{2}{t} \frac{T^{3r}\lr{\frac{2\pi\sqrt{2}}{T}}^{\!9}\Ghalf^9\mathscr{S}\mathscr{S}'\mathscr{S}''\mathscr{S}'''\mathscr{S}''''\mathscr{S}'''''}{\mcsnorm\sqrt{2}^9}\isumset{N} \es{2\sum\limits_{i}\lr{\frac{p_i-\pi\i N_i}{T}}^2} \cdot \nonumber\\
    & \quad\cdot \mathscr{S}'''''' \left[ \left| \frac{\det p}{T^3} \right|^r - \left| \frac{\det p}{T^3} + \frac{\Delta^1_1\lr{p}}{2T} \right|^r \right] ,
\end{align}
with
\begin{align}
    \mathscr{S}'''''' \coloneqq 1- \rreight \frac{\lr{p_6p_7-p_4p_9}^2}{\lr{\det p}^2}T^2
\end{align}
and finally the determinant of the matrix of the $p_i$ denoted as
\begin{align}
    \det p\coloneqq \hat{\mathcal{P}} + p_2p_6p_7-p_2p_4p_9 .
\end{align}

What is now left is combining the two contributions and insert
\begin{equation} \label{eq:statesnorm}
    \mcsnorm = \prod_i\lr{1+K^{(i)}_{t}} \lr{\frac{2\pi\sqrt{\pi}}{T}}^9\es{2\sum\limits_i \lr{\frac{p_i}{T}}^2} \prod_i\lr{1+K^{(i)}_{t}},
\end{equation}
where $K^{(i)}_{t} = \mathcal{O}(t^\infty)$, i.e. $\lim_{t\to 0} \nicefrac{K^{(i)}_{t}}{t^n} = 0 \ \forall n\in\mathbbm{N}$~\cite{Brunnemann2}. The origin is again the damping of all $N_i\neq 0$ contributions, put into $K^{(i)}_{t}$, and that thereby only $\left\{N_i\right\}=0$ remains:
\begin{equation}
    \prod_i\lr{1+K^{(i)}_{t}} \coloneqq \isumset{N} \es{-2\sum\limits_i\lr{\frac{\pi^2N_i{}^2}{T^2}+\frac{2\pi\i p_i N_i}{T^2}}}.
\end{equation}
We already see that the norm's factor $\es{2\sum_{i}\lr{\frac{p_i}{T}}^2}$, as denominator, will cancel $\es{2\sum_{i}\lr{\frac{p_i-\pi\i N_i}{T}}^2}$ of (\ref{beforex2}), considering that $\left\{ N_i \right\} = 0$.

With all that, we finally arrive at
\begin{align}
    \frac{2}{t}\langle \hat{q}^1_1(r) \rangle_{\cs} & \approx \frac{2}{t} \frac{\Ghalf^9 T^{3r} \mathscr{S}\mathscr{S}'\mathscr{S}''\mathscr{S}'''\mathscr{S}''''\mathscr{S}'''''\mathscr{S}''''''}{\sqrt{\pi}^9\prod_i\lr{1+K^{(i)}_{t}}\prod_i\lr{1+K^{(i)}_{t}}}  \left[\left| \frac{\det p}{T^3} \right|^r - \left| \frac{\det p}{T^3}+\frac{\Delta^1_1\lr{p}}{2T} \right|^r \right] . \label{puttingtogether}
\end{align}
The clear structure is pretty much as desired. We have several prefactors of expansion-series-type $(1-\rreight\,f(\left\{p_i\right\})\,T^2)$, the normalisation prefactors completely cancel via $\Ghalf=\sqrt{\pi}$ and a difference in the absolute values of the unshifted and shifted contribution remains. Next, we use the prefactor $T^{3r}$ to cast the square bracket of (\ref{puttingtogether}) into
\begin{align}
    & T^{3r} \left[\left| \frac{\det p}{T^3} \right|^r - \left| \frac{\det p}{T^3}+\frac{\Delta^1_1\lr{p}}{2T} \right|^r \right]  = \left| \det p \right|^r - \left| \det p + \Delta^1_1\lr{p} \frac{T^2}{2} \right|^r \nonumber\\
    & \qquad\qquad\qquad \approx -r \frac{\left|\det p\right|^r\Delta^1_1\lr{p}}{\det p}\frac{T^2}{2} - \frac{r(r-1)}{8} \frac{\left|\det p\right|^r \lr{\Delta^1_1\lr{p}}^2}{\left|\det p\right|^2}T^4 + \mathcal{O}\lr{T^6}. \label{absvaltaylor}
\end{align}

The last step consists of inserting this into (\ref{puttingtogether}) and multiplying all the $\mathscr{S}^{(}{}'{}^{\ldots}{}'{}^{)}$:
\begin{align}
    \frac{2}{t}\langle \hat{q}^1_1(r) \rangle_{\cs} & \approx -  \frac{2}{t} r \frac{|\det p|^r\Delta^1_1\lr{p}}{\det p}\frac{T^2}{2} + \frac{2}{t} \mathscr{F}\;\! T^4 + \frac{2}{t}\mathcal{O}\lr{T^5}.
\end{align}

If we want to compare this to the classical result of the corresponding Poisson bracket, we need to consider the $t^0$ order and using  $t=T^2$.\footnote{And, in order to obtain the volume, set $r=\frac{1}{2}$. Note again that in the course of this calculation, we used an exponent $r$ of the determinant's absolute value --- unlike $\rhalf$ before and in the literature.} We then obtain for the $T^0$ term
\begin{equation}
    r\frac{|\det p|^r\Delta^1_1\lr{p}}{\det p} \nonumber,
\end{equation}{}
which reflects the classical Poisson bracket's differentiation result. It stems from the $T^2$-term of (\ref{absvaltaylor}) multiplied by all $1$s from the $\mathscr{S}^{(}{}'{}^{\ldots}{}'{}^{)}$-series. Luckily, we could en passant determine the next order term $\sim T^4$, too, without the need of further expanding all the $\mathscr{S}^{(}{}'{}^{\ldots}{}'{}^{)}$-series! With all the $\mathscr{S}^{(}{}'{}^{\ldots}{}'{}^{)}$ being series like $1-\ldots T^2+\mathcal{O}\lr{T^3}$ and the expansion of the square bracket above being $\ldots T^2 + \ldots T^4 + \mathcal{O}\lr{T^6}$, the $T^4$-contributions are firstly the square bracket's $T^4$-term multiplied by the 1s from all $\mathscr{S}^{(}{}'{}^{\ldots}{}'{}^{)}$ and, secondly, the square bracket's $T^2$-term multiplied by the sum over all $T^2$-terms of the $\mathscr{S}^{(}{}'{}^{\ldots}{}'{}^{)}$. I.e.
\begin{align}
    \mathscr{F} \coloneqq & - \frac{r(r-1)}{8} \frac{\left|\det p\right|^r \lr{\Delta^1_1\lr{p}}^2}{\left|\det p\right|^2} -\frac{r^2\lr{r-1}}{16} \frac{\left|\det p\right|^r\Delta^1_1\lr{p}}{\det p}  \left[ \frac{1}{p_1{}^2} + \frac{1}{p_5{}^2} + \frac{1}{p_9{}^2} + P +\right. \nonumber\\
    & + \left. \frac{p_1{}^2 p_6{}^2 \tilde{\mathcal{P}}^2+p_1{}^2p_8{}^2\mathcal{P}^2}{\tilde{\mathcal{P}}^2\mathcal{P}^2} + \frac{p_3{}^2p_8{}^2}{\hat{\mathcal{P}}^2} + \frac{\lr{p_6p_7-p_4p_9}^2}{\lr{\det p}^2} \right]  . \label{eq:T4-term}
\end{align}
Each contribution $\sim T^3$ from the factors $\mathscr{S}^{(}{}'{}^{\ldots}{}'{}^{)}$ would therefore at least meet another term $\sim T^2$ and hence contribute as $\mathcal{O}\lr{T^5}$.
Note that $P \overset{\eqref{eq:defP}}{=} \frac{p_7{}^2}{p_1{}^2\,p_9{}^2} + \frac{p_3{}^2}{\lr{p_1p_9-p_3p_7}^2}$, so all addends within the square bracket contribute with $\sim p_i{}^{-2}$. All other occurring $\mathcal{P},\tilde{\mathcal{P}},\hat{\mathcal{P}}$ are $\sim p^3$ and build up to $\det p$. The basic composition of the next-to-leading order term as above can also be seen in the quantum mechanical results of \cite[therein: eq. (3.9) for the AQG result and eq. (3.12) for the one via KCHFs]{paper1}. Namely, there are two contributions that differ by one in the overall exponent in $p$: $\mathscr{F} \sim \ldots p^{3r-2} + \ldots p^{3r-3}$. This reflects one-to-one the quantum mechanical result of \cite{paper1}.

Note that this feature, however, is not present in the result (4.45) of \cite{Towards2}: Collecting all terms of next-to-leading order $t$, they all end up being $\sim p^{3\rhalf-3}$ for $N=1$. The authors think that this is due to the power counting of (4.39) in \cite{Towards2}, where it is argued one could neglect terms $\sim sT$ compared to $\sim s^2$. Now, \cite{Towards2} introduced $s \coloneqq t^{\frac{1}{2}-\alpha}$ to compensate for a rescaling of $p \mapsto q \coloneqq p t^{-\alpha}$ to obtain a quantity ``of order unity'' and is always paired with $q^{-1}$ --- there should be no powers of $t$ including $\alpha$ in the final result. This implies that while $sT = t^{1-\alpha}$ itself is indeed of higher order in $t$ than $s^2 = t^{1-2\alpha}$ (for $\alpha\geq0$), the terms $\sim sT$, in fact, are not: all terms $\sim sT$ also include $q^{-1}$, therefore making them $\sim t$. Likewise, all terms $\sim s^2$ also include $q^{-2}$, making them $\sim t$ as well. This left out term $\sim sT$ contains one $q^{-1}$ less than the collected one, i.e. is of higher order in $p$, which is in line with the fact that their contribution linear in $t$ contained terms $\sim p^{3\rhalf-3}$, while the work at hand obtained additionally one part $\sim p^{3r-2}$ (recap that we started with $r$ instead of $\rhalf$ in \cite{Towards2}).

However, we can say the following: Concerning the powers in $p$, the contribution of $\mathscr{F}$ that includes the square bracket should correspond to the correction terms \cite{Towards2} considers. While this can be motivated by the quantum mechanical or U(1) scenario just like described below \eqref{eq:BBBSanThiemann} (they should be the derivative of the fluctuations of the expectation value of the volume operator), it was not possible to find a close connection of the two paths. Appendix \ref{appendix:ComparisonProcedures} does indeed show such a link for the U(1) case but due to the convoluted successive integrations of the KCHF way versus the all-at-once approach of \cite{Towards2} by Sahlmann and Thiemann, we loose track of a similar link for the higher dimensional case. It is, however, easy to see that the first contribution of $\mathscr{F}$, the one that corresponds to the second derivative, corresponds to the first neglected term $\sim sT$ of the power series (4.39) in \cite{Towards2}.

\section{Estimates}
\label{sec:Estimates}

We present here a short list of the fundamental approximations we use.

\begin{itemize}
    \item The approximation Brunnemann and Thiemann used in~\cite{Brunnemann2},
        \be
            |a|^r-|b|^r \leq ||a|-|b|| \quad (\text{where}\ a,b\in\mathbbm{Z} \ \text{and } r\in\mathbbm{Q}_{[0,1]}
            ) , \label{eq:approxBrunn}
        \ee 
        allowed them to shed the roots completely.
    \item We will make use of
        \be
            |a|^r-|b|^r \leq |a-b|^r \quad (\text{where} \ a,b,r\in\mathbbm{R} \ \text{and } 0\leq r\leq 1)\label{approxdiff}
        \ee
        to get rid of a difference in roots, allowing us to further manipulate the actual form of $(a-b)$ and then integrating it against Gaussians by means of KCHF.
    \item The corresponding approximation for the sum of roots,
        \be
            |a+b|^r\leq|a|^r+|b|^r \quad (\text{where} \ a,b,r\in\mathbbm{R} \ \text{and } 0\leq r\leq 1) , \label{approxsum}
        \ee
        will also be used.
    \item Finally, when trying to establish a good estimate for the U$(1)^3$--case,     we will use
        \be
            |a+C|^r-|(a-1)+C|^r \leq |a|^r - |a-1|^r +2 \quad (\text{where} \ a,C,r\in\mathbbm{R} \ \text{and } 0\leq r\leq 1) .\label{approxfinal}
        \ee
    This estimate can be verified easily using~(\ref{approxdiff}):
    \begin{align}
        |a+C|^r-|a+\delta+C|^r+|a+\delta|^r-|a|^r &\leq |a+C-(a+\delta+C)|^r+|a+\delta-a|^r \nonumber\\
        &=2|\delta|^r .
    \end{align}
\end{itemize}

Note that the estimates~\eqref{approxdiff},~\eqref{approxsum} and~\eqref{approxfinal} still contain via $|\ldots|^r$ information about which roots we are facing, while~\eqref{eq:approxBrunn} does not.

\section{Detailed derivation of the semiclassical continuum limit for graphs of cubic topology}
\label{sec:AppSemLimitCubic}
In this appendix, we present the details of the derivation of the semiclassical continuum limit for graphs of cubic topology that is discussed in subsection \ref{subsec:SemClassCubic} in the main text. There, it is discussed that in the case of the cubic graph and if one considers the limit where both the semiclassical parameter $t\to 0$ as well as the regulator $\epsilon\to 0$, then one obtains the expected classical expression for the Poisson bracket. Particularly, the following identities hold:
\begin{align}
\label{eq:App2Id}
    \lim\limits_{t\to 0}\frac{2\langle \hat{q}^{i_0}_{I_0}(r) \rangle_{\cs}}{t} = - r \frac{\lrabs{\det p^-}^r\Delta^{i_0}_{I_0}\lr{p^-}}{\det p^-} &=
    2\i\:\! h^{i_0}_{I_0}\left\{\left(h^{i_0}_{I_0}\right)^{-1},V^{2r}\lr{R_{\Box_\epsilon}}\right\}\nonumber \\
     \lim\limits_{\epsilon\to 0}\left(  2\i \:\! h^{i_0}_{I_0}\left\{\left(h^{i_0}_{I_0}\right)^{-1},V^{2r}\lr{R_{\Box_\epsilon}}\right\}\right) & = \frac{1}{a^{6r}}\left\{\int_{e_{I_0}}A^{i_0}, V^{2r}\lr{R_x}\right\} .
\end{align}
The prefactors arise because if we substitute ${\rm SU}(2)$ by ${\rm U}(1)^3$, we replace $\tau_j$ by $\i$ and by performing the quantisation step we divide by $\frac{1}{\i\hbar}$. Lastly, since we work with the dimensionless volume, see \eqref{eq:qr}, we have a factor $\frac{1}{a^{6r}}$ involved. The factor of $2$, as already explained in the main text, is necessary in order to obtain the correct semiclassical limit of the Thiemann identity, more details will be presented in \cite{GTtoAppear}.

As a first step, we show the following result for the classical ${\rm U}(1)^3$ model for a graph of cubic topology:
\begin{align}
\label{eq:AppClPB}
 h^{i_0}_{I_0}\left\{\left(h^{i_0}_{I_0}\right)^{-1},V^{2r}\lr{R_{\Box_\epsilon}}\right\} &=-\i\frac{\kappa}{2a^2}r\lrabs{\det p^-}^r\frac{\Delta^{i_0}_{I_0}\lr{p^-}}{\det p^-} \equiv -\i\frac{\kappa}{2a^2}r\lrabs{\det p^-}^r\lr{(p^-)^{-1}}_{I_0i_0}.
\end{align}
The classical Poisson bracket is just given by
\begin{align}
\label{eq:AppPB}
  h^{i_0}_{I_0}\left\{\left(h^{i_0}_{I_0}\right)^{-1},V^{2r}\lr{R_{\Box_\epsilon}}\right\}
 &=
 \kappa  h^{i_0}_{I_0}\int \d^3z \left(\frac{\delta (h^{i_0}_{I_0})^{-1}}{\delta A^k_b(z)}\right) 
 \left(\frac{\delta V^{2r}(p^-)}{\delta E^b_k(z)}\right) .
\end{align}
We then have 
\begin{align}
\label{eq:FuncDerHol}
  h^{i_0}_{I_0}\left(\frac{\delta (h^{i_0}_{I_0})^{-1}}{\delta A^k_b(z)}\right)  
  &=-\i \int\limits_0^1 \d t \, \dot{e}^a_{I_0}(t)\delta^{ki_0}\delta^a_b\delta(t,z)   
\end{align}
and 
\begin{align*}
 \left(\frac{\delta V^{2r}(p^-)}{\delta E^b_k(z)}\right) 
 &=
 r\lrabs{\det p^-}^{r-1}\sgn(p^-)\left(\frac{\delta \det p^-}{\delta E^b_k(z)}\right).
\end{align*}
Next, to compute the functional derivative, we express $p_{I_0i_0}^-$ again in terms of the variables $p_{I_0\sigma_0i_0}$, where $I_0,i_0=1,2,3$ and $\sigma_0=\pm$, which denote the dimensionless fluxes associated with the six edges at a given vertex for a graph of cubic topology. We obtain
\begin{align}
\label{eq:P+-}
p_{I_0i_0}^-
&=
\frac{1}{2}\left(p_{I_0+i_0}-p_{I_0-i_0}\right)=\frac{1}{2a^2}\left(E^{I_0}_{i_0+}- E^{I_0}_{i_0-}\right),\nonumber \\
p_{I_0i_0}^+
&=
\frac{1}{2}\left(p_{I_0+i_0}+p_{I_0-i_0}\right)=\frac{1}{2a^2}\left(E^{I_0}_{i_0+}+ E^{I_0}_{i_0-}\right),
\end{align}
with the fluxes $E^J_{j\sigma}$ defined as 
\begin{align}
 E^J_{j\sigma}\coloneqq \int\limits_{S_{e^\sigma_J}} E^a_j n_a^{S_{e^\sigma_J}},   \label{eq:sigmaFluxes}
\end{align}
where $S_{e^\sigma_J}$ is the surface dual to the edge $e^\sigma_J$ and $n_a^{S_{e^\sigma_J}}$ denotes the conormal of $S_{e^\sigma_J}$, for which we use the abbreviation $n_a^{S_{e^\sigma_J}}\coloneqq n_a^{J\sigma}$. Thus, we get
\begin{align}
\label{eq:FuncDerDet}
\left(\frac{\delta \det p^-}{\delta E^b_k(z)}\right) \!\!\,
= \!\!\,
\frac{\det p^-}{2a^2}\lr{(p^-)^{-1}}_{Jj}
\left( \int\limits_{S_{e^+_J}}\d^2u\, n_a^{J+}\delta_{j}^k\delta^a_b\delta(x(u),z) \right.
\left. -\! \int\limits_{S_{e^-_J}}\d^2u\, n_a^{J-}\delta_{j}^k\delta^a_b\delta(x(u),z) \right) .
\end{align}
Reinserting the results of \eqref{eq:FuncDerHol} and \eqref{eq:FuncDerDet} back into \eqref{eq:AppPB} and performing the integrals involving the delta functions, we end up with
\begin{align}
 h^{i_0}_{I_0}\left\{\left(h^{i_0}_{I_0}\right)^{-1},V^{2r}\lr{R_{\Box_\epsilon}}\right\}
 &=-\frac{\i\kappa}{2a^2}r \lrabs{\det p^-}^{r-1} \lrabs{\det p^-} \lr{(p^-)^{-1}}_{I_0i_0} \nonumber\\
 &=-\frac{\i\kappa}{2a^2}r \lrabs{\det p^-}^{r} \frac{\Delta^{i_0}_{I_0}(p^-)}{\det p^-},
\end{align}
where we used in the last step that $\Delta^{i_0}_{I_0}(p^-)=\frac{1}{2}\epsilon^{i_0k\ell}\epsilon_{I_0KL}p^-_{Kk}p^-_{L\ell}$. The result exactly agrees with \eqref{eq:AppClPB}.

Next, in order to take the limit in which we send the regulator to zero, that is $\epsilon\to 0$, we need to express $p_{I_0i_0}^-$ again in terms of the variables $p_{I_0\sigma_0i_0}$, where $I_0,i_0=1,2,3$ and $\sigma_0=\pm$, which denotes the dimensionless fluxes associated with the six edges at a given vertex for a graph of cubic topology. We have 
\begin{align*}
 p_{I_0+i_0} =    p_{I_0i_0}^+ + p_{I_0i_0}^- \quad \land \quad p_{I_0-i_0} = p_{I_0i_0}^+ - p_{I_0i_0}^-\quad 
  \iff\quad  p_{I_0\sigma_0 i_0} =    p_{I_0i_0}^+ + \sgn(\sigma_0) p_{I_0i_0}^-.
\end{align*}
Using these relations, we further obtain for $\lr{(p^-)^{-1}}_{I_0i_0} \eqqcolon (p_{I_0i_0}^-)^{-1}$ 
\begin{align}
\label{eq:p-Inv}
\sgn(\sigma_0)(p_{I_0i_0}^-)^{-1} &=(\mathbbm{1}-p^{-1}_{I_0\sigma_0 i_0}(p_{I_0i_0}^+))^{-1}p^{-1}_{I_0\sigma_0 i_0},
\end{align}
where we introduced the convention that the superscript $-1$ does not indicate one over the specific matrix element but instead the matrix element of the inverse of the corresponding matrix.
The inverse $p^{-1}_{I_0\sigma_0i_0}$ can be written in terms of $p_{I_0\sigma_0i_0}$ as follows:
\begin{align*}
p^{-1}_{I_0\sigma_0 i_0} &=\frac{1}{2}\epsilon^{i_0mn}\epsilon^{I_0MN}\frac{p_{M\sigma_0 m}p_{N\sigma_0 n}}{\det p_{\sigma_0 }}
= \frac{a^2}{2}\epsilon^{i_0mn}\epsilon^{I_0MN}\frac{E^M_{m\sigma_0}E^N_{n\sigma_0}}{\det E_{\sigma_0}},
\end{align*}
where we analogously to previous notations use $\det E_{\sigma_0} \equiv \det \lr{ E^J_{j\sigma_0}}$ for the determinant of the matrix of the fluxes $E^J_{j\sigma_0}$.
In order to expand the fluxes and the determinant in terms of the regularisation parameter, we consider the following embedding for the cube:
\begin{align*}
X^a_v\colon\left[-\frac{\epsilon}{2},\frac{\epsilon}{2}\right]\to\sigma,\quad (t^1,t^2,t^3)\mapsto X^a_v(t_1,t_2,t_3), \quad X^a_v(0,0,0)=v.   
\end{align*}
Then, the expansion of the fluxes and determinant in terms of powers of the regularisation parameter $\epsilon$ yields
\begin{align*}
  E^J_{j\sigma} &= \epsilon^2 E^a_j(v) n_a^{J\sigma}(v) + \mathcal{O}\lr{\epsilon^3} \quad \text{and}  \\
  \det E_{\sigma}&\equiv \det\lr{E^J_{j\sigma}} = \det\lr{\epsilon^2 E^a_j(v) n_a^{J\sigma}(v) + \mathcal{O}\lr{\epsilon^3}}=\epsilon^6\det\lr{E^a_j(v)} \det\lr{n_a^{J\sigma}(v)} + \mathcal{O}\lr{\epsilon^7} \\
  &= \epsilon^6 \det E\lr{v} \cdot \det n^\sigma\lr{v} + \mathcal{O}\lr{\epsilon^7},
\end{align*}
where $\det E \lr{v} \equiv \det \lr{E^a_j\lr{v}} $ now is the determinant of the matrix of all the fluxes $E^a_j$, linked to the fluxes $E^J_{j\sigma_0}$ via \eqref{eq:sigmaFluxes}, and analogously $\det n^\sigma\lr{v} \equiv \det \lr{n^{J\sigma}_j\lr{v}}$.
Considering $p^{+}_{I_0i_0}$ in \eqref{eq:P+-} together with  $n_a^{J+}(v)=-n_a^{J-}(v)$, we realise that the leading order contribution vanishes and thus the term involving $p^{-1}_{I_0\sigma_0 i_0}p^{+}_{I_0i_0}$ in (\ref{eq:p-Inv}) is at least one power in $\epsilon$ higher than the contribution of the unit matrix. Hence, the term in addition to the unit matrix can be neglected if we are interested in the limit $\epsilon\to 0$. As a consequence, the expansion of (\ref{eq:p-Inv}) yields
\begin{align*}
\sgn(\sigma_0)(p_{I_0i_0}^-)^{-1} &= \sgn(\sigma_0)\frac{a^2}{2}\epsilon^{i_0mn}\epsilon^{I_0MN}\frac{E^a_{m}n_a^{M\sigma_0}E^b_{n}n_b^{N\sigma_0}}{\det E\lr{v} \lrabs{\det n^\sigma\lr{v}}}\left(\frac{1+\mathcal{O}\lr{\epsilon}}{\epsilon^2+\mathcal{O}\lr{\epsilon^3}}\right),
\end{align*}
where the $\sgn$ involved on the right hand side arose because we introduced an absolute value in the denominator. Collecting all intermediate results the expansion of \eqref{eq:AppClPB} is given by
\begin{align}
\label{eq:ClPBExp}
 h^{i_0}_{I_0}\left\{\left(h^{i_0}_{I_0}\right)^{-1},V^{2r}\lr{R_{\Box_\epsilon}}\right\} &=-\i\frac{\kappa}{2a^2}r\lrabs{\frac{\epsilon^6}{a^6}\det E\lr{v} \cdot \det n^\sigma\lr{v}+\mathcal{O}\lr{\epsilon^8}}^r \cdot \nonumber\\
 & \quad\cdot\frac{a^2}{2}\epsilon^{i_0mn}\epsilon^{I_0MN}\frac{E^a_{m}n_a^{M\sigma_0}E^b_{n}n_b^{N\sigma_0}}{\det E\lr{v} \lrabs{\det n^\sigma\lr{v}}}\left(\frac{1+\mathcal{O}\lr{\epsilon}}{\epsilon^2+\mathcal{O}\lr{\epsilon^3}}\right).
\end{align}
In a final step, we consider the limit $\epsilon\to 0$ and therefore keep only the leading order terms in $\epsilon$ in the following. First, we discuss the absolute value on the right hand side of \eqref{eq:ClPBExp}. Considering that $\frac{\partial X^a_v}{\partial t^J}$ are tangent vectors, we have
\begin{align*}
n_a^{J\sigma}\frac{\partial X^b_v}{\partial t^J}=\delta^b_a\sgn(\sigma)\Big|\det\left(\frac{\partial X^c_v}{\partial t^K}\right)\Big| \quad{\rm and}\quad
\det n^\sigma =\Big|\det\left(\frac{\partial X^c_v}{\partial t^K}\right)\Big|^2.
\end{align*}
Thus, the limit $\epsilon\to 0$ of the absolute value in \eqref{eq:ClPBExp} can be rewritten as 
\begin{align}
\label{eq:LimitAbsVal}
\lim\limits_{\epsilon\to 0} \Big|\frac{\epsilon^3}{a^3}\sqrt{\det E}\;\Big|\det\left(\frac{\partial X^c_v}{\partial t^K}\right)\Big|\Big|^{2r}
&=\frac{1}{a^{6r}}\left(\;\int\limits_{\Box_v} \d^3x \sqrt{\det E}(x)\right)^{2r}=\frac{1}{a^{6r}}V^{2r}_{R_x},
\end{align}
where we used that in the limit $\epsilon\to 0$ we have $\epsilon^3\Big|\det\left(\frac{\partial X^c_v}{\partial t^K}\right)\Big|=\int_{\Box_v}\d^3x$. 
In order to take the limit also for the remaining part corresponding to $(p^-_{I_0i_0})^{-1}$ in \eqref{eq:ClPBExp}, we first use 

\begin{align}
\epsilon_{I_0MN}n_a^{M\sigma_0}n_b^{N\sigma_0} & =
\det n^\sigma \frac{\sgn(\sigma_0)}{\Big|\det\left(\frac{\partial X^d_v}{\partial t^K}\right)\Big|}\epsilon_{abc}\frac{\partial X^c_v}{\partial t^{I_0}}
=\Big|\det\left(\frac{\partial X^d_v}{\partial t^K}\right)\Big|\sgn(\sigma_0)\epsilon_{abc}\frac{\partial X^c_v}{\partial t^{I_0}} .
\end{align}
Further, we expand the fraction by $\epsilon$ yielding a linear power in the numerator and a cubic one in the denominator. The contribution for $(p^-_{I_0i_0})^{-1}$ then reads
\begin{align}
\label{eq:Limitp-}
\lim\limits_{\epsilon\to 0}(p^-_{I_0i_0})^{-1} & =
\lim\limits_{\epsilon\to 0}\left(\frac{a^2}{2}\frac{\epsilon\sgn(\sigma_0)\epsilon_{i_0mn}\epsilon_{abc}\frac{E^a_m(v)E^b_n(v)}{\sqrt{\det E}(v)}\frac{\partial X^c_v}{\partial t^{I_0}}}{\epsilon^3\sqrt{\det E}(v)\Big|\det\left(\frac{\partial X^d_v}{\partial t^K}\right)\Big|}\right) \\ \nonumber 
&= 
\frac{\int\limits_{0}^1 \d t\frac{1}{2}\epsilon_{i_0mn}\epsilon_{abc}\frac{E^a_mE^b_n}{\sqrt{\det E}}\frac{\partial X^c_v}{\partial t^{I_0}}}{\int\limits_{\Box_v}\d^3 x\sqrt{\det E}(x)} 
= \frac{2a^2}{\kappa} V^{-1}_{R_x}\int\limits_0^1 \d t \left\{A^{i_0}_{a}(e(t)),V_{R_x}\right\}(\dot{e}_{I_0}^{\sigma_0})^a \nonumber \\
&=
\frac{2a^2}{\kappa}V^{-1}_{R_x}\left\{ \int_{e^{\sigma_0}_{I_0}} A^{i_0},V_{R_x}\right\}.
\end{align}
We collect the results of \eqref{eq:LimitAbsVal} and \eqref{eq:Limitp-} and combine them with the result of \eqref{eq:AppClPB} and get
\begin{align}
2\i\cdot\lim\limits_{\epsilon\to 0} h^{i_0}_{I_0}\left\{\left(h^{i_0}_{I_0}\right)^{-1},V^{2r}\lr{R_{\Box_\epsilon}}\right\} 
&=2\frac{r}{2}\frac{2}{a^{6r}}V^{2r-1}_{R_x}
\left\{ \int_{e^{\sigma_0}_{I_0}} A^{i_0},V_{R_x}\right\} \nonumber \\
&= \frac{1}{a^{6r}}\left\{ \int_{e^{\sigma_0}_{I_0}} A^{i_0},V^{2r}_{R_x}\right\}.
\end{align}
Hence, we have shown the two identities in \eqref{eq:App2Id}. We realise again how crucial the factor $2$ whose relevance will be discussed in detail in \cite{GTtoAppear} is in order to get the correct semiclassical limit.

\section{Comparison of the KCHF procedure and the one used by Sahlmann and Thiemann applied to the U(1)~case}
\label{appendix:ComparisonProcedures}

The aim of this appendix is to show how the procedure used by Sahlmann and Thiemann in \cite{Towards2} is linked at least in the U(1) case with the KCHF approach presented in the work at hand. We choose the following integral as the starting point:
\begin{align}
    \mathfrak{I} = \norm{\Psi}^{-2} T^r \e{\frac{p^2}{T^2}} \infint x \es{-x^2} \lrabs{x+\frac{p}{T}}^r, \label{eq:appendix:ComparisonStart}
\end{align}
where the norm of the state is essentially $\norm{\Psi}^2 = \sqrt{\pi}\es{\frac{p^2}{T^2}}$.\footnote{I.e. we only consider the $N=0$ contribution right from the start, as motivated by \cite{Towards2}.}\\
\par
\underline{The method used by Sahlmann and Thiemann in \cite{Towards2}:}\\
The procedure used by Sahlmann and Thiemann then applies the power series expansion of the absolute value:
\begin{align}
    \lrabs{x+\frac{p}{T}}^r = \lrabs{\frac{p}{T}}^r\cdot\lrabs{1+p\inv Tx}^r \approx \lrabs{\frac{p}{T}}^r \lr{1+rp\inv Tx + \frac{r(r-1)}{2}\lr{p\inv Tx}^2 + \mathcal{O}\lr{T^3}}. \label{eq:AppendixTaylor}
\end{align}
Inserting this into \eqref{eq:appendix:ComparisonStart}, we see that we can directly neglect the term $\sim x$ as the integration against the Gaussian vanishes. We then obtain
\begin{align}
    \mathfrak{I} &= \lrabs{p}^r \lr{1 + \frac{1}{\sqrt{\pi}} \frac{r(r-1)}{2} \frac{T^2}{p^2} \infint x\es{-x^2}x^2 + \mathcal{O}\lr{T^3} } \label{eq:AppendixSeriesST} \\
    & = \lrabs{p}^r \lr{1 + \frac{r(r-1)}{4} \frac{T^2}{p^2}+ \mathcal{O}\lr{T^3} }
\end{align}\\
\par
\underline{The KCHF method:}\\
Going the KCHF way means
\begin{align}
    \mathfrak{I} & = \norm{\Psi}^{-2} T^r \e{\frac{p^2}{T^2}} \infint x \es{-\lr{x-\frac{p}{T}}^2} \lrabs{x}^r \\
    & = \frac{T^r}{\sqrt{\pi}} \Gamma\lr{\tfrac{r+1}{2}} \kchf{-\rhalf,\frac{1}{2},-\lr{\frac{p}{T}}^2} \\
    & = \frac{T^r}{\sqrt{\pi}} \Gamma\lr{\tfrac{r+1}{2}} \es{-\frac{p^2}{T^2}} \kchf{\frac{r+1}{2},\frac{1}{2},\lr{\frac{p}{T}}^2} \\
    & \simeq \lrabs{p}^r \sum_{n=0}^\infty \frac{\lr{-\rhalf}_n \lr{-\rhalf+\frac{1}{2}}_n}{n!} \lr{p\inv T}^{2n} \label{eq:AppendixKummerExpansion} \\
    & = \lrabs{p}^r \lr{1 + \frac{r(r-1)}{4}\frac{T^2}{p^2} + \mathcal{O}\lr{T^3}}.
\end{align}
Therein, we used the Kummer transformation \eqref{KummerTrafo} from the second to the third line and the asymptotic expansion \eqref{expansion} from the third to the fourth line, realising that again only one of its series contributes due to the Gaussian prefactor.

While we already see that both ways result in the same expression, we may compare them better by reformulating the Taylor series in the course of the procedure used by Sahlmann and Thiemann in \cite{Towards2}, \eqref{eq:AppendixTaylor}. The numerical prefactors in there are of course the generalised binomial coefficients
\begin{align}
    \binom{r}{n} = \frac{\lr{r}_n}{n!}.
\end{align}
As only even $n$ survive the integration against the Gaussian and to better adapt it to the sum that arises during the KCHF method due to the asymptotic expansion, we go over to using $2n$ and can then associate the numerical prefactors of the two procedures (as in \eqref{eq:AppendixSeriesST} and \eqref{eq:AppendixKummerExpansion}) via
\begin{align}
    \frac{1}{\sqrt{\pi}} \frac{\lr{r}_{2n}}{\lr{2n}!} \infint{x} \es{-x^2} x^{2n} = \frac{\lr{-\frac{r}{2}}_n \lr{-\rhalf+\frac{1}{2}}_n}{n!} .
\end{align}
This equality can be shown to be correct by using 
\begin{equation}
    \infint{x}\es{-x^2}x^{2n} = \Gamma\lr{n+\tfrac{1}{2}} = \frac{\lr{2n}!}{n!4^n}\sqrt{\pi}.
\end{equation}

All the above then, ultimately, allows us to rewrite the asymptotic expansion of the KCHF by means of the power series expansion in terms of contributions used by Sahlmann and Thiemann,
\begin{align}
    T^r\e{-\frac{p^2}{T^2}}\kchf{-\frac{r+1}{2},\frac{1}{2},\frac{p^2}{T^2}} \simeq \lrabs{p}^r \cdot \sum_{n=0}^\infty \frac{1}{\sqrt{\pi}}\frac{\lr{r}_{2n}}{\lr{2n}!}\lr{p\inv T}^{2n}  \infint{x}\es{-x^2}x^{2n},
\end{align}
and thereby show that the two paths are just different routes one can pursue in \textit{this} scenario. If one wants to consider the case of small $p$, however, factoring out $p$ can cause trouble as then $p\inv$ becomes large and thereby destroying the smallness of the term $p\inv Tx$ of the approach followed by Sahlmann and Thiemann. KCHFs, meanwhile, can then just be evaluated by their defining series \eqref{eq:defKCHF1}, which again results in a power series for small arguments.

However, going over to the higher-dimensional case of U$(1)^3$, we lose track of such a close link between the two approaches. This is mainly because of the KCHF method performing the intertwined integrations iteratively, while the procedure by Sahlmann and Thiemann in \cite{Towards2} evaluates the more unravelled integrals that result from the power series expansion all at once.

\newpage
\bibliography{CoherentStates}
\bibliographystyle{unsrtnat} 

\end{document}